\documentclass[prb,twocolumn,english,superscriptaddress,amsmath,amssymb,floatfix]{revtex4-1}
\usepackage{float}
\usepackage{graphicx}
\usepackage{dcolumn}
\usepackage{bm}

\graphicspath{{graphs/}}

\begin{document}

\title{Anisotropic signatures of the electronic correlations in the electrical resistivity of UTe$_2$}

\author{T. Thebault}
 \affiliation{Laboratoire National des Champs Magn\'{e}tiques Intenses - EMFL, CNRS, Univ. Grenoble Alples, INSA-T, Univ. Toulouse 3, 31400 Toulouse, France}
\author{W. Knafo}
 \affiliation{Laboratoire National des Champs Magn\'{e}tiques Intenses - EMFL, CNRS, Univ. Grenoble Alples, INSA-T, Univ. Toulouse 3, 31400 Toulouse, France}
\author{M. Vali\v{s}ka}
 \affiliation{Univ. Grenoble Alpes, CEA, Grenoble INP, IRIG, PHELIQS, 38000, Grenoble, France}
 \affiliation{Charles University, Faculty of Mathematics and Physics, Department of Condensed Matter Physics, Ke Karlovu 5, Prague 2, 121 16, Czech Republic}
\author{G. Lapertot}
 \affiliation{Univ. Grenoble Alpes, CEA, Grenoble INP, IRIG, PHELIQS, 38000, Grenoble, France}
\author{A. Pourret}
 \affiliation{Univ. Grenoble Alpes, CEA, Grenoble INP, IRIG, PHELIQS, 38000, Grenoble, France}
\author{D. Aoki}
 \affiliation{Institute for Materials Research, Tohoku University, Ikaraki 311-1313, Japan}
\author{G. Knebel}
 \affiliation{Univ. Grenoble Alpes, CEA, Grenoble INP, IRIG, PHELIQS, 38000, Grenoble, France}
\author{D. Braithwaite}
 \affiliation{Univ. Grenoble Alpes, CEA, Grenoble INP, IRIG, PHELIQS, 38000, Grenoble, France}

\date{\today}

\begin{abstract}

Multiple unconventional superconducting phases are suspected to be driven by magnetic fluctuations in the heavy-fermion paramagnet UTe$_2$, and a challenge is to identify the signatures of the electronic correlations, including the magnetic fluctuations, in the bulk physical quantities. Here, we investigate thoroughly the anisotropy of the electrical resistivity of UTe$_2$ under intense magnetic fields up to 70~T, for different electrical-current and magnetic-field configurations. Two characteristic temperatures and an anisotropic low-temperature Fermi-liquid-like coefficient $A$, controlled by the electronic correlations, are extracted. Their critical behavior near the metamagnetic transition induced at $\mu_0H_m\simeq35$~T for $\mathbf{H}\parallel\mathbf{b}$ is characterized. Anisotropic scattering processes are evidenced and magnetic fluctuations are proposed to contribute, via a Kondo hybridization, to the electrical resistivity. Our work appeals for a microscopic modeling of the anisotropic contributions to the electrical resistivity as a milestone for understanding magnetically-mediated superconductivity in UTe$_2$.

\end{abstract}

\maketitle

\section{Introduction}
\label{Intro}

Magnetic fluctuations are suspected to be at the origin of the superconducting pairing mechanism in many unconventional superconductors \cite{Monthoux2007}. However, a difficulty is to reveal experimentally the intimate relationship between the magnetic and superconducting properties. The recent discovery of unconventional superconductivity in the paramagnetic compound UTe$_2$ [\onlinecite{Ran2019a},\onlinecite{Aoki2019b},\onlinecite{Aoki2022}] opened a new route to investigate the interplay between magnetism and unconventional superconductivity. A spin-triplet nature of superconducting pairing has been proposed for this compound initially presented as a nearly-ferromagnetic system. Spin-triplet superconductivity in UTe$_2$ is supported by the observation of a critical superconducting field exceeding the Pauli limitation expected for the three crystallographic directions \cite{Ran2019a,Ran2019b} and by nuclear-magnetic-resonance (NMR) knight-shift experiments made in the superconducting state \cite{Ran2019a,Nakamine2019,Fujibayashi2022}. Superconductivity is reinforced near a first-order metamagnetic transition induced at a magnetic field $\mu_0H_m=35$~T applied along the hard direction $\mathbf{b}$ [\onlinecite{Ran2019a},\onlinecite{Ran2019b},\onlinecite{Knafo2019},\onlinecite{Miyake2019},\onlinecite{Knebel2019}]. A phase transition between the low-field superconducting phase SC1 and a high-field superconducting phase SC2 was identified by heat capacity at ambient pressure \cite{Rosuel2022} and by tunnel-diode-oscillator technique under pressure \cite{Lin2020}. The phase SC2 suddenly collapses in fields higher than $H_m$ [\onlinecite{Ran2019b},\onlinecite{Knebel2019},\onlinecite{Knafo2021a}], where an abrupt Fermi-surface change was also reported \cite{Niu2020b}. The metamagnetic transition turns into a crossover at temperatures higher than $T_{CEP}\simeq5-7$~K, which corresponds to a critical end-point \cite{Knafo2019,Knafo2021a}. The crossover vanishes near the temperature $T_{\chi_b}^{max}=35$~K, where the magnetic susceptibility measured for $\mathbf{H}\parallel\mathbf{b}$ shows a broad maximum at low fields \cite{Ran2019a,Knafo2019,Miyake2019,Ikeda2006}. As in many heavy-fermion paramagnets, $H_m$ and $T_{\chi_b}^{max}$ delimitate a correlated paramagnetic (CPM) regime and a polarized paramagnetic regime is established for $H>H_m$ \cite{Aoki2013,Knafo2021c}. Field-induced reinforcement of superconductivity rapidly disappears when the magnetic field is tilted away from $\mathbf{b}$, and a second field-induced superconducting phase, labeled SC-PPM, develops in the PPM regime close to $\mu_0H_m\simeq45$~T in a magnetic field tilted by 30~$^{\circ}$ from $\mathbf{b}$ to $\mathbf{c}$ [\onlinecite{Knebel2019},\onlinecite{Ran2019b},\onlinecite{Knafo2021a}]. Multiple superconducting and magnetic phases have also been evidenced under pressure, leading to complex three-dimensional pressure - magnetic field - temperature phase diagrams illustrating the subtle interplay between magnetism and superconductivity in UTe$_2$ \cite{Braithwaite2019,Knebel2020,Ran2020,Thomas2020,Aoki2020,Lin2020,Li2021,Ran2021,Aoki2021,Valiska2021}.

\begin{figure*}[ht]
\includegraphics[width=\textwidth]{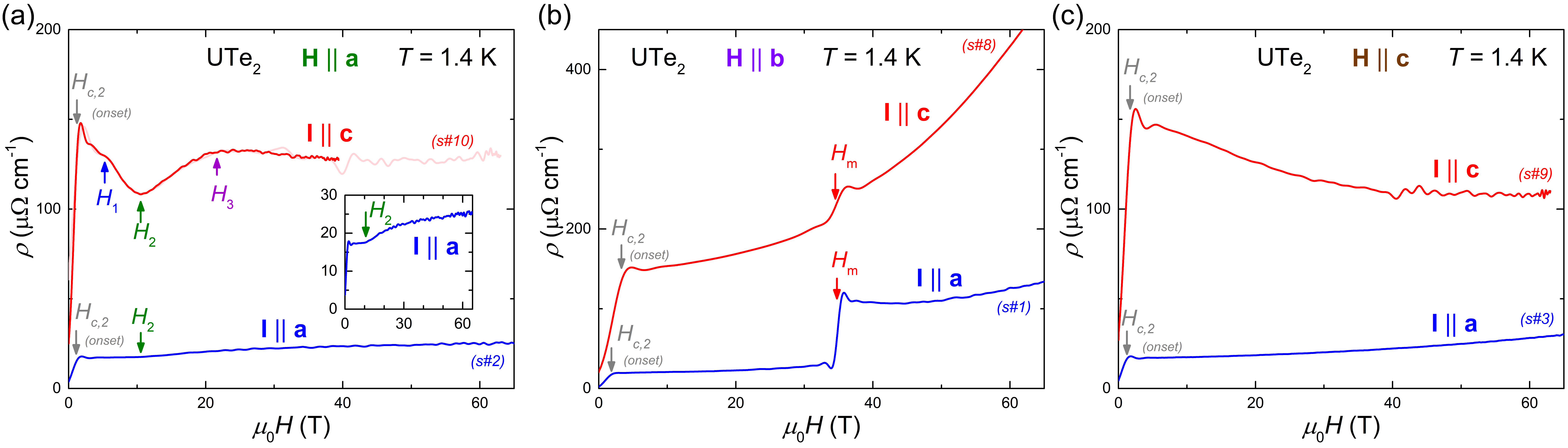}
\caption{\label{Fig1} Magnetic-field dependence of the electrical resistivities $\rho_{xx}$ and  $\rho_{zz}$, measured with currents $\mathbf{I}\parallel\mathbf{a}$ and $\mathbf{I}\parallel\mathbf{c}$, respectively, of UTe$_2$ at the temperature $T=1.4$~K under magnetic fields (a) $\mathbf{H}\parallel\mathbf{a}$,  (b) $\mathbf{H}\parallel\mathbf{b}$, (c) $\mathbf{H}\parallel\mathbf{c}$.}
\end{figure*}

\begin{figure*}[ht]
\includegraphics[width=01\textwidth]{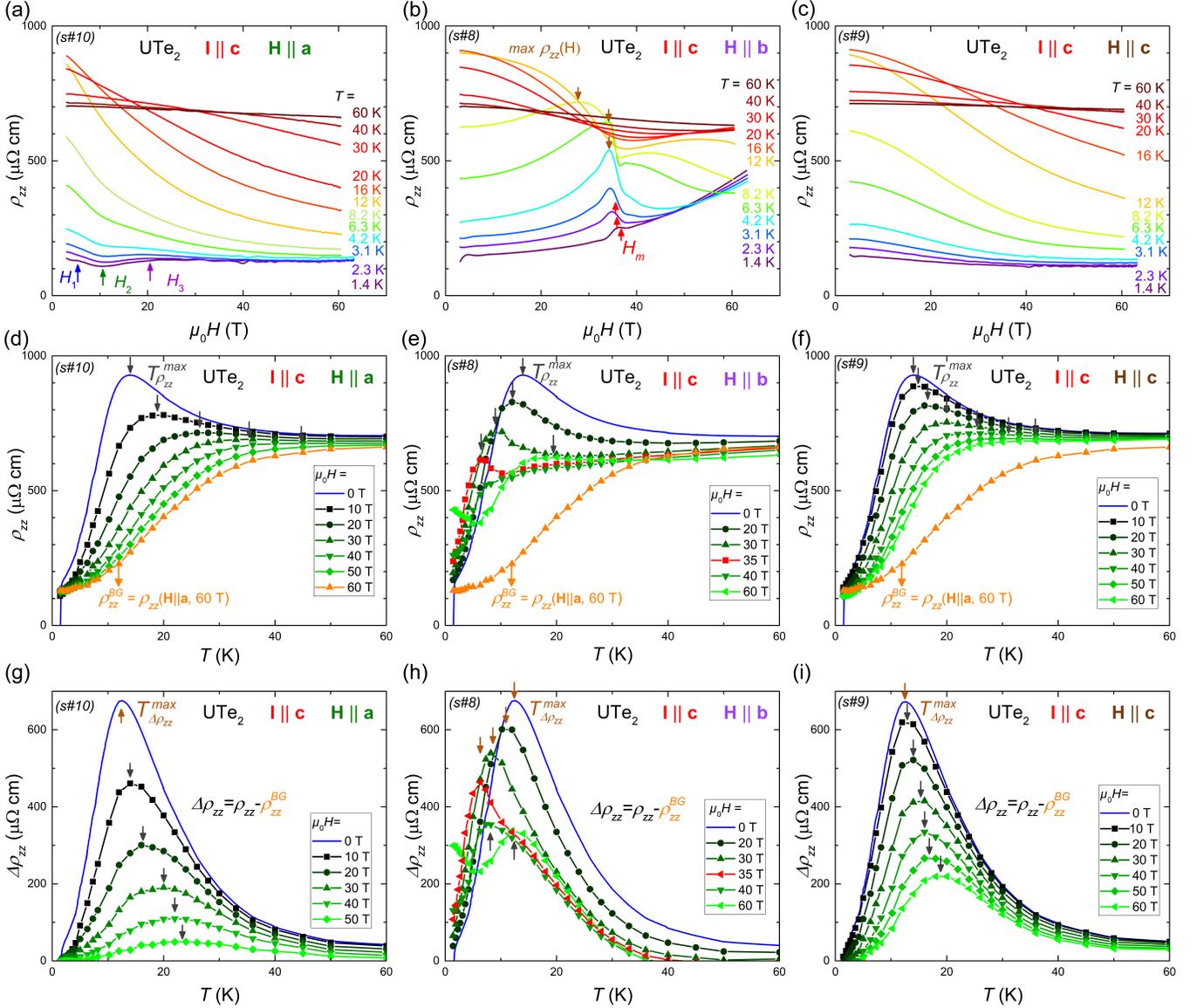}
\caption{\label{Fig2} Magnetic-field dependence of the electrical resistivity $\rho_{zz}$, measured with $\mathbf{I}\parallel\mathbf{c}$, of UTe$_2$ at temperatures from 1.4 to 60~K under magnetic fields (a) $\mathbf{H}\parallel\mathbf{a}$, (b) $\mathbf{H}\parallel\mathbf{b}$, and (c) $\mathbf{H}\parallel\mathbf{c}$. Temperature dependance of $\rho_{zz}$ at constant magnetic fields (d) $\mu_0\mathbf{H}\parallel\mathbf{a}$, (e) $\mu_0\mathbf{H}\parallel\mathbf{b}$, and (f) $\mu_0\mathbf{H}\parallel\mathbf{c}$ up to 60~T). Temperature dependance of $\Delta\rho_{zz}$, extracted after subtraction of a background estimated as $\rho_{zz}(\mathbf{H}\parallel\mathbf{a},60~\rm{T})$, at constant magnetic fields (g) $\mu_0\mathbf{H}\parallel\mathbf{a}$, (h) $\mu_0\mathbf{H}\parallel\mathbf{b}$, and (i) $\mu_0\mathbf{H}\parallel\mathbf{c}$ up to 60~T).}
\end{figure*}

\begin{figure*}[ht]
\includegraphics[width=\textwidth]{Fig3}
\caption{\label{Fig3} Magnetic-field dependence of the electrical resistivity $\rho_{xx}$, measured with $\mathbf{I}\parallel\mathbf{a}$, of UTe$_2$ at temperatures from 1.4 to 60~K under magnetic fields (a) $\mathbf{H}\parallel\mathbf{a}$, (b) $\mathbf{H}\parallel\mathbf{b}$, and (c) $\mathbf{H}\parallel\mathbf{c}$. Temperature dependance of $\rho_{xx}$ at constant magnetic fields (d) $\mu_0\mathbf{H}\parallel\mathbf{a}$, (e) $\mu_0\mathbf{H}\parallel\mathbf{b}$, and (f) $\mu_0\mathbf{H}\parallel\mathbf{c}$ up to 60~T. Temperature dependance of $\Delta\rho_{x}$, extracted after subtraction of a background estimated as $\rho_{xx}(\mathbf{H}\parallel\mathbf{b},49~\rm{T},2.25~\rm{GPa})$, at constant magnetic fields (g) $\mu_0\mathbf{H}\parallel\mathbf{a}$, (h) $\mu_0\mathbf{H}\parallel\mathbf{b}$, and (i) $\mu_0\mathbf{H}\parallel\mathbf{c}$ up to 60~T.}
\end{figure*}

Contrary to early expectations \cite{Ran2019a}, inelastic-neutron-scattering experiments showed the presence of low-dimensional antiferromagnetic fluctuations, but no indication for ferromagnetic fluctuations so far \cite{Duan2020,Knafo2021b,Butch2021}. These antiferromagnetic fluctuations, peaked at the incommensurate wavevector $\mathbf{k}_1=(0,0.57,0)$, saturate below the characteristic temperature $T_1^*=15$~K [\onlinecite{Knafo2021b}] and become gapped in the superconducting phase [\onlinecite{Duan2021},\onlinecite{Raymond2021}]. They may, therefore, play a role for the superconducting pairing mechanism. Electrical-resistivity measurements further revealed a broad maximum at the temperature $T_{\rho_{zz}}^{max}\simeq15$~K for a current $\mathbf{I}\parallel\mathbf{c}$ and the possible role of magnetic fluctuations was emphasized \cite{Eo2021}. Higher temperature scales $T_{\rho_{xx}}^{max}\simeq T_{\rho_{yy}}^{max}\simeq60-70$~K were also revealed at broad maxima of the electrical resistivity measured with $\mathbf{I}$ applied along $\mathbf{a}$ and $\mathbf{b}$, respectively \cite{Eo2021}. The anisotropy of the electrical resistivity presumably results from the combination of anisotropic magnetic and Fermi-surface properties. Anisotropic magnetic fluctuations from moments along the easy magnetic axis $\mathbf{a}$ were evidenced by NMR relaxation-rate measurements \cite{Tokunaga2019}, and cylindrical Fermi surfaces along the direction $\mathbf{c}$ expected from electronic-structure calculations have been confirmed by angle-resolved-photo-emission spectroscopy \cite{Xu2019,Ishizuka2019,Miao2020}. A challenge is now to determine how the magnetic fluctuations and the band structure are modified in a magnetic field, in particular when field-induced superconductivity is stabilized.

In this work, we focus on a systematic investigation of the electrical resistivity of UTe$_2$ under high magnetic fields $\mu_0\mathbf{H}$ up to 70~T applied along the three crystallographic directions $\mathbf{a}$, $\mathbf{b}$, and $\mathbf{c}$. We compare new sets of data corresponding to the configuration $\mathbf{I}\parallel\mathbf{c}$ to data with the configuration $\mathbf{I}\parallel\mathbf{a}$, published initially in [\onlinecite{Knafo2019}] and re-analyzed here. The evolution of the resistivity maxima observed for $\mathbf{I}\parallel\mathbf{a},\mathbf{c}$ are characterized. Two temperatures characterizing the electronic correlations are identified and their magnetic-field evolutions are determined, allowing to construct magnetic-field-temperature phase diagrams for the three directions of magnetic field. At low temperatures, Fermi-liquid-like fits to the resistivity data let us extracting the quadratic coefficient $A$, whose anisotropic behavior in high magnetic field is evidenced. The possible roles played by magnetic fluctuations and Kondo hybridization are discussed.

\section{Methods}
\label{Methods}

Single crystals were grown by chemical vapor deposition \cite{Ran2019a,Aoki2019b}. Their orientation was checked at room temperature using a Laue diffractometer. Oriented plates were prepared using an electrical-spark cutter. 15-$\mu$m gold wires have been spot-welded along the larger dimension of each sample. Samples measured with an electrical current $\mathbf{I}\parallel\mathbf{c}$ had dimensions of approximately 1-2~mm along $\mathbf{c}$, 0.3-0.5~mm along $\mathbf{a}$ and 0.1-0.3~mm along $\mathbf{b}$. Samples measured with an electrical current $\mathbf{I}\parallel\mathbf{a}$ had dimensions of approximately 1-2~mm along $\mathbf{a}$, 0.5~mm along $\mathbf{b}$ and 0.1-0.3~mm along $\mathbf{c}$. Electrical-resistivity measurements have been performed using 70-T pulsed-field magnets at the Laboratoire National des Champs Magn\'{e}tiques Intenses in Toulouse. A 6-MJ generator was used to generate pulses of 30-ms-rise and 150-ms-fall durations. Temperature and magnetic-field variations of the resistivity are presented for the up-sweep and the variations of the quadratic coefficient $A$ are presented for both up- and down-sweeps (more data from up- and down-sweeps are shown in the Supplemental Material \cite{SM}). Pulsed-field experiments were performed at constant temperatures from 1.4 to 80~K using a $^4$He cryostat. Resistivity was measured by the four-point technique, with electrical currents at frequencies between 20-40~kHz and a digital lock-in analysis. For each current direction, measurements under magnetic field directions $\mathbf{H}\parallel\mathbf{a}$, $\mathbf{b}$, and $\mathbf{c}$ were made on three samples simultaneously (samples $\sharp1$, $\sharp2$, and $\sharp3$ with $\mathbf{I}\parallel\mathbf{a}$ and samples $\sharp8$, $\sharp9$, and $\sharp10$ with $\mathbf{I}\parallel\mathbf{c}$). The resistivity data were normalized to absolute values following measurements made at the CEA-Grenoble on samples with well-defined geometrical shapes \cite{Knebel2022}. The resistivity data of samples with $\mathbf{I}\parallel\mathbf{a}$, initially published in [\onlinecite{Knafo2019}], have been reanalyzed using a geometric factor consistent with data published in [\onlinecite{Knafo2021a}].

\section{Results}
\label{Results}

\subsection{Low-temperature electrical resistivity}
\label{Low_T}

Fig. \ref{Fig1} compares the electrical resistivity versus magnetic field of UTe$_2$ at $T=1.4$~K for $\mathbf{I}\parallel\mathbf{a}$ and $\mathbf{c}$, for the three magnetic field directions $\mathbf{H}\parallel\mathbf{a}$, $\mathbf{b}$, and $\mathbf{c}$. The temperature $T=1.4$~K is slightly below the superconducting temperature $T_{sc}\simeq1.6-1.7$~K and all curves show the superconducting-to-normal-state transition at an upper critical field $\mu_0H_{c2}$ of a few T, defined here at the kink preceding the restoration of the normal state. Fig. \ref{Fig1}(a) shows that, in a magnetic field $\mathbf{H}\parallel\mathbf{a}$, three anomalies characterize broad crossovers for $\mathbf{I}\parallel\mathbf{c}$ at the magnetic fields $\mu_0H_1\simeq6$~T, $\mu_0H_2\simeq10$~T and $\mu_0H_3\simeq20$~T, in agreement with a previous report \cite{Niu2020a}, while only a broad kink is observed at $H_2$ for $\mathbf{I}\parallel\mathbf{a}$. Fig. \ref{Fig1}(b) shows that, in a magnetic field $\mathbf{H}\parallel\mathbf{b}$, the first-order metamagnetic transition at $\mu_0H_m=35$~T induces a much larger jump for $\mathbf{I}\parallel\mathbf{a}$ than for $\mathbf{I}\parallel\mathbf{c}$. Fig. \ref{Fig1}(c) shows that, for $\mathbf{H}\parallel\mathbf{c}$, beyond the superconducting transition, no anomaly is induced by a magnetic field for both current directions, and $\rho_{xx}$ monotonously increases with $H$ while $\rho_{zz}$ monotonously decreases with $H$. A $H^2$ increase of the resistivity found for two transverse configurations ($\mathbf{I}\parallel\mathbf{c}$, $\mathbf{H}\parallel\mathbf{b}$) and ($\mathbf{I}\parallel\mathbf{a}$, $\mathbf{H}\parallel\mathbf{c}$) is ascribed to field-induced contributions from charge carriers [see Figs. \ref{Fig1}(b-c) and Supplemental Material \cite{SM}]. In the following, we present a detailed study of the temperature dependence of electrical-resistivity-versus-magnetic-field data of UTe$_2$ measured within the six configurations considered here.

\subsection{\label{Ic}Electrical resistivity with $\mathbf{I}\parallel\mathbf{c}$}

Fig. \ref{Fig2} presents electrical-resistivity data of UTe$_2$ measured with a current $\mathbf{I}\parallel\mathbf{c}$, either as function of magnetic field for different temperatures from 1.4 to 60~K [panels (a-c)], or as function of temperature for different magnetic fields up to 60~T [panels (d-i)], for the three configurations with $\mathbf{H}\parallel\mathbf{a}$, $\mathbf{b}$, and $\mathbf{c}$.

For $\mathbf{H}\parallel\mathbf{a}$ [Fig. \ref{Fig2}(a)], the anomalies at the fields $H_1$, $H_2$ and $H_3$ show signatures at temperatures up to $T\approx5$~K (see Supplemental Material \cite{SM}). At temperatures $T>5$~K,  $\rho_{zz}$ monotonously decreases with $H$. For $\mathbf{H}\parallel\mathbf{b}$ [Fig. \ref{Fig2}(b)], the low-temperature step-like increase of $\rho_{zz}$ observed at the metamagnetic field $H_m$ changes into a decrease at higher temperatures. The resistivity jump is accompanied by a hysteresis emphasized in field-derivative $\partial\rho_{zz}/\partial H$ plots (shown for up- and down-sweeps in the Supplemental Material \cite{SM}). We lose the trace of the hysteresis at a temperature of 6~K, where the critical-end-point can be defined, and beyond which the metamagnetic first-order transition at $H_m$ characterized by a step variation in $\rho_{zz}$ turns into a crossover characterized by a broad maximum in $\rho_{zz}(H)$. The crossover field decreases with increasing temperature and disappears for $T\geq12$~K. In this transversal configuration, an increase of $\rho_{zz}$ is observed at low temperature for $\mu_0H\geq40$~T. In this regime, $\rho_{zz}$ follows a $H^2$ behavior, which is presumably controlled by a field-induced cyclotron motion of carriers (see Supplemental Material \cite{SM}). For $\mathbf{H}\parallel\mathbf{c}$ [Fig. \ref{Fig2}(c)], no anomalies are detected and the resistivity monotonously decreases at all temperatures.

Figs. \ref{Fig2}(d-f) show resistivity $\rho_{zz}$ versus $T$ data at constant magnetic field. These data were extracted from $\rho_{zz}$ versus $H$ data measured at constant temperature [see Figs. \ref{Fig2}(a-c)]. For $\mathbf{H}\parallel\mathbf{a}$ [Fig. \ref{Fig2}(d)] and $\mathbf{H}\parallel\mathbf{c}$ [Fig. \ref{Fig2}(f)], the maximum of $\rho_{zz}$ is shifted to higher temperature from $T_{\rho_{zz}}^{max}\approx14$~K at $H=0$ to $T_{\rho_{zz}}^{max}>40$~K at $\mu_0H\gtrsim50$~T. The maximal value of $\rho_{zz}(T)$ decreases when the magnetic field increases and the decrease is faster for $\mathbf{H}\parallel\mathbf{a}$ than for $\mathbf{H}\parallel\mathbf{c}$. For $\mathbf{H}\parallel\mathbf{b}$ [Fig. \ref{Fig2}(e)], the temperature $T_{\rho_{zz}}^{max}$ decreases for $H<H_m$, it cannot be defined at fields slightly higher than $H_m$, and it increases for $H>H_m$. The low-temperature increase of $\rho_{xx}$ at high fields, presumably controlled by the cyclotron motion of carriers, is emphasized in Fig. \ref{Fig2}(e).

\begin{figure*}[ht]
\includegraphics[width=\textwidth]{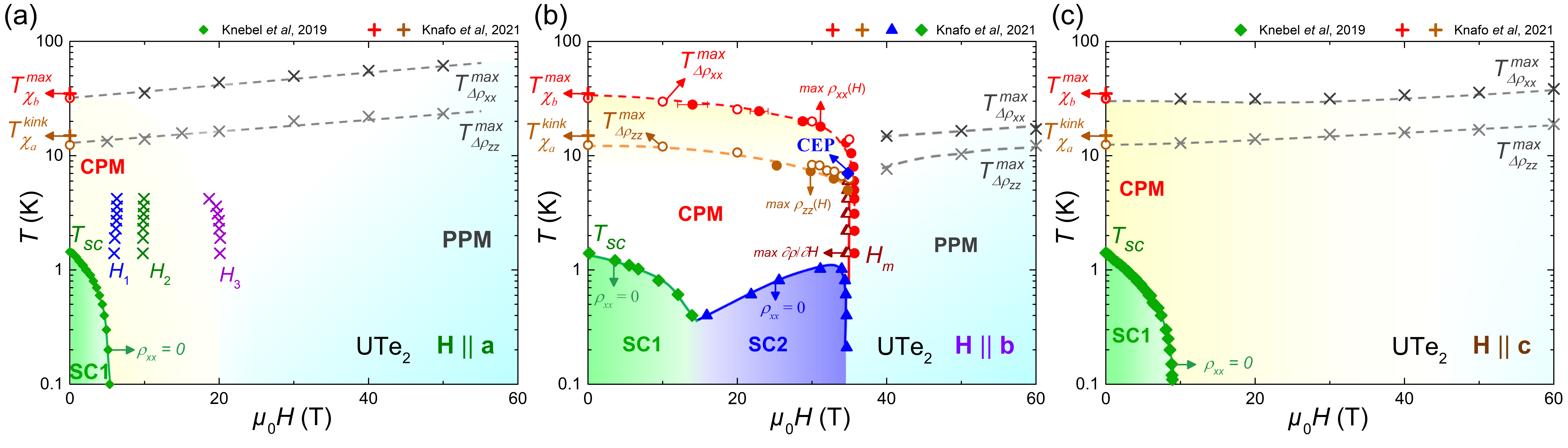}
\caption{\label{Fig4}Magnetic-field-temperature phase diagram of UTe$_2$ under a magnetic field: (a) $\mathbf{H}\parallel\mathbf{a}$, (b) $\mathbf{H}\parallel\mathbf{b}$, and (c) $\mathbf{H}\parallel\mathbf{c}$. CPM labels the correlated paramagnetic regime and PPM labels the polarized paramagnetic regime. SC1 labels the low-field superconducting phase and SC2 labels the superconducting phase induced by a magnetic field for $\mathbf{H}\parallel\mathbf{b}$. The superconducting boundaries correspond to the onset of zero resistivity and were extracted from [\onlinecite{Knebel2019}] in the insets (a,c), and from [\onlinecite{Knafo2021a}] in the inset (b), where a scaling of the magnetic-field scale by a factor 1.02 was applied (to correct a small misorientation between the different sets of data).}
\end{figure*}

For the three directions of magnetic field, a maximum in $\rho_{zz}$ versus $T$ cannot be defined in fields above 30-40~T. Here, we propose to subtract from $\rho_{zz}$ a background term $\rho_{zz}^{BG}$, estimated as $\rho_{zz}(\mathbf{H}\parallel\mathbf{a},60\rm{T})$ measured in a magnetic field  $\mu_0\mathbf{H}\parallel\mathbf{a}$ of 60~T. This background is characteristic of the high-field PPM regime. In a magnetic field $\mu_0\mathbf{H}\parallel\mathbf{a}$ of 60~T, a saturation of the low-temperature magnetization \cite{Miyake2021} indicates that most of the magnetic fluctuations have been quenched, which possibly drives the loss of the low-temperature contribution to the electrical resistivity observed here (see Supplemental Material \cite{SM} and Footnote \footnote{\label{note} The pertinence of the background subtractions done here is supported by the findings i) that the fields at the maxima of $\rho_{zz}$ versus $H$ extracted at constant temperatures coincide with the temperatures $T_{\Delta\rho_{zz}}^{max}$ at the maxima of $\Delta\rho_{zz}$ versus $T$ extracted at constant fields and ii) that the fields at the maxima of $\rho_{xx}$ versus $H$ extracted at constant temperatures coincide with the temperatures $T_{\Delta\rho_{xx}}^{max}$ at the maxima of $\Delta\rho_{xx}$ versus $T$ extracted at constant fields (see phase diagrams in Fig. \ref{Fig4}). Similar background-substraction procedure was done to analyze $\rho_{xx}$ data measured under pressure combined with magnetic fields in Ref. [\onlinecite{Valiska2021}].}\newcounter{fnnumber}\setcounter{fnnumber}{\thefootnote}). Figs. \ref{Fig2}(g-i) show the temperature dependence of $\Delta\rho_{zz}$, for $\mathbf{H}\parallel\mathbf{a}$, $\mathbf{b}$, and $\mathbf{c}$, respectively, estimated after subtraction of the resistivity background $\rho_{zz}^{BG}$. The temperature $T_{\Delta\rho_{zz}}^{max}$ (= 12.5~K at $H=0$) defined at the maximum of $\Delta\rho_{zz}$ versus $T$ is slightly smaller than $T_{\rho_{zz}}^{max}$ (= 14~K at $H=0$) defined at the maximum of $\rho_{zz}$ versus $T$. For $\mathbf{H}\parallel\mathbf{a}$ [Fig. \ref{Fig2}(g)], the fast decrease of $\Delta\rho_{zz}$ with increasing $\mu_0H$ up to 50~T is accompanied by an increase of $T_{\Delta\rho_{zz}}^{max}$ by almost a factor two. For $\mathbf{H}\parallel\mathbf{c}$ [Fig. \ref{Fig2}(i)], a slower decrease of $\Delta\rho_{zz}$ is accompanied by a slower increase of $T_{\Delta\rho_{zz}}^{max}$. For $\mathbf{H}\parallel\mathbf{b}$ [Fig. \ref{Fig2}(h)], both the CPM regime for $H<H_m$ and the PPM regime for $H>H_m$ are characterized by a maximum of $\Delta\rho_{zz}$ versus $T$, and the metamagnetic transition at $H_m$ is accompanied by a minimal value of $T_{\Delta\rho_{zz}}^{max}$.

\subsection{\label{Ia}Electrical resistivity with $\mathbf{I}\parallel\mathbf{a}$}

Fig. \ref{Fig3} presents electrical-resistivity data of UTe$_2$ measured with a current $\mathbf{I}\parallel\mathbf{a}$, either as function of magnetic field for different temperatures from 1.4 to 80~K [panels (a-c)], or as function of temperature for different magnetic fields up to 60~T [panels (d-i)], for the three configurations with $\mathbf{H}\parallel\mathbf{a}$, $\mathbf{b}$, and $\mathbf{c}$ (data initially published in [\onlinecite{Knafo2019}] and re-analyzed here).

Figs. \ref{Fig3}(a-c) show $\rho_{xx}$ versus $H$ measured at constant temperatures with a magnetic field along $\mathbf{a}$, $\mathbf{b}$, and $\mathbf{c}$, respectively. For $\mathbf{H}\parallel\mathbf{a}$ [Fig. \ref{Fig3}(a)], a kink is observed at $H_2$ for temperatures $T\lesssim4$~K (a zoom is provided in the Supplemental Material \cite{SM}). For $\mathbf{H}\parallel\mathbf{b}$ [Fig. \ref{Fig3}(b)], the step-like increase of $\rho_{xx}$ at $H_m$ turns into a broad maximum at temperatures above $T_{CEP}\simeq7$~K. This maximum shifts to lower magnetic fields when the temperature is further increased above $T_{CEP}$, before vanishing at temperatures $T>30$~K. The hysteresis of the first-order metamagnetic transition is characterized through $\partial\rho_{xx}/\partial H$ versus $H$ plots for up- and down-sweeps (see Supplemental Material \cite{SM}). Finally, for $\mathbf{H}\parallel\mathbf{c}$ [Fig. \ref{Fig3}(c)], the $\rho_{xx}$ versus $H$ curves monotonously increase for $T<5$~K and they monotonously decrease for $T>5$~K. For this transversal configuration, $\rho_{xx}$ follows a $H^2$ behavior at low temperature, which may possibly be controlled by a field-induced cyclotron motion of carriers (see Supplemental Material \cite{SM}). The low-temperature enhancement of $\rho_{xx}$ visible in fields  $\mu_0\mathbf{H}\parallel\mathbf{b}$ higher than 40~T [Fig. \ref{Fig3}(b)] may also be controlled by cyclotron motion of carriers.

Figs. \ref{Fig3}(d-f) show the resistivity $\rho_{xx}$ versus $T$ at constant magnetic fields extracted from field-scans of $\rho_{xx}$ at constant temperatures [see Figs. \ref{Fig3}(a-c)]. The temperature $T_{\rho_{xx}}^{max}$ at the maximum of $\rho_{xx}$ equals $\sim65$~K at zero field and increases with increasing magnetic fields $\mathbf{H}\parallel\mathbf{a}$ and $\mathbf{H}\parallel\mathbf{c}$. For $\mathbf{H}\parallel\mathbf{b}$, $T_{\rho_{xx}}^{max}$ decreases with $H$ for $H<H_m$ and increases with $H$ for $H>H_m$.

The maximum in $\rho_{xx}$ versus $T$ is not well-defined in high magnetic fields as the anomaly becomes broader and less intense. To characterize this crossover, we propose to subtract from $\rho_{xx}$ a background term $\rho_{xx}^{BG}$, estimated as the resistivity $\rho_{xx}(\mathbf{H}\parallel\mathbf{b},49\rm{T},2.25\rm{GPa})$ measured in a magnetic field $\mu_0\mathbf{H}\parallel\mathbf{b}$ of 49~T combined with a pressure of 2.25 GPa and published in [\onlinecite{Valiska2021}]. This background is representative from a state deep inside the PPM regime \cite{Li2021}, where most of the magnetic fluctuations and electronic correlations have been quenched (see Supplemental Material \cite{SM}). Figs. \ref{Fig3}(g-i) show $\Delta\rho_{xx}$ versus $T$ data for magnetic fields along $\mathbf{a}$, $\mathbf{b}$, and $\mathbf{c}$, respectively, extracted after subtraction of the estimated background $\rho_{xx}^{BG}$. The temperature $T_{\Delta\rho_{xx}}^{max}$ defined at the maximum of $\Delta\rho_{xx}$ versus $T$ is much smaller than $T_{\rho_{xx}}^{max}$ defined at the maximum of $\rho_{xx}$ versus $T$ ($T_{\Delta\rho_{xx}}^{max}\simeq35$~K~$<T_{\rho_{xx}}^{max}\simeq65$~K at $H=0$).\footnotemark[\thefnnumber] Under magnetic fields, $T_{\Delta\rho_{xx}}^{max}$ varies in a similar manner than $T_{\Delta\rho_{zz}}^{max}$ extracted in Section \ref{Ic}: it monotonously increases with $H$ for $\mathbf{H}\parallel\mathbf{a}$ [Fig. \ref{Fig3}(g)] and $\mathbf{H}\parallel\mathbf{c}$ [Fig. \ref{Fig3}(i)], the increase being faster for $\mathbf{H}\parallel\mathbf{a}$, and it passes through a minimal value at the metamagnetic field $H_m$ for $\mathbf{H}\parallel\mathbf{b}$ [Fig. \ref{Fig3}(h)]. For the three field directions, the amplitude of the anomaly in $\Delta\rho_{xx}$ is strongly reduced in high magnetic fields, being compatible with a field-induced loss of the electronic correlations.

\begin{figure*}[ht]
\includegraphics[width=\textwidth]{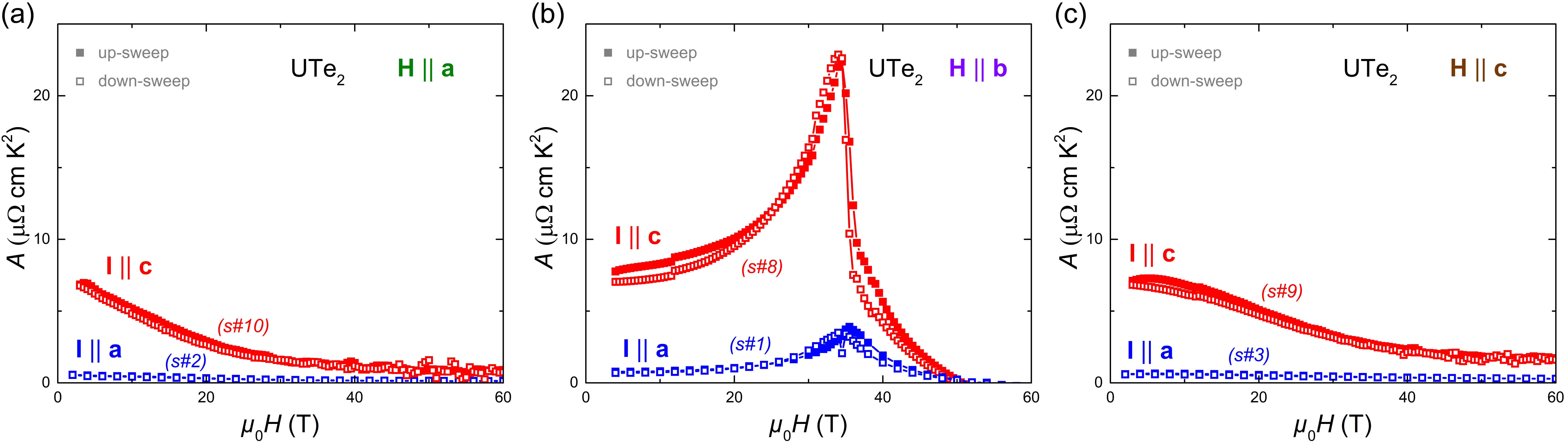}
\caption{\label{Fig5} Magnetic-field dependence of the quadratic coefficient $A$ extracted using $T^2$ fits to the electrical resistivities $\rho_{xx}$ and $\rho_{zz}$, measured respectively with currents $\mathbf{I}\parallel\mathbf{a}$ and $\mathbf{I}\parallel\mathbf{c}$, for (a) $\mathbf{H}\parallel\mathbf{a}$, (b) $\mathbf{H}\parallel\mathbf{b}$, and (c) $\mathbf{H}\parallel\mathbf{c}$.}
\end{figure*}

\subsection{\label{Phdiag}Phase diagrams and quantum critical properties}

Fig. \ref{Fig4} presents the magnetic-field-temperature phase diagrams of UTe$_2$ for $\mathbf{H}\parallel\mathbf{a}$, $\mathbf{b}$, and $\mathbf{c}$, constructed using data from this work and data from Refs. [\onlinecite{Knebel2019},\onlinecite{Knafo2021a},\onlinecite{Li2021}]. The zero-field values of $T_{\Delta\rho_{xx}}^{max}\simeq35$~K and $T_{\Delta\rho_{zz}}^{max}\simeq12.5$~K coincide with the temperatures $T_{\chi_{b}}^{max}\simeq35$~K and $T_{\chi_{a}}^{kink}\simeq15$~K, where anomalies are observed in the low-field magnetic susceptibilities $\chi_b$ and $\chi_a$, respectively [\onlinecite{Ikeda2006},\onlinecite{Li2021},\onlinecite{Knafo2021a},\onlinecite{Aoki2022}]. For the three magnetic-field directions, the two temperatures $T_{\Delta\rho_{xx}}^{max}$ and $T_{\Delta\rho_{zz}}^{max}$ further have similar field dependences, indicating related associated phenomena. Under a magnetic field $\mathbf{H}$ applied along the easy magnetic axis $\mathbf{a}$ [Fig. \ref{Fig4}(a)], the fast polarization of the magnetic moments is accompanied by three successive anomalies in the low-temperature electrical resistivity, at $\mu_0H_1\simeq6$~T, $\mu_0H_2\simeq10$~T and $\mu_0H_3\simeq20$~T. $H_1$ coincides with a maximum in the field-derivative of the magnetization \cite{Miyake2019} and a minimum in the thermoelectric power, which was identified as a signature of a Lifshitz transition \cite{Niu2020a}. The polarization of the magnetic moments is accompanied by a monotonous increase with $H$ of the two temperature scales $T_{\Delta\rho_{xx}}^{max}$ and $T_{\Delta\rho_{zz}}^{max}$ characterizing the electronic correlations. Under magnetic field $\mathbf{H}$ applied along the hard magnetic axis $\mathbf{c}$ [Fig. \ref{Fig4}(c)], a slower magnetic polarization \cite{Miyake2021} is accompanied by a larger upper critical field $\mu_0H_{c2}\simeq10$~T of the low-temperature superconducting phase, but there are no additional field-induced anomalies in resistivity or magnetization data. A monotonous increase with $H$ of $T_{\Delta\rho_{xx}}^{max}$ and $T_{\Delta\rho_{zz}}^{max}$ is also observed. For $\mathbf{H}\parallel\mathbf{b}$, which is the hardest magnetic axis [Fig. \ref{Fig4}(b)], $T_{\Delta\rho_{xx}}^{max}$ and $T_{\Delta\rho_{zz}}^{max}$ both decrease with $H$ and merge near to the critical end point at the temperature $T_{CEP}\simeq7$~K, below which a first-order transition at the metamagnetic field $\mu_0H_m\simeq35$~T marks the onset of the PPM regime. For $H<H_m$, $T_{\Delta\rho_{xx}}^{max}$ and $T_{\Delta\rho_{zz}}^{max}$ are characteristic temperature scales of the CPM regime. For $H>H_m$, $T_{\Delta\rho_{xx}}^{max}$ and $T_{\Delta\rho_{zz}}^{max}$ increase and can then be considered as characteristic temperature scales of the PPM regime. While the low-temperature superconducting phase SC1 is delimited by an upper critical field  $\mu_0H_{c2}\simeq15$~T [\onlinecite{Rosuel2022}], the field-induced superconducting phase SC2, stabilized near $H_m$, vanishes at fields higher than $H_m$, in the PPM regime.

Within a Fermi-liquid behavior, the electrical resistivity follows a quadratic temperature dependence  $\rho(T)=\rho_0+AT^2$, where $A$ is related to the effective mass $m^*$ by $A\propto m^{*2}$. In heavy-fermion compounds, the effective mass $m^*$ is large, typically of the order of hundred times the free-electron mass, and the enhancement of $m^*$ is generally driven by magnetic fluctuations, consequence of nearby quantum magnetic instabilities. The electrical resistivity of UTe$_2$, measured here at temperatures down to 1.4~K and in magnetic fields up to 70~T, was fitted by a Fermi-liquid formula for the six configurations of electrical current and magnetic field. Depending on the magnetic-field direction and strength, the temperature window of the fit was adjusted from 4.2~K down to 1.4~K in the absence of superconductivity and down to temperatures $>T_{SC}$ when superconductivity is established. Fig. \ref{Fig5} presents the magnetic-field variations of $A$ extracted for both current directions $\mathbf{I}\parallel\mathbf{a}$ and $\mathbf{I}\parallel\mathbf{c}$ and for the three magnetic-field directions $\mathbf{H}\parallel\mathbf{a}$, $\mathbf{H}\parallel\mathbf{b}$ and $\mathbf{H}\parallel\mathbf{c}$ \footnote{Details about the $T^2$ fits to the data are shown in the Supplemental Material for $\mathbf{I}\parallel\mathbf{c}$ and in the Supplementary Materials of Ref. [\onlinecite{Knafo2019}] for $\mathbf{I}\parallel\mathbf{a}$}. The coefficient $A$ is anisotropic and is an order magnitude larger for $\mathbf{I}\parallel\mathbf{c}$ than for $\mathbf{I}\parallel\mathbf{a}$. For $\mathbf{H}\parallel\mathbf{a}$ [Fig. \ref{Fig5}(a)] and $\mathbf{H}\parallel\mathbf{c}$ [Fig. \ref{Fig5}(c)], $A$ decreases monotonously for both current directions. For $\mathbf{H}\parallel\mathbf{b}$, $A$ presents a sharp maximum at $H_m$ for both current directions. A striking difference between the two current-direction configurations is visible: the maximum in the field variation of $A$ is symmetrical around $H_m$ for $\mathbf{I}\parallel\mathbf{a}$ but $A$ rapidly drops above $H_m$, by a factor 2 in a 2-T window, for $\mathbf{I}\parallel\mathbf{c}$. The anisotropy of $A$, which is strong in the CPM regime for $H<H_m$, is reduced in the PPM regime for $H>H_m$. In the high-field limit, contributions controlled by field-effects on the Fermi surface (for instance controlled by the field-induced cyclotron motion of carriers in transversal configurations) lead to deviations from a Fermi-liquid picture and to non-physical negative coefficients $A$.

\section {\label{Discussion}Discussion}

\subsection{Energy scales of the electronic correlations}

Two temperatures $T_{\Delta\rho_{xx}}^{max}\simeq35$~K and $T_{\Delta\rho_{zz}}^{max}\simeq12.5$~K characterizing the electronic correlations have been extracted from anomalies in the electrical resistivity with applied currents $\mathbf{I}\parallel\mathbf{a}$ and $\mathbf{I}\parallel\mathbf{c}$, respectively. A question is whether these characteristic energies are manifestations of one or two different phenomena. To address this question, a comparison with other physical quantities is of interest. We have seen that $T_{\Delta\rho_{xx}}^{max}$ and $T_{\Delta\rho_{zz}}^{max}$ coincide with the temperatures $T_{\chi_b}^{max}\simeq35$~K and $T_{\chi_a}^{kink}\simeq15$~K, at which a maximum in the magnetic susceptibility $\chi_b$ measured with $\mathbf{H}\parallel\mathbf{b}$ and a kink in the magnetic susceptibility $\chi_a$ measured with $\mathbf{H}\parallel\mathbf{a}$ [\onlinecite{Ikeda2006},\onlinecite{Knafo2021b}] (related with a minimum of $\partial\chi_a/\partial T$ [\onlinecite{Aoki2022}]), are respectively observed. Anomalies are also visible in other physical quantities at these two temperature scales: a maximum in the electronic heat capacity, a minimum in the thermal expansion measured with lengths $\mathbf{L}\parallel\mathbf{b},\mathbf{c}$ [\onlinecite{Thomas2021},\onlinecite{Willa2021}], and a minimum in the thermoelectric power measured with a current $\mathbf{I}\parallel\mathbf{a}$ [\onlinecite{Niu2020a}] were observed at a temperature $\simeq15$~K, while a maximum in the Hall effect measured with a current $\mathbf{I}\parallel\mathbf{a}$ and a magnetic field $\mathbf{H}\parallel\mathbf{b}$ was observed at a temperature $\simeq35$~K [\onlinecite{Niu2020b},\onlinecite{Niu2020a}]. These coincidences may support the picture of two characteristic temperatures related with two different energy scales \footnote{In Ref. [\onlinecite{Willa2021}], it has been alternatively proposed that the magnetic susceptibility $\chi_a$, the electronic heat capacity, and the thermal expansion should not be compared to $\rho_{xx}$ or $\rho_{zz}$, but that they should be compared to $\partial\rho_{xx}/\partial T$, where a maximum at $\simeq15$~K is also observed. Within this picture, a set of three energy scales may be needed to describe the transport and thermodynamic properties of UTe$_2$: the first one of $\simeq15$~K already identified, the second one of $\simeq7$~K defined at the maximum of $\partial\rho_{zz}/\partial T$, and the third one of $\simeq35$~K defined at the maximum of $\chi_b$. In a recent NMR investigation \cite{Tokunaga2022}, three temperatures scales were defined from the variation of the spin-spin-relaxation-rate $1/T_2$ measurements: $T_H=30$~K at the onset of low-temperature increase of $1/T_2$, $T_P=16$~K at a maximum of $1/T_2$, and $T_L=7$~K at the onset of a lower-temperature increase of $1/T_2$. These three temperatures were proposed to be respectively related with the temperatures $T_{\chi_b}^{max}$, $T^*$ and a third temperature $T_\mu=5$~K, below which the muon spin relaxation rate was found to increase \cite{Sundar2019}. }. In the following, we discuss the possible microscopic origin of the two energy scales identified here.

Signatures of fluctuating magnetic moments $\mathbf{\mu}\parallel\mathbf{a}$ associated with a characteristic temperature $T^*\simeq15-20$~K were observed by NMR relaxation-rate measurements [\onlinecite{Tokunaga2019},\onlinecite{Tokunaga2022}]. These fluctuations were found to develop progressively at temperatures below 30-40~K [\onlinecite{Tokunaga2022}]. Inelastic-neutron-scattering experiments have further shown that antiferromagnetic fluctuations, peaked at the incommensurate wavevector $\mathbf{k}_1=(0,0.57,0)$, saturate below a similar temperature $T_1^*\simeq15$~K [\onlinecite{Knafo2021b}]. A quasi-two-dimensional (2D) character of the magnetic fluctuations was found and related with the low-dimensional structure of the magnetic U atoms in UTe$_2$: two-legs ladders with legs along $\mathbf{a}$ and rungs along $\mathbf{c}$, these ladders being weakly coupled along $\mathbf{b}$ and not coupled along $\mathbf{c}$. These antiferromagnetic fluctuations may drive the anomalies at the first characteristic temperature of 15~K in a large set of physical properties, including $\rho_{zz}$ investigated here.

The relation between $T_{\chi_b}^{max}=35$~K and the metamagnetic field $\mu_0H_m=35$~T was emphasized \cite{Knafo2019,Miyake2019}, indicating a standard heavy-fermion-paramagnet behavior \cite{Aoki2013,Knafo2021c}. In several prototypical heavy-fermion systems, $T_\chi^{max}$ and $H_m$ are the boundaries (crossover or phase transition) delimiting a CPM regime, which is the place of antiferromagnetic fluctuations (see for instance CeRu$_2$Si$_2$ and CeCu$_6$ [\onlinecite{RossatMignod1988}]). Therefore, in UTe$_2$ we expect that the second characteristic temperature of $\simeq35$~K may be driven by antiferromagnetic fluctuations too. However, no magnetic-fluctuation mode with a temperature scale of $\simeq35$~K was observed so far by inelastic neutron scattering. A few scenarios can be considered. Within a first scenario, quasi-one-dimensional (1D) magnetic fluctuations from non-interacting ladders, i.e., controlled only by the two nearest-distance interactions within the ladders, would progressively develop near $\simeq35$~K. Quasi-1D fluctuations would then be transformed in quasi-2D fluctuations below 15~K, once the magnetic interaction between ladders would be activated. Within a second scenario, the 35-K anomaly may be related to a second magnetic-fluctuations mode with a different wavevector and/or a different moment direction. Within a third scenario, a single-site Kondo crossover would drive localized $f$ electrons at temperatures $T>35$~K to itinerant $f$ electrons at temperatures $T\lesssim35$~K (see proposition in Ref. [\onlinecite{Eo2021}]). New experiments are needed to precise which description is pertinent.

\subsection{Effect of a magnetic field}

The fact that, in a magnetic field $\mathbf{H}\parallel\mathbf{b}$, $T_{\Delta\rho_{xx}}^{max}$ and $T_{\Delta\rho_{zz}}^{max}$ collapse in a similar manner when $H_m$ is approached indicates that both are controlled by a common parameter \footnote{Similar falls of two temperatures scales, $T_\chi^{max}$ at the maximum of the magnetic susceptibility, and $T_0$ at the onset of an 'hidden-order' phase transition, were observed in the vicinity of metamagnetism in the heavy-fermion paramagnet URu$_2$Si$_2$ [\onlinecite{Knafo2020}].}. The phenomenon associated with the 35-K temperature scale is a precursor of the phenomenon, identified here as quasi-2D antiferromagnetic fluctuations, associated with the 15-K temperature scale. This may be compatible with the first and third scenarios mentioned above. For $\mathbf{H}\parallel\mathbf{b}$ and $H>H_m$, the PPM regime is also characterized by broad maxima in $\Delta\rho_{xx}$ and $\Delta\rho_{zz}$ at the temperatures $T_{\Delta\rho_{xx}}^{max}$ and $T_{\Delta\rho_{zz}}^{max}$, respectively, which both then increase with $H$. For $\mathbf{H}\parallel\mathbf{a}$ and $\mathbf{H}\parallel\mathbf{c}$, there is no metamagnetic transition and $T_{\Delta\rho_{xx}}^{max}$ and $T_{\Delta\rho_{zz}}^{max}$ monotonously increase with $H$. The change from the CPM to the PPM regimes is then smooth and very progressive.

Under a magnetic field $\mathbf{H}\parallel\mathbf{b}$, the coefficient $A$ passes through a maximum at $H_m$, indicating the presence of magnetic-field-induced critical magnetic fluctuations. The nature of these critical fluctuations is unknown. They could possibly be ferromagnetic, as observed at the metamagnetic transition of the heavy-fermion paramagnet CeRu$_2$Si$_2$  [\onlinecite{Raymond1998},\onlinecite{Flouquet2004},\onlinecite{Sato2004}], or antiferromagnetic, as observed at the metamagnetic transition of Sr$_3$Ru$_2$O$_7$ [\onlinecite{Lester2021}]. The strong anisotropy of $A$, which is an order of magnitude larger for $\mathbf{I}\parallel\mathbf{c}$ than for $\mathbf{I}\parallel\mathbf{a}$ at zero field, indicates an anisotropic scattering process of the conduction electrons by the fluctuating magnetic moments. Interestingly, for $\mathbf{H}\parallel\mathbf{b}$ the field-variation of $A$ measured with a current $\mathbf{I}\parallel\mathbf{c}$ has a similar asymmetric variation around $H_m$ than the field variation of the Sommerfeld coefficient extracted from heat-capacity measurements \cite{Rosuel2022}. Therefore, the coefficient $A$ from the configuration with  $\mathbf{I}\parallel\mathbf{c}$ may capture the magnetic-fluctuations phenomena driving the entropy change near $H_m$.

The anisotropy of $A$ is partly released in the PPM regime reached beyond $H_m$. For both current directions, a faster decrease of $A$ observed for $\mathbf{H}\parallel\mathbf{a}$ than for $\mathbf{H}\parallel\mathbf{c}$ indicates a faster quench of the magnetic fluctuations. In relation with the decrease of $A$, the intensity of the resistivity anomalies in $\Delta\rho_{xx}$ and $\Delta\rho_{zz}$ decrease with $H$ due to the field-induced quench of the magnetic fluctuations. This is consistent with the observation by magnetization measurements of a faster magnetic polarization for $\mathbf{H}\parallel\mathbf{a}$ than for $\mathbf{H}\parallel\mathbf{c}$ [\onlinecite{Miyake2021}].

\subsection{Perspectives}

For a deeper understanding, a modeling of the contribution from magnetic fluctuations to electrical-transport properties is mandatory. Several theoretical studies have examined the contribution of magnetic fluctuations to the low-temperature resistivity regime. In an early work from Mills and Lederer, the contribution to the electrical resistivity, via $s-d$ exchange, of magnetic fluctuations from localized electrons was modeled and was shown to lead to a $T^2$ Fermi-liquid behavior \cite{Mills1966}. A bit later, Jullien, B\'eal-Monod, and Coqblin described the contribution from magnetic fluctuations to the electrical resistivity over a broad temperature range \cite{Jullien1974}. Further works analyzed the dominant contribution of critical magnetic fluctuations near a quantum magnetic phase transition at low temperatures, leading to a non-fermi-liquid behavior in the electrical resistivity (see for instance Refs. [\onlinecite{Moriya1995}]). The effect of disorder and of Fermi-surface hot spots in nearly-antiferromagnets was also emphasized \cite{Rosch1999}. These approaches were however essentially done for isotropic cases. The challenge to reproduce the anisotropic features reported in the electrical resistivity and in the magnetic susceptibility of UTe$_2$ requires a new generation of modeling, possibly based on anisotropic Kondo hybridization processes \cite{Cooper1985}, and taking into account the magnetic and Fermi-surface anisotropies. In particular, we have seen here that, at ambient pressure, $\rho_{zz}$ seems to capture the 15-K anomaly revealed in $\chi_a$, while $\rho_{xx}$ seems to capture the 35-K anomaly revealed in $\chi_b$. Under pressures higher than the critical pressure $p_c\simeq1.5-1.7$~GPa, it was also shown that $\rho_{xx}$ captures an anomaly at $\simeq15-20$~K, which is possibly related with maxima in $\chi_a$ or in $\chi_c$ [\onlinecite{Li2021},\onlinecite{Valiska2021},\onlinecite{Kinjo2022}]. The possibility to quantitatively, or at least qualitatively, describe the anisotropic contributions from the magnetic fluctuations to the electrical resistivity may be useful to understand their relationship with the stabilization of multiple unconventional superconducting phases in UTe$_2$. Indeed, magnetic fluctuations are suspected to contribute to the superconducting pairing mechanisms in UTe$_2$, and magnetic-field- or pressure-induced modifications of the nature and strength of the magnetic fluctuations may be related to the stabilization or destabilization of the superconducting phases. The anisotropy of the critical field $H_{c2}$ of the low-field superconducting phase SC1 was found to be inversely related with the anisotropy of the low-temperature magnetic susceptibility \cite{Knafo2021a}: for instance, for $\mathbf{H}\parallel\mathbf{a}$, $H_{c2}$ is minimal and the low-temperature magnetic susceptibility $\chi$ is maximal, the magnetic polarization is faster (evidenced from magnetization measurements \cite{Miyake2019,Miyake2021}) and the quench of the low-temperature contribution to $\rho_{zz}$ and $\rho_{xx}$ is reached at lower fields. These phenomena are presumably related to the fast quench of magnetic fluctuations. The anisotropic variation of the coefficient $A$ in a magnetic field $\mathbf{H}\parallel\mathbf{b}$ is also presently unexplained and may result from peculiar field-induced variations of the magnetic excitation spectrum, with possible feedback to superconductivity. The modeling of the magnetic fluctuations and their anisotropic contributions to the electrical resistivity of a real crystal, as the unconventional superconductor UTe$_2$ studied here, may request considering the anisotropy of the electronic bands and constitutes a theoretical challenge for the coming years.

\section*{Acknowledgements}

We acknowledge financial support from the French National Research Agency collaborative research project FRESCO No. ANR-20-CE30-0020 and from the JSPS KAKENHI Grants Nos. JP19H00646, 19K03756, JP20H00130, 20H01864, JP20K20889, JP20KK0061, and 21H04987.

%\bibliography{UTe2_Thebault_2022}

%merlin.mbs apsrev4-1.bst 2010-07-25 4.21a (PWD, AO, DPC) hacked
%Control: key (0)
%Control: author (8) initials jnrlst
%Control: editor formatted (1) identically to author
%Control: production of article title (-1) disabled
%Control: page (0) single
%Control: year (1) truncated
%Control: production of eprint (0) enabled
%

\newpage

\onecolumngrid

\setcounter{figure}{0}
\setcounter{page}{1}
\renewcommand{\theequation}{S\arabic{equation}}
\renewcommand{\thefigure}{S\arabic{figure}}
\renewcommand{\thetable}{S\arabic{table}}
\renewcommand{\thepage}{S\arabic{page}}
\renewcommand{\bibnumfmt}[1]{[S#1]}
\renewcommand{\citenumfont}[1]{S#1}

\vspace{15cm}
\begin{center}
\large {\textbf {Supplemental Material:\\ Anisotropic signatures of the electronic correlations in the electrical resistivity of UTe$_2$}}
\end{center}
\vspace{1cm}

In this Supplemental Material, we present complementary graphs about the electrical-resistivity data accumulated for this study. Details about the construction of the phase diagrams and about the extraction of the Fermi-liquid coefficient $A$ are given.

\newpage

\textbf{Contribution to the electrical resistivity from field-induced motion of carriers.}

\begin{figure*}[t]
\includegraphics[width=\textwidth]{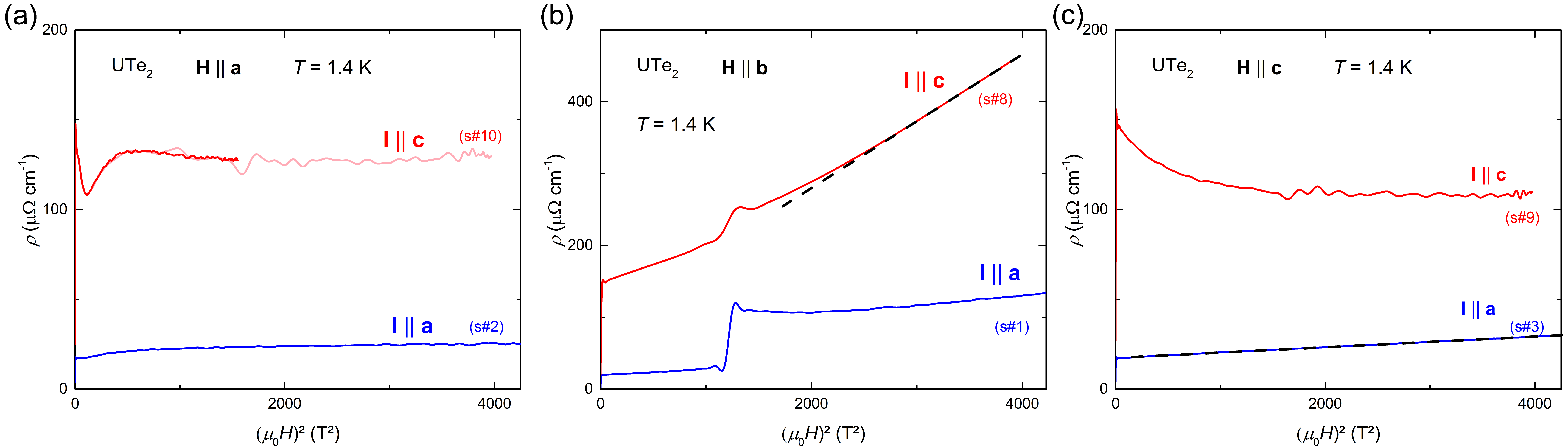}
\caption{\label{FigS1} Electrical resistivities $\rho_{xx}$ and  $\rho_{zz}$, measured with currents $\mathbf{I}\parallel\mathbf{a}$ and $\mathbf{I}\parallel\mathbf{c}$, respectively, of UTe$_2$ versus square of the magnetic field $H^2$ at the temperature $T=1.4$~K under magnetic fields (a) $\mathbf{H}\parallel\mathbf{a}$,  (b) $\mathbf{H}\parallel\mathbf{b}$, (c) $\mathbf{H}\parallel\mathbf{c}$. Dashed lines indicates $H^2$ variations observed for the configurations $\mathbf{I}\parallel\mathbf{c}$,$\mathbf{H}\parallel\mathbf{b}$ and  $\mathbf{I}\parallel\mathbf{a}$,$\mathbf{H}\parallel\mathbf{c}$.}
\end{figure*}

Fig. \ref{FigS1} presents plots of the electrical resistivity $\rho$ versus the square of the magnetic field $H^2$ at $T=1.4$~K, for both current directions $\mathbf{I}\parallel\mathbf{a}$ and $\mathbf{I}\parallel\mathbf{c}$ and the three field directions  $\mathbf{H}\parallel\mathbf{a}$ [Inset (a)],  $\mathbf{H}\parallel\mathbf{b}$ [Inset (b)], $\mathbf{H}\parallel\mathbf{c}$ [Inset (c)]. A high-field $H^2$ variation of the electrical resistivity is observed for the two transverse configurations $\mathbf{I}\parallel\mathbf{c}$, $\mathbf{H}\parallel\mathbf{b}$ and $\mathbf{I}\parallel\mathbf{a}$, $\mathbf{H}\parallel\mathbf{c}$. A high-field increase of the electrical resistivity, which does not follow a $H^2$ law, is also observed for the transverse configuration $\mathbf{I}\parallel\mathbf{a}$, $\mathbf{H}\parallel\mathbf{b}$.

\begin{table}[t]
\begin{center}
\caption{High-field resistivity for different magnetic-field and electrical-current configurations, expected for a cylindrical Fermi surface \cite{Onuki2018_sm} and observed here for UTe$_2$.}
\begin{tabular}{c|c|c|c}
  \hline
  Magnetic field & Electrical current & Expected for a cylindrical Fermi surface & Observed for UTe$_2$\\
  \hline
  $\mathbf{H}\parallel\mathbf{a}$ & $\mathbf{I}\parallel\mathbf{a}$ & saturation & saturation \\
  & $\mathbf{I}\parallel\mathbf{b}$ & saturation & \textit{(not measured)} \\
  & $\mathbf{I}\parallel\mathbf{c}$ & $H^2$ & saturation \\
  \hline
  $\mathbf{H}\parallel\mathbf{b}$ & $\mathbf{I}\parallel\mathbf{a}$ & saturation & small increase \\
  & $\mathbf{I}\parallel\mathbf{b}$ & saturation & \textit{(not measured)} \\
  & $\mathbf{I}\parallel\mathbf{c}$ & $H^2$ & $H^2$ \\
  \hline
  $\mathbf{H}\parallel\mathbf{c}$ & $\mathbf{I}\parallel\mathbf{a}$ & $H^2$ & $H^2$ \\
  & $\mathbf{I}\parallel\mathbf{b}$ & $H^2$ & \textit{(not measured)} \\
  & $\mathbf{I}\parallel\mathbf{c}$ & saturation & saturation \\
  \hline
\end{tabular}
\label{table}
\end{center}
\end{table}

Knowing that UTe$_2$ is a compensated metal and assuming that cylindrical Fermi surfaces along the direction $\mathbf{c}$ dominate the high-field electrical resistivity \cite{Xu2019_sm,Ishizuka2019_sm,Miao2020_sm}, one can expect $H^2$ variations for certain transverse magnetic-field and electrical-current configurations \cite{Onuki2018_sm}:
\begin{itemize}
  \item For $\mathbf{H}\parallel\mathbf{c}$, cyclotron orbits have their axis $\parallel\mathbf{H}$ and appear as closed orbits. In a high magnetic field, the contribution from carriers to the resistivity in the transverse configurations with $\mathbf{I}\parallel\mathbf{a},\mathbf{b}$ is expected to follow a $H^2$ law, while that in the longitudinal configuration with $\mathbf{I}\parallel\mathbf{c}$ is expected to saturate.
  \item For $\mathbf{H}\perp\mathbf{c}$, cyclotron orbits appear as open orbits. In a high magnetic field, the contribution from carriers to the resistivity in the transverse configuration with $\mathbf{I}\parallel\mathbf{c}$ is expected to follow a $H^2$ law, while that in the transverse and longitudinal configurations with $\mathbf{I}\perp\mathbf{c}$ are expected to saturate.
\end{itemize}
Table \ref{table} summarizes the behaviors expected for cylindrical Fermi surfaces and those observed here for UTe$_2$. Four of the six considered configurations indicate that a first approximation of cylindrical Fermi surfaces seems appropriate for UTe$_2$. However, deviations from the two-dimensional cylindrical case are found for two configurations: for $\mathbf{H}\parallel\mathbf{a}$ and $\mathbf{I}\parallel\mathbf{c}$, where the resistivity saturates, and for $\mathbf{H}\parallel\mathbf{b}$ and $\mathbf{I}\parallel\mathbf{a}$, where the resistivity slightly increases. These differences may be due to a three-dimensional character, i.e., a non-perfectly cylindrical shape, of the Fermi surfaces dominating the high-field resistivity of UTe$_2$.

\begin{figure*}[t]
\includegraphics[width=0.7\textwidth]{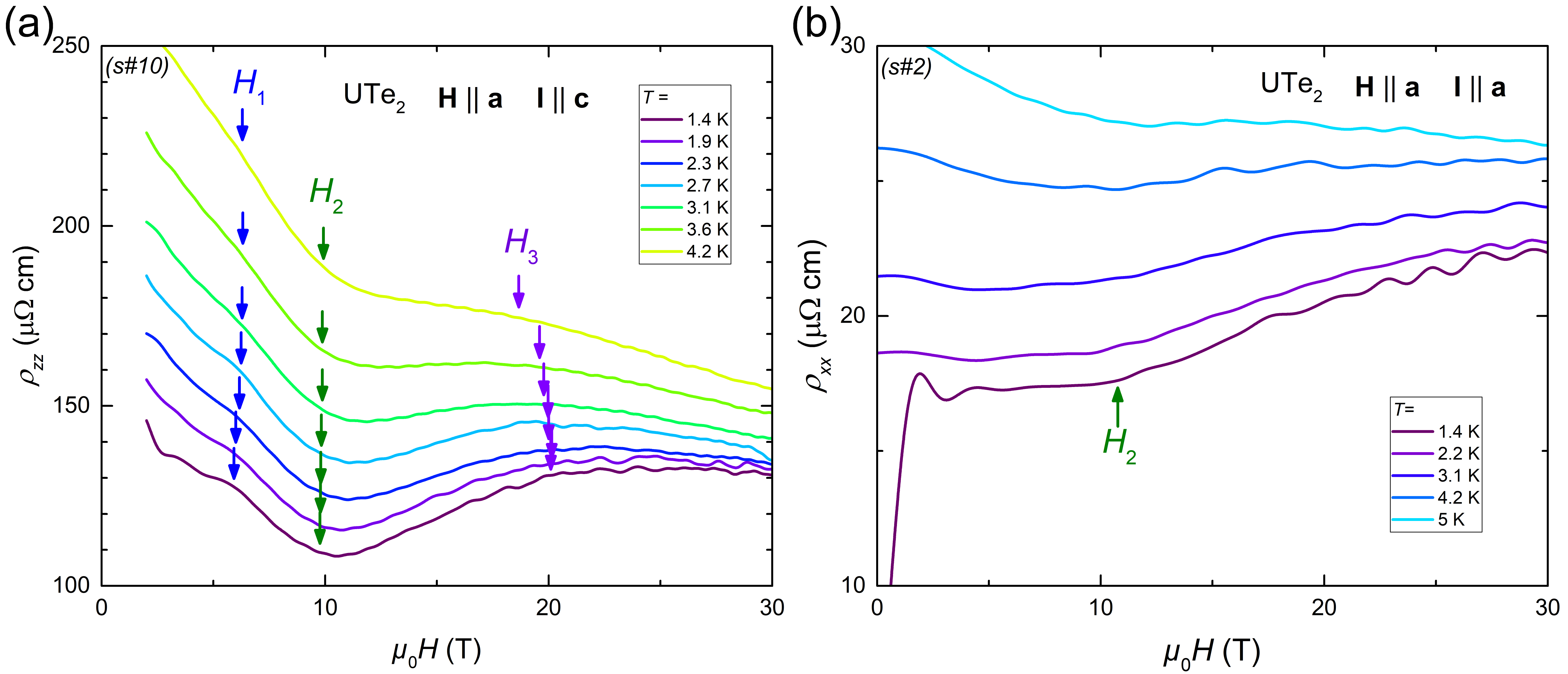}
\caption{\label{FigS2} Magnetic-field dependence of the electrical resistivities (a) $\rho_{xx}$ and (b) $\rho_{zz}$, measured with currents $\mathbf{I}\parallel\mathbf{a}$ and $\mathbf{I}\parallel\mathbf{c}$, respectively, at temperatures $T\leq5$~K and magnetic fields $\mu_0H\leq30$~T applied along $\mathbf{a}$.}
\end{figure*}

\begin{figure*}[t]
\includegraphics[width=\textwidth]{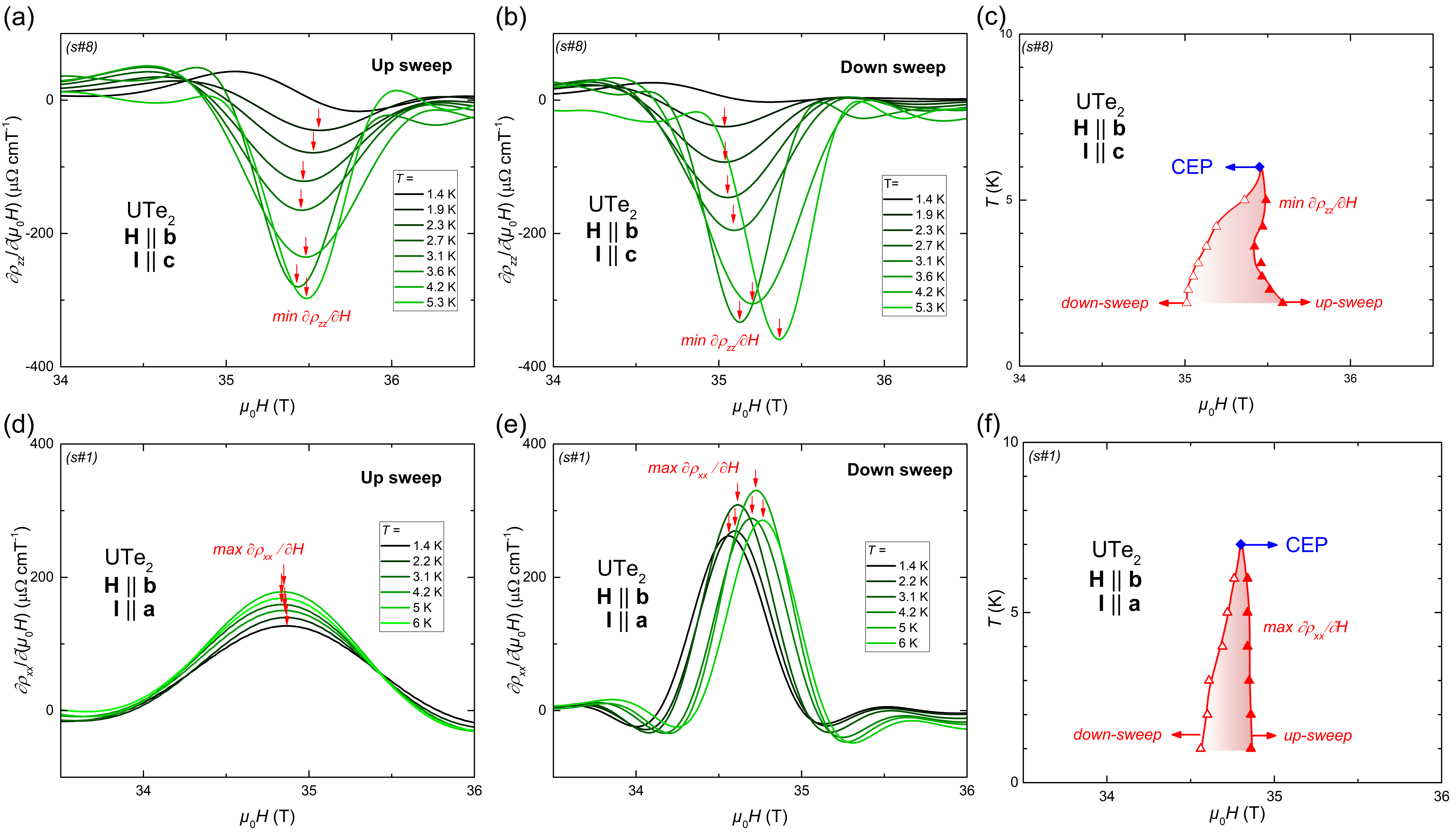}
\caption{\label{FigS3} Field-derivative of the electrical resistivity $\partial\rho_{zz}/\partial H$ versus magnetic field $\mathbf{H}\parallel\mathbf{b}$ measured for $\mathbf{I}\parallel\mathbf{c}$ at temperatures $T\leq5.3$~K: (a) up-sweep and (b) down-sweep, and (c) corresponding magnetic-field-temperature phase diagram showing the hysteresis at $H_m$, and field-derivative of the electrical resistivity $\partial\rho_{xx}/\partial H$ versus magnetic field measured $\mathbf{H}\parallel\mathbf{b}$ for $\mathbf{I}\parallel\mathbf{a}$ at temperatures $T\leq6$~K: (d) up-sweep and (e) down-sweep, and (f) corresponding magnetic-field-temperature phase diagram showing the hysteresis at $H_m$.}
\end{figure*}

\newpage

\textbf{Anomalies at the magnetic fields $H_1$, $H_2$, $H_3$, and $H_m$.}

Fig. \ref{FigS2} shows a zoom on the magnetic-field dependence of the electrical resistivities $\rho_{xx}$ and $\rho_{zz}$, measured with currents $\mathbf{I}\parallel\mathbf{a}$ and $\mathbf{I}\parallel\mathbf{c}$, respectively, at temperatures $T\leq5$~K and magnetic fields $\mu_0H\leq30$~T applied along $\mathbf{a}$. Signatures of magnetic crossover are observed at the fields $\mu_0H_1\simeq6$~T, $\mu_0H_2\simeq10$~T and $\mu_0H_3\simeq20$~T, in agreement with a previous report where they were attributed to Fermi-surface reconstructions \cite{Niu2020a_sm}, while only a broad kink is observed at $H_2$ for $\mathbf{I}\parallel\mathbf{a}$. The trace of these anomalies is lost when the temperature is increased beyond 5~K.

Fig. \ref{FigS3} presents details concerning the determination of the metamagnetic field $H_m$ for $\mathbf{H}\parallel\mathbf{b}$. The first-order character of the transition is indicated by an extremum of the field-derivative of the electrical resistivity, at which $H_m$ is defined, and by an hysteresis. For $\mathbf{I}\parallel\mathbf{c}$, we lose the signatures of the first-order transition at temperatures beyond 6~K, where the critical endpoint can be defined. For $\mathbf{I}\parallel\mathbf{a}$, we lose the signatures of the first-order transition at temperatures beyond 7~K, where the critical endpoint can be defined. The differences in the temperature dependences of the hysteresis at $H_m$ may result from deviations from isothermal conditions in the two setups, due to Eddy currents and magnetocaloric effect in the samples. Such deviations may differ for the two configurations, in relation with different geometries of the samples and different couplings to the sapphire sample holder.\\

\textbf{Fermi-liquid fits to the data.}

\begin{figure*}[t]
\includegraphics[width=\textwidth]{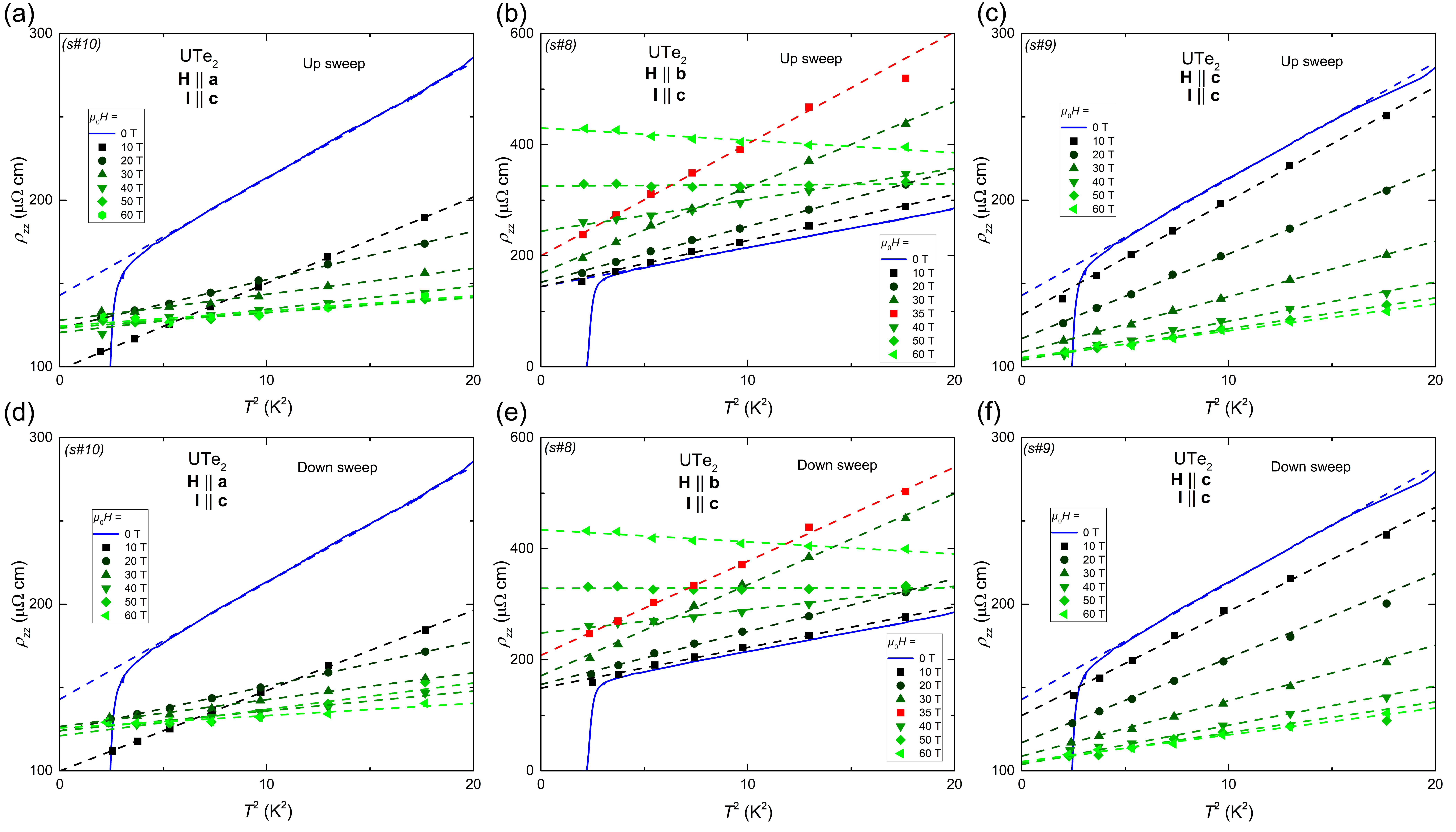}
\caption{\label{FigS4} Electrical resistivity $\rho_{zz}$, measured for $\mathbf{I}\parallel\mathbf{c}$, versus square of the temperature, and $T^2$ fits (dashed lines) to the data, for up-sweep with (a) $\mathbf{H}\parallel\mathbf{a}$,  (b) $\mathbf{H}\parallel\mathbf{b}$, (c) $\mathbf{H}\parallel\mathbf{c}$, and down-sweep with (d) $\mathbf{H}\parallel\mathbf{a}$,  (e) $\mathbf{H}\parallel\mathbf{b}$, (f) $\mathbf{H}\parallel\mathbf{c}$.}
\end{figure*}

Fig. \ref{FigS4} shows details about the Fermi-liquid fits to the electrical-resistivity data, by formula $\rho=\rho_0+AT^2$, used to extract the quadratic coefficient $A$. For the six configurations of electrical current and magnetic field investigated here, resistivity data follow a $T^2$ behavior at low temperatures, when superconductivity is not observed. Fits to the data were done here from 4.2~K down to 1.4~K, or to a temperature higher than the superconducting transition temperature $T_{sc}$ at fields where superconductivity is established at low temperature. Fermi-liquid fits were also done near the metamagnetic field $H_m$ for $\mathbf{H}\parallel\mathbf{b}$, where $A$ is maximal for the two current directions. Non-physical negative coefficients $A$ are obtained in high fields for the two transversal configurations measured with $\mathbf{H}\parallel\mathbf{b}$, where a low-temperature increase of the resistivity at high field is presumably driven by cyclotron motion of the carriers.\\

\begin{figure*}[t]
\includegraphics[width=0.7\textwidth]{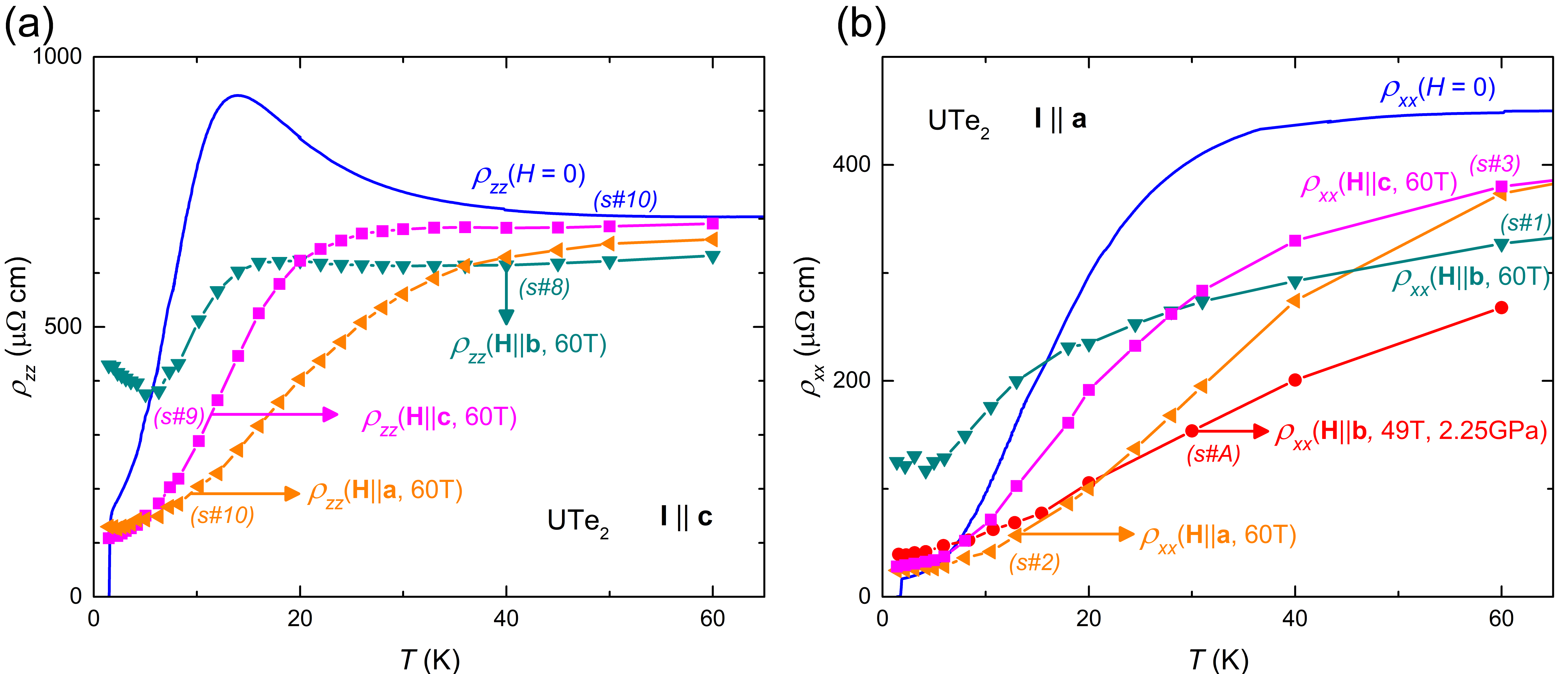}
\caption{\label{FigS5} (a) Electrical resistivity $\rho_{zz}$, measured for $\mathbf{I}\parallel\mathbf{c}$, versus temperature, for $H=0$ and $\mu_0H=60$~T applied along $\mathbf{a}$, $\mathbf{b}$, and $\mathbf{c}$. (b) Electrical resistivity $\rho_{xx}$, measured for $\mathbf{I}\parallel\mathbf{a}$, versus temperature, for $H=0$, $\mu_0H=60$~T applied along $\mathbf{a}$, $\mathbf{b}$, and $\mathbf{c}$, and for $\mu_0H=60$~T applied along $\mathbf{b}$ at the pressure $p=2.25$~GPa.}
\end{figure*}

\textbf{Estimation of a background for the electrical-resistivity data.}

Fig. \ref{FigS5} compares the temperature dependences of the electrical resistivities $\rho_{zz}$ with $\mathbf{I}\parallel\mathbf{c}$ and $\rho_{xx}$ with $\mathbf{I}\parallel\mathbf{a}$ measured at zero field and in a magnetic field of 60~T for the three directions of field $\mathbf{H}\parallel\mathbf{a}$, $\mathbf{H}\parallel\mathbf{b}$, and $\mathbf{H}\parallel\mathbf{c}$, and at 49~T for $\mathbf{I}\parallel\mathbf{a}$, $\mathbf{H}\parallel\mathbf{b}$ within a pressure $p=2.25$~GPa.
\begin{itemize}
  \item In the configuration with $\mathbf{I}\parallel\mathbf{c}$, $\rho_{zz}(\mathbf{H}\parallel\mathbf{a},60\rm{T})$ corresponds to the smallest resistivity in the temperature range [1.5;60~K] considered here. The low-temperature anomaly reported at $\simeq15$~K at zero field has almost completely vanished. Knowing that the magnetization saturates beyond 20~T at low temperature when the magnetic field is applied along the easy magnetic axis $\mathbf{a}$ \cite{Miyake2019_sm,Miyake2021_sm}, we can reasonably infer that most of the low-temperature electronic correlations, including the magnetic fluctuations, have been quenched by a magnetic field of 60~T applied along $\mathbf{a}$. Oppositely, for $\mathbf{H}\parallel\mathbf{c}$ the magnetization is not fully polarized at 60~T at low temperature \cite{Miyake2021_sm}, and a contribution to the electrical resistivity $\rho_{zz}(\mathbf{H}\parallel\mathbf{c},60\rm{T})$ is visible at $\simeq20$~K, indicating that part of the electronic correlations have not been quenched. Within first approximation $\rho_{zz}(\mathbf{H}\parallel\mathbf{a},60\rm{T})$ can thus be considered as a good estimate of a background for the electrical resistivity $\rho_{zz}$.
  \item In the configuration with $\mathbf{I}\parallel\mathbf{a}$, $\rho_{xx}(\mathbf{H}\parallel\mathbf{b},49\rm{T},2.25\rm{GPa})$ measured for $\mu_0H=60$~T applied along $\mathbf{b}$ and at the pressure $p=2.25$~GPa  is smaller than $\rho_{xx}(\mathbf{H}\parallel\mathbf{a},60\rm{T})$ measured in a magnetic field  $\mu_0\mathbf{H}\parallel\mathbf{a}$ of 60~T. It offers a better estimation of the background in the temperature range [1.5;80~K] considered here. For $\mathbf{H}\parallel\mathbf{b}$, a strong decrease of $\mu_0H_m$ down to 15~T is observed under pressures near the critical pressure $p_c\simeq1.5-1.7$~GPa, beyond which long-range magnetic order is stabilized \cite{Braithwaite2019_sm,Knebel2020_sm,Valiska2021_sm}, implying that at $p=2.25$~GPa a field of 49~T leads to a state deep inside the PPM regime, where most of the electronic correlations have vanished. Near the critical pressure $p_c$, the magnetic susceptibilities for $\mathbf{H}\parallel\mathbf{a}$ and $\mathbf{H}\parallel\mathbf{b}$ become similar in magnitude, and for $p>p_c$ the direction $\mathbf{b}$ becomes the easy magnetic axis \cite{Li2021_sm}. This confirms that a highly polarized regime, where most of the electronic correlations have vanished, is expected in a magnetic field $\mu_0\mathbf{H}\parallel\mathbf{b}$ of 49~T combined with a pressure of 2.25 GPa.
\end{itemize}

%\bibliography{UTe2_Thebault_2022}

\begin{thebibliography}{60}%
\makeatletter
\providecommand \@ifxundefined [1]{%
 \@ifx{#1\undefined}
}%
\providecommand \@ifnum [1]{%
 \ifnum #1\expandafter \@firstoftwo
 \else \expandafter \@secondoftwo
 \fi
}%
\providecommand \@ifx [1]{%
 \ifx #1\expandafter \@firstoftwo
 \else \expandafter \@secondoftwo
 \fi
}%
\providecommand \natexlab [1]{#1}%
\providecommand \enquote  [1]{``#1''}%
\providecommand \bibnamefont  [1]{#1}%
\providecommand \bibfnamefont [1]{#1}%
\providecommand \citenamefont [1]{#1}%
\providecommand \href@noop [0]{\@secondoftwo}%
\providecommand \href [0]{\begingroup \@sanitize@url \@href}%
\providecommand \@href[1]{\@@startlink{#1}\@@href}%
\providecommand \@@href[1]{\endgroup#1\@@endlink}%
\providecommand \@sanitize@url [0]{\catcode `\\12\catcode `\$12\catcode
  `\&12\catcode `\#12\catcode `\^12\catcode `\_12\catcode `\%12\relax}%
\providecommand \@@startlink[1]{}%
\providecommand \@@endlink[0]{}%
\providecommand \url  [0]{\begingroup\@sanitize@url \@url }%
\providecommand \@url [1]{\endgroup\@href {#1}{\urlprefix }}%
\providecommand \urlprefix  [0]{URL }%
\providecommand \Eprint [0]{\href }%
\providecommand \doibase [0]{http://dx.doi.org/}%
\providecommand \selectlanguage [0]{\@gobble}%
\providecommand \bibinfo  [0]{\@secondoftwo}%
\providecommand \bibfield  [0]{\@secondoftwo}%
\providecommand \translation [1]{[#1]}%
\providecommand \BibitemOpen [0]{}%
\providecommand \bibitemStop [0]{}%
\providecommand \bibitemNoStop [0]{.\EOS\space}%
\providecommand \EOS [0]{\spacefactor3000\relax}%
\providecommand \BibitemShut  [1]{\csname bibitem#1\endcsname}%
\let\auto@bib@innerbib\@empty
%</preamble>
\bibitem [{\citenamefont {Monthoux}\ \emph {et~al.}(2007)\citenamefont
  {Monthoux}, \citenamefont {Pines},\ and\ \citenamefont
  {Lonzarich}}]{Monthoux2007}%
  \BibitemOpen
  \bibfield  {author} {\bibinfo {author} {\bibfnamefont {P.}~\bibnamefont
  {Monthoux}}, \bibinfo {author} {\bibfnamefont {D.}~\bibnamefont {Pines}}, \
  and\ \bibinfo {author} {\bibfnamefont {G.}~\bibnamefont {Lonzarich}},\
  }\href@noop {} {\bibfield  {journal} {\bibinfo  {journal} {Nature}\ }\textbf
  {\bibinfo {volume} {450}},\ \bibinfo {pages} {1177} (\bibinfo {year}
  {2007})}\BibitemShut {NoStop}%
\bibitem [{\citenamefont {Ran}\ \emph {et~al.}(2019{\natexlab{a}})\citenamefont
  {Ran}, \citenamefont {Eckberg}, \citenamefont {Ding}, \citenamefont
  {Furukawa}, \citenamefont {Metz}, \citenamefont {Saha}, \citenamefont {Liu},
  \citenamefont {Zic}, \citenamefont {Kim}, \citenamefont {Paglione},\ and\
  \citenamefont {Butch}}]{Ran2019a}%
  \BibitemOpen
  \bibfield  {author} {\bibinfo {author} {\bibfnamefont {S.}~\bibnamefont
  {Ran}}, \bibinfo {author} {\bibfnamefont {C.}~\bibnamefont {Eckberg}},
  \bibinfo {author} {\bibfnamefont {Q.-P.}\ \bibnamefont {Ding}}, \bibinfo
  {author} {\bibfnamefont {Y.}~\bibnamefont {Furukawa}}, \bibinfo {author}
  {\bibfnamefont {T.}~\bibnamefont {Metz}}, \bibinfo {author} {\bibfnamefont
  {S.~R.}\ \bibnamefont {Saha}}, \bibinfo {author} {\bibfnamefont {I.-L.}\
  \bibnamefont {Liu}}, \bibinfo {author} {\bibfnamefont {M.}~\bibnamefont
  {Zic}}, \bibinfo {author} {\bibfnamefont {H.}~\bibnamefont {Kim}}, \bibinfo
  {author} {\bibfnamefont {J.}~\bibnamefont {Paglione}}, \ and\ \bibinfo
  {author} {\bibfnamefont {N.~P.}\ \bibnamefont {Butch}},\ }\href@noop {}
  {\bibfield  {journal} {\bibinfo  {journal} {Science}\ }\textbf {\bibinfo
  {volume} {365}},\ \bibinfo {pages} {684} (\bibinfo {year}
  {2019}{\natexlab{a}})}\BibitemShut {NoStop}%
\bibitem [{\citenamefont {Aoki}\ \emph {et~al.}(2019)\citenamefont {Aoki},
  \citenamefont {Nakamura}, \citenamefont {Honda}, \citenamefont {Li},
  \citenamefont {Homma}, \citenamefont {Shimizu}, \citenamefont {Sato},
  \citenamefont {Knebel}, \citenamefont {Brison}, \citenamefont {Pourret},
  \citenamefont {Braithwaite}, \citenamefont {Lapertot}, \citenamefont {Niu},
  \citenamefont {Vališka}, \citenamefont {Harima},\ and\ \citenamefont
  {Flouquet}}]{Aoki2019b}%
  \BibitemOpen
  \bibfield  {author} {\bibinfo {author} {\bibfnamefont {D.}~\bibnamefont
  {Aoki}}, \bibinfo {author} {\bibfnamefont {A.}~\bibnamefont {Nakamura}},
  \bibinfo {author} {\bibfnamefont {F.}~\bibnamefont {Honda}}, \bibinfo
  {author} {\bibfnamefont {D.}~\bibnamefont {Li}}, \bibinfo {author}
  {\bibfnamefont {Y.}~\bibnamefont {Homma}}, \bibinfo {author} {\bibfnamefont
  {Y.}~\bibnamefont {Shimizu}}, \bibinfo {author} {\bibfnamefont {Y.~J.}\
  \bibnamefont {Sato}}, \bibinfo {author} {\bibfnamefont {G.}~\bibnamefont
  {Knebel}}, \bibinfo {author} {\bibfnamefont {J.-P.}\ \bibnamefont {Brison}},
  \bibinfo {author} {\bibfnamefont {A.}~\bibnamefont {Pourret}}, \bibinfo
  {author} {\bibfnamefont {D.}~\bibnamefont {Braithwaite}}, \bibinfo {author}
  {\bibfnamefont {G.}~\bibnamefont {Lapertot}}, \bibinfo {author}
  {\bibfnamefont {Q.}~\bibnamefont {Niu}}, \bibinfo {author} {\bibfnamefont
  {M.}~\bibnamefont {Vališka}}, \bibinfo {author} {\bibfnamefont
  {H.}~\bibnamefont {Harima}}, \ and\ \bibinfo {author} {\bibfnamefont
  {J.}~\bibnamefont {Flouquet}},\ }\href@noop {} {\bibfield  {journal}
  {\bibinfo  {journal} {J. Phys. Soc. Jpn.}\ }\textbf {\bibinfo {volume}
  {88}},\ \bibinfo {pages} {043702} (\bibinfo {year} {2019})}\BibitemShut
  {NoStop}%
\bibitem [{\citenamefont {Aoki}\ \emph {et~al.}(2022)\citenamefont {Aoki},
  \citenamefont {Brison}, \citenamefont {Flouquet}, \citenamefont {Ishida},
  \citenamefont {Knebel}, \citenamefont {Tokunaga},\ and\ \citenamefont
  {Yanase}}]{Aoki2022}%
  \BibitemOpen
  \bibfield  {author} {\bibinfo {author} {\bibfnamefont {D.}~\bibnamefont
  {Aoki}}, \bibinfo {author} {\bibfnamefont {J.-P.}\ \bibnamefont {Brison}},
  \bibinfo {author} {\bibfnamefont {J.}~\bibnamefont {Flouquet}}, \bibinfo
  {author} {\bibfnamefont {K.}~\bibnamefont {Ishida}}, \bibinfo {author}
  {\bibfnamefont {G.}~\bibnamefont {Knebel}}, \bibinfo {author} {\bibfnamefont
  {Y.}~\bibnamefont {Tokunaga}}, \ and\ \bibinfo {author} {\bibfnamefont
  {Y.}~\bibnamefont {Yanase}},\ }\href@noop {} {\bibfield  {journal} {\bibinfo
  {journal} {J. Phys.: Condens. Matter}\ }\textbf {\bibinfo {volume} {34}},\
  \bibinfo {pages} {243002} (\bibinfo {year} {2022})}\BibitemShut {NoStop}%
\bibitem [{\citenamefont {Ran}\ \emph {et~al.}(2019{\natexlab{b}})\citenamefont
  {Ran}, \citenamefont {Liu}, \citenamefont {Eo}, \citenamefont {Campbell},
  \citenamefont {Neves}, \citenamefont {Fuhrman}, \citenamefont {Saha},
  \citenamefont {Eckberg}, \citenamefont {Kim}, \citenamefont {Paglione},
  \citenamefont {Graf}, \citenamefont {Singleton},\ and\ \citenamefont
  {Butch}}]{Ran2019b}%
  \BibitemOpen
  \bibfield  {author} {\bibinfo {author} {\bibfnamefont {S.}~\bibnamefont
  {Ran}}, \bibinfo {author} {\bibfnamefont {I.-L.}\ \bibnamefont {Liu}},
  \bibinfo {author} {\bibfnamefont {Y.~S.}\ \bibnamefont {Eo}}, \bibinfo
  {author} {\bibfnamefont {D.~J.}\ \bibnamefont {Campbell}}, \bibinfo {author}
  {\bibfnamefont {P.}~\bibnamefont {Neves}}, \bibinfo {author} {\bibfnamefont
  {W.~T.}\ \bibnamefont {Fuhrman}}, \bibinfo {author} {\bibfnamefont {S.~R.}\
  \bibnamefont {Saha}}, \bibinfo {author} {\bibfnamefont {C.}~\bibnamefont
  {Eckberg}}, \bibinfo {author} {\bibfnamefont {H.}~\bibnamefont {Kim}},
  \bibinfo {author} {\bibfnamefont {J.}~\bibnamefont {Paglione}}, \bibinfo
  {author} {\bibfnamefont {D.}~\bibnamefont {Graf}}, \bibinfo {author}
  {\bibfnamefont {J.}~\bibnamefont {Singleton}}, \ and\ \bibinfo {author}
  {\bibfnamefont {N.~P.}\ \bibnamefont {Butch}},\ }\href@noop {} {\bibfield
  {journal} {\bibinfo  {journal} {Nat. Phys.}\ }\textbf {\bibinfo {volume}
  {15}},\ \bibinfo {pages} {1250} (\bibinfo {year}
  {2019}{\natexlab{b}})}\BibitemShut {NoStop}%
\bibitem [{\citenamefont {Nakamine}\ \emph {et~al.}(2019)\citenamefont
  {Nakamine}, \citenamefont {Kitagawa}, \citenamefont {Ishida}, \citenamefont
  {Tokunaga}, \citenamefont {Sakai}, \citenamefont {Kambe}, \citenamefont
  {Nakamura}, \citenamefont {Shimizu}, \citenamefont {Homma}, \citenamefont
  {Li}, \citenamefont {Honda},\ and\ \citenamefont {Aoki}}]{Nakamine2019}%
  \BibitemOpen
  \bibfield  {author} {\bibinfo {author} {\bibfnamefont {G.}~\bibnamefont
  {Nakamine}}, \bibinfo {author} {\bibfnamefont {S.}~\bibnamefont {Kitagawa}},
  \bibinfo {author} {\bibfnamefont {K.}~\bibnamefont {Ishida}}, \bibinfo
  {author} {\bibfnamefont {Y.}~\bibnamefont {Tokunaga}}, \bibinfo {author}
  {\bibfnamefont {H.}~\bibnamefont {Sakai}}, \bibinfo {author} {\bibfnamefont
  {S.}~\bibnamefont {Kambe}}, \bibinfo {author} {\bibfnamefont
  {A.}~\bibnamefont {Nakamura}}, \bibinfo {author} {\bibfnamefont
  {Y.}~\bibnamefont {Shimizu}}, \bibinfo {author} {\bibfnamefont
  {Y.}~\bibnamefont {Homma}}, \bibinfo {author} {\bibfnamefont
  {D.}~\bibnamefont {Li}}, \bibinfo {author} {\bibfnamefont {F.}~\bibnamefont
  {Honda}}, \ and\ \bibinfo {author} {\bibfnamefont {D.}~\bibnamefont {Aoki}},\
  }\href@noop {} {\bibfield  {journal} {\bibinfo  {journal} {J. Phys. Soc.
  Jpn.}\ }\textbf {\bibinfo {volume} {88}},\ \bibinfo {pages} {113703}
  (\bibinfo {year} {2019})}\BibitemShut {NoStop}%
\bibitem [{\citenamefont {Fujibayashi}\ \emph {et~al.}(2022)\citenamefont
  {Fujibayashi}, \citenamefont {Nakamine}, \citenamefont {Kinjo}, \citenamefont
  {Kitagawa}, \citenamefont {Ishida}, \citenamefont {Tokunaga}, \citenamefont
  {Sakai}, \citenamefont {Kambe}, \citenamefont {Nakamura}, \citenamefont
  {Shimizu}, \citenamefont {Homma}, \citenamefont {Li}, \citenamefont {Honda},\
  and\ \citenamefont {Aoki}}]{Fujibayashi2022}%
  \BibitemOpen
  \bibfield  {author} {\bibinfo {author} {\bibfnamefont {H.}~\bibnamefont
  {Fujibayashi}}, \bibinfo {author} {\bibfnamefont {G.}~\bibnamefont
  {Nakamine}}, \bibinfo {author} {\bibfnamefont {K.}~\bibnamefont {Kinjo}},
  \bibinfo {author} {\bibfnamefont {S.}~\bibnamefont {Kitagawa}}, \bibinfo
  {author} {\bibfnamefont {K.}~\bibnamefont {Ishida}}, \bibinfo {author}
  {\bibfnamefont {Y.}~\bibnamefont {Tokunaga}}, \bibinfo {author}
  {\bibfnamefont {H.}~\bibnamefont {Sakai}}, \bibinfo {author} {\bibfnamefont
  {S.}~\bibnamefont {Kambe}}, \bibinfo {author} {\bibfnamefont
  {A.}~\bibnamefont {Nakamura}}, \bibinfo {author} {\bibfnamefont
  {Y.}~\bibnamefont {Shimizu}}, \bibinfo {author} {\bibfnamefont
  {Y.}~\bibnamefont {Homma}}, \bibinfo {author} {\bibfnamefont
  {D.}~\bibnamefont {Li}}, \bibinfo {author} {\bibfnamefont {F.}~\bibnamefont
  {Honda}}, \ and\ \bibinfo {author} {\bibfnamefont {D.}~\bibnamefont {Aoki}},\
  }\href {\doibase 10.7566/JPSJ.91.043705} {\bibfield  {journal} {\bibinfo
  {journal} {J. Phys. Soc. Jpn.}\ }\textbf {\bibinfo {volume} {91}},\ \bibinfo
  {pages} {043705} (\bibinfo {year} {2022})}\BibitemShut {NoStop}%
\bibitem [{\citenamefont {Knafo}\ \emph {et~al.}(2019)\citenamefont {Knafo},
  \citenamefont {Vali\v{s}ka}, \citenamefont {Braithwaite}, \citenamefont
  {Lapertot}, \citenamefont {Knebel}, \citenamefont {Pourret}, \citenamefont
  {Brison}, \citenamefont {Flouquet},\ and\ \citenamefont {Aoki}}]{Knafo2019}%
  \BibitemOpen
  \bibfield  {author} {\bibinfo {author} {\bibfnamefont {W.}~\bibnamefont
  {Knafo}}, \bibinfo {author} {\bibfnamefont {M.}~\bibnamefont {Vali\v{s}ka}},
  \bibinfo {author} {\bibfnamefont {D.}~\bibnamefont {Braithwaite}}, \bibinfo
  {author} {\bibfnamefont {G.}~\bibnamefont {Lapertot}}, \bibinfo {author}
  {\bibfnamefont {G.}~\bibnamefont {Knebel}}, \bibinfo {author} {\bibfnamefont
  {A.}~\bibnamefont {Pourret}}, \bibinfo {author} {\bibfnamefont {J.-P.}\
  \bibnamefont {Brison}}, \bibinfo {author} {\bibfnamefont {J.}~\bibnamefont
  {Flouquet}}, \ and\ \bibinfo {author} {\bibfnamefont {D.}~\bibnamefont
  {Aoki}},\ }\href@noop {} {\bibfield  {journal} {\bibinfo  {journal} {J. Phys.
  Soc. Jpn.}\ }\textbf {\bibinfo {volume} {88}},\ \bibinfo {pages} {063705}
  (\bibinfo {year} {2019})}\BibitemShut {NoStop}%
\bibitem [{\citenamefont {Miyake}\ \emph {et~al.}(2019)\citenamefont {Miyake},
  \citenamefont {Shimizu}, \citenamefont {Sato}, \citenamefont {Li},
  \citenamefont {Nakamura}, \citenamefont {Homma}, \citenamefont {Honda},
  \citenamefont {Flouquet}, \citenamefont {Tokunaga},\ and\ \citenamefont
  {Aoki}}]{Miyake2019}%
  \BibitemOpen
  \bibfield  {author} {\bibinfo {author} {\bibfnamefont {A.}~\bibnamefont
  {Miyake}}, \bibinfo {author} {\bibfnamefont {Y.}~\bibnamefont {Shimizu}},
  \bibinfo {author} {\bibfnamefont {Y.~J.}\ \bibnamefont {Sato}}, \bibinfo
  {author} {\bibfnamefont {D.}~\bibnamefont {Li}}, \bibinfo {author}
  {\bibfnamefont {A.}~\bibnamefont {Nakamura}}, \bibinfo {author}
  {\bibfnamefont {Y.}~\bibnamefont {Homma}}, \bibinfo {author} {\bibfnamefont
  {F.}~\bibnamefont {Honda}}, \bibinfo {author} {\bibfnamefont
  {J.}~\bibnamefont {Flouquet}}, \bibinfo {author} {\bibfnamefont
  {M.}~\bibnamefont {Tokunaga}}, \ and\ \bibinfo {author} {\bibfnamefont
  {D.}~\bibnamefont {Aoki}},\ }\href@noop {} {\bibfield  {journal} {\bibinfo
  {journal} {J. Phys. Soc. Jpn.}\ }\textbf {\bibinfo {volume} {88}},\ \bibinfo
  {pages} {063706} (\bibinfo {year} {2019})}\BibitemShut {NoStop}%
\bibitem [{\citenamefont {Knebel}\ \emph {et~al.}(2019)\citenamefont {Knebel},
  \citenamefont {Knafo}, \citenamefont {Pourret}, \citenamefont {Niu},
  \citenamefont {Vali\v{s}ka}, \citenamefont {Braithwaite}, \citenamefont
  {Lapertot}, \citenamefont {Nardone}, \citenamefont {Zitouni}, \citenamefont
  {Mishra}, \citenamefont {Sheikin}, \citenamefont {Seyfarth}, \citenamefont
  {Brison}, \citenamefont {Aoki},\ and\ \citenamefont {Flouquet}}]{Knebel2019}%
  \BibitemOpen
  \bibfield  {author} {\bibinfo {author} {\bibfnamefont {G.}~\bibnamefont
  {Knebel}}, \bibinfo {author} {\bibfnamefont {W.}~\bibnamefont {Knafo}},
  \bibinfo {author} {\bibfnamefont {A.}~\bibnamefont {Pourret}}, \bibinfo
  {author} {\bibfnamefont {Q.}~\bibnamefont {Niu}}, \bibinfo {author}
  {\bibfnamefont {M.}~\bibnamefont {Vali\v{s}ka}}, \bibinfo {author}
  {\bibfnamefont {D.}~\bibnamefont {Braithwaite}}, \bibinfo {author}
  {\bibfnamefont {G.}~\bibnamefont {Lapertot}}, \bibinfo {author}
  {\bibfnamefont {M.}~\bibnamefont {Nardone}}, \bibinfo {author} {\bibfnamefont
  {A.}~\bibnamefont {Zitouni}}, \bibinfo {author} {\bibfnamefont
  {S.}~\bibnamefont {Mishra}}, \bibinfo {author} {\bibfnamefont
  {I.}~\bibnamefont {Sheikin}}, \bibinfo {author} {\bibfnamefont
  {G.}~\bibnamefont {Seyfarth}}, \bibinfo {author} {\bibfnamefont {J.-P.}\
  \bibnamefont {Brison}}, \bibinfo {author} {\bibfnamefont {D.}~\bibnamefont
  {Aoki}}, \ and\ \bibinfo {author} {\bibfnamefont {J.}~\bibnamefont
  {Flouquet}},\ }\href@noop {} {\bibfield  {journal} {\bibinfo  {journal} {J.
  Phys. Soc. Jpn.}\ }\textbf {\bibinfo {volume} {88}},\ \bibinfo {pages}
  {063707} (\bibinfo {year} {2019})}\BibitemShut {NoStop}%
\bibitem [{\citenamefont {Rosuel}\ \emph {et~al.}()\citenamefont {Rosuel},
  \citenamefont {Marcenat}, \citenamefont {Knebel}, \citenamefont {Klein},
  \citenamefont {Pourret}, \citenamefont {Marquardt}, \citenamefont {Niu},
  \citenamefont {Rousseau}, \citenamefont {Demuer}, \citenamefont {Seyfarth},
  \citenamefont {Lapertot}, \citenamefont {Aoki}, \citenamefont {Braithwaite},
  \citenamefont {Flouquet},\ and\ \citenamefont {Brison}}]{Rosuel2022}%
  \BibitemOpen
  \bibfield  {author} {\bibinfo {author} {\bibfnamefont {A.}~\bibnamefont
  {Rosuel}}, \bibinfo {author} {\bibfnamefont {C.}~\bibnamefont {Marcenat}},
  \bibinfo {author} {\bibfnamefont {G.}~\bibnamefont {Knebel}}, \bibinfo
  {author} {\bibfnamefont {T.}~\bibnamefont {Klein}}, \bibinfo {author}
  {\bibfnamefont {A.}~\bibnamefont {Pourret}}, \bibinfo {author} {\bibfnamefont
  {N.}~\bibnamefont {Marquardt}}, \bibinfo {author} {\bibfnamefont
  {Q.}~\bibnamefont {Niu}}, \bibinfo {author} {\bibfnamefont {S.}~\bibnamefont
  {Rousseau}}, \bibinfo {author} {\bibfnamefont {A.}~\bibnamefont {Demuer}},
  \bibinfo {author} {\bibfnamefont {G.}~\bibnamefont {Seyfarth}}, \bibinfo
  {author} {\bibfnamefont {G.}~\bibnamefont {Lapertot}}, \bibinfo {author}
  {\bibfnamefont {D.}~\bibnamefont {Aoki}}, \bibinfo {author} {\bibfnamefont
  {D.}~\bibnamefont {Braithwaite}}, \bibinfo {author} {\bibfnamefont
  {J.}~\bibnamefont {Flouquet}}, \ and\ \bibinfo {author} {\bibfnamefont
  {J.-P.}\ \bibnamefont {Brison}},\ }\href@noop {} {\bibinfo  {journal}
  {arxiv:2205.04524}\ }\BibitemShut {NoStop}%
\bibitem [{\citenamefont {Lin}\ \emph {et~al.}(2020)\citenamefont {Lin},
  \citenamefont {Campbell}, \citenamefont {Ran}, \citenamefont {Liu},
  \citenamefont {Kim}, \citenamefont {Nevidomskyy}, \citenamefont {Graf},
  \citenamefont {Butch},\ and\ \citenamefont {Paglione}}]{Lin2020}%
  \BibitemOpen
\bibfield  {journal} {  }\bibfield  {author} {\bibinfo {author} {\bibfnamefont
  {W.~C.}\ \bibnamefont {Lin}}, \bibinfo {author} {\bibfnamefont {D.~J.}\
  \bibnamefont {Campbell}}, \bibinfo {author} {\bibfnamefont {S.}~\bibnamefont
  {Ran}}, \bibinfo {author} {\bibfnamefont {I.-L.}\ \bibnamefont {Liu}},
  \bibinfo {author} {\bibfnamefont {H.}~\bibnamefont {Kim}}, \bibinfo {author}
  {\bibfnamefont {A.~H.}\ \bibnamefont {Nevidomskyy}}, \bibinfo {author}
  {\bibfnamefont {D.}~\bibnamefont {Graf}}, \bibinfo {author} {\bibfnamefont
  {N.~P.}\ \bibnamefont {Butch}}, \ and\ \bibinfo {author} {\bibfnamefont
  {J.}~\bibnamefont {Paglione}},\ }\href@noop {} {\bibfield  {journal}
  {\bibinfo  {journal} {npj Quantum Mater.}\ }\textbf {\bibinfo {volume} {5}},\
  \bibinfo {pages} {68} (\bibinfo {year} {2020})}\BibitemShut {NoStop}%
\bibitem [{\citenamefont {Knafo}\ \emph
  {et~al.}(2021{\natexlab{a}})\citenamefont {Knafo}, \citenamefont {Nardone},
  \citenamefont {Vali\v{s}ka}, \citenamefont {Zitouni}, \citenamefont
  {Lapertot}, \citenamefont {Aoki}, \citenamefont {Knebel},\ and\ \citenamefont
  {Braithwaite}}]{Knafo2021a}%
  \BibitemOpen
  \bibfield  {author} {\bibinfo {author} {\bibfnamefont {W.}~\bibnamefont
  {Knafo}}, \bibinfo {author} {\bibfnamefont {M.}~\bibnamefont {Nardone}},
  \bibinfo {author} {\bibfnamefont {M.}~\bibnamefont {Vali\v{s}ka}}, \bibinfo
  {author} {\bibfnamefont {A.}~\bibnamefont {Zitouni}}, \bibinfo {author}
  {\bibfnamefont {G.}~\bibnamefont {Lapertot}}, \bibinfo {author}
  {\bibfnamefont {D.}~\bibnamefont {Aoki}}, \bibinfo {author} {\bibfnamefont
  {G.}~\bibnamefont {Knebel}}, \ and\ \bibinfo {author} {\bibfnamefont
  {D.}~\bibnamefont {Braithwaite}},\ }\href@noop {} {\bibfield  {journal}
  {\bibinfo  {journal} {Commun. Phys.}\ }\textbf {\bibinfo {volume} {4}},\
  \bibinfo {pages} {40} (\bibinfo {year} {2021}{\natexlab{a}})}\BibitemShut
  {NoStop}%
\bibitem [{\citenamefont {Niu}\ \emph {et~al.}(2020{\natexlab{a}})\citenamefont
  {Niu}, \citenamefont {Knebel}, \citenamefont {Braithwaite}, \citenamefont
  {Aoki}, \citenamefont {Lapertot}, \citenamefont {Vali\v{s}ka}, \citenamefont
  {Seyfarth}, \citenamefont {Knafo}, \citenamefont {Helm}, \citenamefont
  {Brison}, \citenamefont {Flouquet},\ and\ \citenamefont
  {Pourret}}]{Niu2020b}%
  \BibitemOpen
  \bibfield  {author} {\bibinfo {author} {\bibfnamefont {Q.}~\bibnamefont
  {Niu}}, \bibinfo {author} {\bibfnamefont {G.}~\bibnamefont {Knebel}},
  \bibinfo {author} {\bibfnamefont {D.}~\bibnamefont {Braithwaite}}, \bibinfo
  {author} {\bibfnamefont {D.}~\bibnamefont {Aoki}}, \bibinfo {author}
  {\bibfnamefont {G.}~\bibnamefont {Lapertot}}, \bibinfo {author}
  {\bibfnamefont {M.}~\bibnamefont {Vali\v{s}ka}}, \bibinfo {author}
  {\bibfnamefont {G.}~\bibnamefont {Seyfarth}}, \bibinfo {author}
  {\bibfnamefont {W.}~\bibnamefont {Knafo}}, \bibinfo {author} {\bibfnamefont
  {T.}~\bibnamefont {Helm}}, \bibinfo {author} {\bibfnamefont {J.-P.}\
  \bibnamefont {Brison}}, \bibinfo {author} {\bibfnamefont {J.}~\bibnamefont
  {Flouquet}}, \ and\ \bibinfo {author} {\bibfnamefont {A.}~\bibnamefont
  {Pourret}},\ }\href@noop {} {\bibfield  {journal} {\bibinfo  {journal} {Phys.
  Rev. Research}\ }\textbf {\bibinfo {volume} {2}},\ \bibinfo {pages} {033179}
  (\bibinfo {year} {2020}{\natexlab{a}})}\BibitemShut {NoStop}%
\bibitem [{\citenamefont {Ikeda}\ \emph {et~al.}(2006)\citenamefont {Ikeda},
  \citenamefont {Sakai}, \citenamefont {Aoki}, \citenamefont {Homma},
  \citenamefont {Yamamoto}, \citenamefont {Nakamura}, \citenamefont {Shiokawa},
  \citenamefont {Haga},\ and\ \citenamefont {\={O}nuki}}]{Ikeda2006}%
  \BibitemOpen
  \bibfield  {author} {\bibinfo {author} {\bibfnamefont {S.}~\bibnamefont
  {Ikeda}}, \bibinfo {author} {\bibfnamefont {H.}~\bibnamefont {Sakai}},
  \bibinfo {author} {\bibfnamefont {D.}~\bibnamefont {Aoki}}, \bibinfo {author}
  {\bibfnamefont {Y.}~\bibnamefont {Homma}}, \bibinfo {author} {\bibfnamefont
  {E.}~\bibnamefont {Yamamoto}}, \bibinfo {author} {\bibfnamefont
  {A.}~\bibnamefont {Nakamura}}, \bibinfo {author} {\bibfnamefont
  {Y.}~\bibnamefont {Shiokawa}}, \bibinfo {author} {\bibfnamefont
  {Y.}~\bibnamefont {Haga}}, \ and\ \bibinfo {author} {\bibfnamefont
  {Y.}~\bibnamefont {\={O}nuki}},\ }\href@noop {} {\bibfield  {journal}
  {\bibinfo  {journal} {J. Phys. Soc. Jpn.}\ }\textbf {\bibinfo {volume}
  {75}},\ \bibinfo {pages} {116} (\bibinfo {year} {2006})}\BibitemShut
  {NoStop}%
\bibitem [{\citenamefont {Aoki}\ \emph {et~al.}(2013)\citenamefont {Aoki},
  \citenamefont {Knafo},\ and\ \citenamefont {Sheikin}}]{Aoki2013}%
  \BibitemOpen
  \bibfield  {author} {\bibinfo {author} {\bibfnamefont {D.}~\bibnamefont
  {Aoki}}, \bibinfo {author} {\bibfnamefont {W.}~\bibnamefont {Knafo}}, \ and\
  \bibinfo {author} {\bibfnamefont {I.}~\bibnamefont {Sheikin}},\ }\href@noop
  {} {\bibfield  {journal} {\bibinfo  {journal} {C. R. Physique}\ }\textbf
  {\bibinfo {volume} {14}},\ \bibinfo {pages} {53} (\bibinfo {year}
  {2013})}\BibitemShut {NoStop}%
\bibitem [{\citenamefont {Knafo}(3458)}]{Knafo2021c}%
  \BibitemOpen
  \bibfield  {author} {\bibinfo {author} {\bibfnamefont {W.}~\bibnamefont
  {Knafo}},\ }\href@noop {} {\emph {\bibinfo {title} {Habilitation \`{a}
  Diriger des Recherches}}}\ (\bibinfo  {publisher} {University of Toulouse},\
  \bibinfo {year} {2021, arXiv:2107.13458})\BibitemShut {NoStop}%
\bibitem [{\citenamefont {Braithwaite}\ \emph {et~al.}(2019)\citenamefont
  {Braithwaite}, \citenamefont {Vali\v{s}ka}, \citenamefont {Knebel},
  \citenamefont {Lapertot}, \citenamefont {Brison}, \citenamefont {Pourret},
  \citenamefont {Zhitomirsky}, \citenamefont {Flouquet}, \citenamefont
  {Honda},\ and\ \citenamefont {Aoki}}]{Braithwaite2019}%
  \BibitemOpen
  \bibfield  {author} {\bibinfo {author} {\bibfnamefont {D.}~\bibnamefont
  {Braithwaite}}, \bibinfo {author} {\bibfnamefont {M.}~\bibnamefont
  {Vali\v{s}ka}}, \bibinfo {author} {\bibfnamefont {G.}~\bibnamefont {Knebel}},
  \bibinfo {author} {\bibfnamefont {G.}~\bibnamefont {Lapertot}}, \bibinfo
  {author} {\bibfnamefont {J.-P.}\ \bibnamefont {Brison}}, \bibinfo {author}
  {\bibfnamefont {A.}~\bibnamefont {Pourret}}, \bibinfo {author} {\bibfnamefont
  {M.~E.}\ \bibnamefont {Zhitomirsky}}, \bibinfo {author} {\bibfnamefont
  {J.}~\bibnamefont {Flouquet}}, \bibinfo {author} {\bibfnamefont
  {F.}~\bibnamefont {Honda}}, \ and\ \bibinfo {author} {\bibfnamefont
  {D.}~\bibnamefont {Aoki}},\ }\href@noop {} {\bibfield  {journal} {\bibinfo
  {journal} {Commun. Phys.}\ }\textbf {\bibinfo {volume} {2}},\ \bibinfo
  {pages} {147} (\bibinfo {year} {2019})}\BibitemShut {NoStop}%
\bibitem [{\citenamefont {Knebel}\ \emph {et~al.}(2020)\citenamefont {Knebel},
  \citenamefont {Kimata}, \citenamefont {Vali\v{s}ka}, \citenamefont {Honda},
  \citenamefont {Li}, \citenamefont {Braithwaite}, \citenamefont {Lapertot},
  \citenamefont {Knafo}, \citenamefont {Pourret}, \citenamefont {Sato},
  \citenamefont {Shimizu}, \citenamefont {Kihara}, \citenamefont {Brison},
  \citenamefont {Flouquet},\ and\ \citenamefont {Aoki}}]{Knebel2020}%
  \BibitemOpen
  \bibfield  {author} {\bibinfo {author} {\bibfnamefont {G.}~\bibnamefont
  {Knebel}}, \bibinfo {author} {\bibfnamefont {M.}~\bibnamefont {Kimata}},
  \bibinfo {author} {\bibfnamefont {M.}~\bibnamefont {Vali\v{s}ka}}, \bibinfo
  {author} {\bibfnamefont {F.}~\bibnamefont {Honda}}, \bibinfo {author}
  {\bibfnamefont {D.}~\bibnamefont {Li}}, \bibinfo {author} {\bibfnamefont
  {D.}~\bibnamefont {Braithwaite}}, \bibinfo {author} {\bibfnamefont
  {G.}~\bibnamefont {Lapertot}}, \bibinfo {author} {\bibfnamefont
  {W.}~\bibnamefont {Knafo}}, \bibinfo {author} {\bibfnamefont
  {A.}~\bibnamefont {Pourret}}, \bibinfo {author} {\bibfnamefont {Y.~J.}\
  \bibnamefont {Sato}}, \bibinfo {author} {\bibfnamefont {Y.}~\bibnamefont
  {Shimizu}}, \bibinfo {author} {\bibfnamefont {T.}~\bibnamefont {Kihara}},
  \bibinfo {author} {\bibfnamefont {J.-P.}\ \bibnamefont {Brison}}, \bibinfo
  {author} {\bibfnamefont {J.}~\bibnamefont {Flouquet}}, \ and\ \bibinfo
  {author} {\bibfnamefont {D.}~\bibnamefont {Aoki}},\ }\href@noop {} {\bibfield
   {journal} {\bibinfo  {journal} {J. Phys. Soc. Jpn.}\ }\textbf {\bibinfo
  {volume} {89}},\ \bibinfo {pages} {053707} (\bibinfo {year}
  {2020})}\BibitemShut {NoStop}%
\bibitem [{\citenamefont {Ran}\ \emph {et~al.}(2020)\citenamefont {Ran},
  \citenamefont {Kim}, \citenamefont {Liu}, \citenamefont {Saha}, \citenamefont
  {Hayes}, \citenamefont {Metz}, \citenamefont {Eo}, \citenamefont {Paglione},\
  and\ \citenamefont {Butch}}]{Ran2020}%
  \BibitemOpen
  \bibfield  {author} {\bibinfo {author} {\bibfnamefont {S.}~\bibnamefont
  {Ran}}, \bibinfo {author} {\bibfnamefont {H.}~\bibnamefont {Kim}}, \bibinfo
  {author} {\bibfnamefont {I.-L.}\ \bibnamefont {Liu}}, \bibinfo {author}
  {\bibfnamefont {S.~R.}\ \bibnamefont {Saha}}, \bibinfo {author}
  {\bibfnamefont {I.}~\bibnamefont {Hayes}}, \bibinfo {author} {\bibfnamefont
  {T.}~\bibnamefont {Metz}}, \bibinfo {author} {\bibfnamefont {Y.~S.}\
  \bibnamefont {Eo}}, \bibinfo {author} {\bibfnamefont {J.}~\bibnamefont
  {Paglione}}, \ and\ \bibinfo {author} {\bibfnamefont {N.~P.}\ \bibnamefont
  {Butch}},\ }\href@noop {} {\bibfield  {journal} {\bibinfo  {journal} {Phys.
  Rev. B}\ }\textbf {\bibinfo {volume} {101}},\ \bibinfo {pages} {140503(R)}
  (\bibinfo {year} {2020})}\BibitemShut {NoStop}%
\bibitem [{\citenamefont {Thomas}\ \emph {et~al.}(2020)\citenamefont {Thomas},
  \citenamefont {Santos}, \citenamefont {Christensen}, \citenamefont {Asaba},
  \citenamefont {Ronning}, \citenamefont {Thompson}, \citenamefont {Bauer},
  \citenamefont {Fernandes}, \citenamefont {Fabbris},\ and\ \citenamefont
  {Rosa}}]{Thomas2020}%
  \BibitemOpen
  \bibfield  {author} {\bibinfo {author} {\bibfnamefont {S.~M.}\ \bibnamefont
  {Thomas}}, \bibinfo {author} {\bibfnamefont {F.~B.}\ \bibnamefont {Santos}},
  \bibinfo {author} {\bibfnamefont {M.~H.}\ \bibnamefont {Christensen}},
  \bibinfo {author} {\bibfnamefont {T.}~\bibnamefont {Asaba}}, \bibinfo
  {author} {\bibfnamefont {F.}~\bibnamefont {Ronning}}, \bibinfo {author}
  {\bibfnamefont {J.~D.}\ \bibnamefont {Thompson}}, \bibinfo {author}
  {\bibfnamefont {E.~D.}\ \bibnamefont {Bauer}}, \bibinfo {author}
  {\bibfnamefont {R.~M.}\ \bibnamefont {Fernandes}}, \bibinfo {author}
  {\bibfnamefont {G.}~\bibnamefont {Fabbris}}, \ and\ \bibinfo {author}
  {\bibfnamefont {P.~F.~S.}\ \bibnamefont {Rosa}},\ }\href@noop {} {\bibfield
  {journal} {\bibinfo  {journal} {Sci. Adv.}\ }\textbf {\bibinfo {volume} {6}}
  (\bibinfo {year} {2020})}\BibitemShut {NoStop}%
\bibitem [{\citenamefont {Aoki}\ \emph {et~al.}(2020)\citenamefont {Aoki},
  \citenamefont {Honda}, \citenamefont {Knebel}, \citenamefont {Braithwaite},
  \citenamefont {Nakamura}, \citenamefont {Li}, \citenamefont {Homma},
  \citenamefont {Shimizu}, \citenamefont {Sato}, \citenamefont {Brison},\ and\
  \citenamefont {Flouquet}}]{Aoki2020}%
  \BibitemOpen
  \bibfield  {author} {\bibinfo {author} {\bibfnamefont {D.}~\bibnamefont
  {Aoki}}, \bibinfo {author} {\bibfnamefont {F.}~\bibnamefont {Honda}},
  \bibinfo {author} {\bibfnamefont {G.}~\bibnamefont {Knebel}}, \bibinfo
  {author} {\bibfnamefont {D.}~\bibnamefont {Braithwaite}}, \bibinfo {author}
  {\bibfnamefont {A.}~\bibnamefont {Nakamura}}, \bibinfo {author}
  {\bibfnamefont {D.}~\bibnamefont {Li}}, \bibinfo {author} {\bibfnamefont
  {Y.}~\bibnamefont {Homma}}, \bibinfo {author} {\bibfnamefont
  {Y.}~\bibnamefont {Shimizu}}, \bibinfo {author} {\bibfnamefont {Y.~J.}\
  \bibnamefont {Sato}}, \bibinfo {author} {\bibfnamefont {J.-P.}\ \bibnamefont
  {Brison}}, \ and\ \bibinfo {author} {\bibfnamefont {J.}~\bibnamefont
  {Flouquet}},\ }\href@noop {} {\bibfield  {journal} {\bibinfo  {journal} {J.
  Phys. Soc. Jpn.}\ }\textbf {\bibinfo {volume} {89}},\ \bibinfo {pages}
  {053705} (\bibinfo {year} {2020})}\BibitemShut {NoStop}%
\bibitem [{\citenamefont {Li}\ \emph {et~al.}(2021)\citenamefont {Li},
  \citenamefont {Nakamura}, \citenamefont {Honda}, \citenamefont {Sato},
  \citenamefont {Homma}, \citenamefont {Shimizu}, \citenamefont {Ishizuka},
  \citenamefont {Yanase}, \citenamefont {Knebel}, \citenamefont {Flouquet},\
  and\ \citenamefont {Aoki}}]{Li2021}%
  \BibitemOpen
  \bibfield  {author} {\bibinfo {author} {\bibfnamefont {D.}~\bibnamefont
  {Li}}, \bibinfo {author} {\bibfnamefont {A.}~\bibnamefont {Nakamura}},
  \bibinfo {author} {\bibfnamefont {F.}~\bibnamefont {Honda}}, \bibinfo
  {author} {\bibfnamefont {Y.~J.}\ \bibnamefont {Sato}}, \bibinfo {author}
  {\bibfnamefont {Y.}~\bibnamefont {Homma}}, \bibinfo {author} {\bibfnamefont
  {Y.}~\bibnamefont {Shimizu}}, \bibinfo {author} {\bibfnamefont
  {J.}~\bibnamefont {Ishizuka}}, \bibinfo {author} {\bibfnamefont
  {Y.}~\bibnamefont {Yanase}}, \bibinfo {author} {\bibfnamefont
  {G.}~\bibnamefont {Knebel}}, \bibinfo {author} {\bibfnamefont
  {J.}~\bibnamefont {Flouquet}}, \ and\ \bibinfo {author} {\bibfnamefont
  {D.}~\bibnamefont {Aoki}},\ }\href {\doibase 10.7566/JPSJ.90.073703}
  {\bibfield  {journal} {\bibinfo  {journal} {J. Phys. Soc. Jpn.}\ }\textbf
  {\bibinfo {volume} {90}},\ \bibinfo {pages} {073703} (\bibinfo {year}
  {2021})}\BibitemShut {NoStop}%
\bibitem [{\citenamefont {Ran}\ \emph {et~al.}(2021)\citenamefont {Ran},
  \citenamefont {Saha}, \citenamefont {Liu}, \citenamefont {Graf},
  \citenamefont {Paglione},\ and\ \citenamefont {Butch}}]{Ran2021}%
  \BibitemOpen
  \bibfield  {author} {\bibinfo {author} {\bibfnamefont {S.}~\bibnamefont
  {Ran}}, \bibinfo {author} {\bibfnamefont {S.~R.}\ \bibnamefont {Saha}},
  \bibinfo {author} {\bibfnamefont {I.-L.}\ \bibnamefont {Liu}}, \bibinfo
  {author} {\bibfnamefont {D.}~\bibnamefont {Graf}}, \bibinfo {author}
  {\bibfnamefont {J.}~\bibnamefont {Paglione}}, \ and\ \bibinfo {author}
  {\bibfnamefont {N.~P.}\ \bibnamefont {Butch}},\ }\href@noop {} {\bibfield
  {journal} {\bibinfo  {journal} {npj Quantum Mater.}\ }\textbf {\bibinfo
  {volume} {6}},\ \bibinfo {pages} {75} (\bibinfo {year} {2021})}\BibitemShut
  {NoStop}%
\bibitem [{\citenamefont {Aoki}\ \emph {et~al.}(2021)\citenamefont {Aoki},
  \citenamefont {Kimata}, \citenamefont {Sato}, \citenamefont {Knebel},
  \citenamefont {Honda}, \citenamefont {Nakamura}, \citenamefont {Li},
  \citenamefont {Homma}, \citenamefont {Shimizu}, \citenamefont {Knafo},
  \citenamefont {Braithwaite}, \citenamefont {Vali\v{s}ka}, \citenamefont
  {Pourret}, \citenamefont {Brison},\ and\ \citenamefont
  {Flouquet}}]{Aoki2021}%
  \BibitemOpen
  \bibfield  {author} {\bibinfo {author} {\bibfnamefont {D.}~\bibnamefont
  {Aoki}}, \bibinfo {author} {\bibfnamefont {M.}~\bibnamefont {Kimata}},
  \bibinfo {author} {\bibfnamefont {Y.~J.}\ \bibnamefont {Sato}}, \bibinfo
  {author} {\bibfnamefont {G.}~\bibnamefont {Knebel}}, \bibinfo {author}
  {\bibfnamefont {F.}~\bibnamefont {Honda}}, \bibinfo {author} {\bibfnamefont
  {A.}~\bibnamefont {Nakamura}}, \bibinfo {author} {\bibfnamefont
  {D.}~\bibnamefont {Li}}, \bibinfo {author} {\bibfnamefont {Y.}~\bibnamefont
  {Homma}}, \bibinfo {author} {\bibfnamefont {Y.}~\bibnamefont {Shimizu}},
  \bibinfo {author} {\bibfnamefont {W.}~\bibnamefont {Knafo}}, \bibinfo
  {author} {\bibfnamefont {D.}~\bibnamefont {Braithwaite}}, \bibinfo {author}
  {\bibfnamefont {M.}~\bibnamefont {Vali\v{s}ka}}, \bibinfo {author}
  {\bibfnamefont {A.}~\bibnamefont {Pourret}}, \bibinfo {author} {\bibfnamefont
  {J.-P.}\ \bibnamefont {Brison}}, \ and\ \bibinfo {author} {\bibfnamefont
  {J.}~\bibnamefont {Flouquet}},\ }\href@noop {} {\bibfield  {journal}
  {\bibinfo  {journal} {J. Phys. Soc. Jpn.}\ }\textbf {\bibinfo {volume}
  {90}},\ \bibinfo {pages} {074705} (\bibinfo {year} {2021})}\BibitemShut
  {NoStop}%
\bibitem [{\citenamefont {Vali{\v{s}}ka}\ \emph {et~al.}(2021)\citenamefont
  {Vali{\v{s}}ka}, \citenamefont {Knafo}, \citenamefont {Knebel}, \citenamefont
  {Lapertot}, \citenamefont {Aoki},\ and\ \citenamefont
  {Braithwaite}}]{Valiska2021}%
  \BibitemOpen
  \bibfield  {author} {\bibinfo {author} {\bibfnamefont {M.}~\bibnamefont
  {Vali{\v{s}}ka}}, \bibinfo {author} {\bibfnamefont {W.}~\bibnamefont
  {Knafo}}, \bibinfo {author} {\bibfnamefont {G.}~\bibnamefont {Knebel}},
  \bibinfo {author} {\bibfnamefont {G.}~\bibnamefont {Lapertot}}, \bibinfo
  {author} {\bibfnamefont {D.}~\bibnamefont {Aoki}}, \ and\ \bibinfo {author}
  {\bibfnamefont {D.}~\bibnamefont {Braithwaite}},\ }\href@noop {} {\bibfield
  {journal} {\bibinfo  {journal} {Phys. Rev. B}\ }\textbf {\bibinfo {volume}
  {104}},\ \bibinfo {pages} {214507} (\bibinfo {year} {2021})}\BibitemShut
  {NoStop}%
\bibitem [{\citenamefont {Duan}\ \emph {et~al.}(2020)\citenamefont {Duan},
  \citenamefont {Sasmal}, \citenamefont {Maple}, \citenamefont {Podlesnyak},
  \citenamefont {Zhu}, \citenamefont {Si},\ and\ \citenamefont
  {Dai}}]{Duan2020}%
  \BibitemOpen
  \bibfield  {author} {\bibinfo {author} {\bibfnamefont {C.}~\bibnamefont
  {Duan}}, \bibinfo {author} {\bibfnamefont {K.}~\bibnamefont {Sasmal}},
  \bibinfo {author} {\bibfnamefont {M.~B.}\ \bibnamefont {Maple}}, \bibinfo
  {author} {\bibfnamefont {A.}~\bibnamefont {Podlesnyak}}, \bibinfo {author}
  {\bibfnamefont {J.-X.}\ \bibnamefont {Zhu}}, \bibinfo {author} {\bibfnamefont
  {Q.}~\bibnamefont {Si}}, \ and\ \bibinfo {author} {\bibfnamefont
  {P.}~\bibnamefont {Dai}},\ }\href@noop {} {\bibfield  {journal} {\bibinfo
  {journal} {Phys. Rev. Lett.}\ }\textbf {\bibinfo {volume} {125}},\ \bibinfo
  {pages} {237003} (\bibinfo {year} {2020})}\BibitemShut {NoStop}%
\bibitem [{\citenamefont {Knafo}\ \emph
  {et~al.}(2021{\natexlab{b}})\citenamefont {Knafo}, \citenamefont {Knebel},
  \citenamefont {Steffens}, \citenamefont {Kaneko}, \citenamefont {Rosuel},
  \citenamefont {Brison}, \citenamefont {Flouquet}, \citenamefont {Aoki},
  \citenamefont {Lapertot},\ and\ \citenamefont {Raymond}}]{Knafo2021b}%
  \BibitemOpen
  \bibfield  {author} {\bibinfo {author} {\bibfnamefont {W.}~\bibnamefont
  {Knafo}}, \bibinfo {author} {\bibfnamefont {G.}~\bibnamefont {Knebel}},
  \bibinfo {author} {\bibfnamefont {P.}~\bibnamefont {Steffens}}, \bibinfo
  {author} {\bibfnamefont {K.}~\bibnamefont {Kaneko}}, \bibinfo {author}
  {\bibfnamefont {A.}~\bibnamefont {Rosuel}}, \bibinfo {author} {\bibfnamefont
  {J.-P.}\ \bibnamefont {Brison}}, \bibinfo {author} {\bibfnamefont
  {J.}~\bibnamefont {Flouquet}}, \bibinfo {author} {\bibfnamefont
  {D.}~\bibnamefont {Aoki}}, \bibinfo {author} {\bibfnamefont {G.}~\bibnamefont
  {Lapertot}}, \ and\ \bibinfo {author} {\bibfnamefont {S.}~\bibnamefont
  {Raymond}},\ }\href@noop {} {\bibfield  {journal} {\bibinfo  {journal} {Phys.
  Rev. B}\ }\textbf {\bibinfo {volume} {104}},\ \bibinfo {pages} {L100409}
  (\bibinfo {year} {2021}{\natexlab{b}})}\BibitemShut {NoStop}%
\bibitem [{\citenamefont {Butch}\ \emph {et~al.}(2022)\citenamefont {Butch},
  \citenamefont {Ran}, \citenamefont {Saha}, \citenamefont {Neves},
  \citenamefont {Zic}, \citenamefont {Paglione}, \citenamefont {Gladchenko},
  \citenamefont {Ye},\ and\ \citenamefont {Rodriguez}}]{Butch2021}%
  \BibitemOpen
  \bibfield  {author} {\bibinfo {author} {\bibfnamefont {N.~P.}\ \bibnamefont
  {Butch}}, \bibinfo {author} {\bibfnamefont {S.}~\bibnamefont {Ran}}, \bibinfo
  {author} {\bibfnamefont {S.~R.}\ \bibnamefont {Saha}}, \bibinfo {author}
  {\bibfnamefont {P.~M.}\ \bibnamefont {Neves}}, \bibinfo {author}
  {\bibfnamefont {M.~P.}\ \bibnamefont {Zic}}, \bibinfo {author} {\bibfnamefont
  {J.}~\bibnamefont {Paglione}}, \bibinfo {author} {\bibfnamefont
  {S.}~\bibnamefont {Gladchenko}}, \bibinfo {author} {\bibfnamefont
  {Q.}~\bibnamefont {Ye}}, \ and\ \bibinfo {author} {\bibfnamefont {J.~A.}\
  \bibnamefont {Rodriguez}},\ }\href@noop {} {\bibfield  {journal} {\bibinfo
  {journal} {npj Quantum Mater.}\ }\textbf {\bibinfo {volume} {7}},\ \bibinfo
  {pages} {39} (\bibinfo {year} {2022})}\BibitemShut {NoStop}%
\bibitem [{\citenamefont {Duan}\ \emph {et~al.}(2021)\citenamefont {Duan},
  \citenamefont {Baumbach}, \citenamefont {Podlesnyak}, \citenamefont {Deng},
  \citenamefont {Moir}, \citenamefont {Breindel}, \citenamefont {Maple},
  \citenamefont {Nica}, \citenamefont {Si},\ and\ \citenamefont
  {Dai}}]{Duan2021}%
  \BibitemOpen
  \bibfield  {author} {\bibinfo {author} {\bibfnamefont {C.}~\bibnamefont
  {Duan}}, \bibinfo {author} {\bibfnamefont {R.}~\bibnamefont {Baumbach}},
  \bibinfo {author} {\bibfnamefont {A.}~\bibnamefont {Podlesnyak}}, \bibinfo
  {author} {\bibfnamefont {Y.}~\bibnamefont {Deng}}, \bibinfo {author}
  {\bibfnamefont {C.}~\bibnamefont {Moir}}, \bibinfo {author} {\bibfnamefont
  {A.~J.}\ \bibnamefont {Breindel}}, \bibinfo {author} {\bibfnamefont {M.~B.}\
  \bibnamefont {Maple}}, \bibinfo {author} {\bibfnamefont {E.}~\bibnamefont
  {Nica}}, \bibinfo {author} {\bibfnamefont {Q.}~\bibnamefont {Si}}, \ and\
  \bibinfo {author} {\bibfnamefont {P.}~\bibnamefont {Dai}},\ }\href@noop {}
  {\bibfield  {journal} {\bibinfo  {journal} {Nature}\ }\textbf {\bibinfo
  {volume} {600}},\ \bibinfo {pages} {636} (\bibinfo {year}
  {2021})}\BibitemShut {NoStop}%
\bibitem [{\citenamefont {Raymond}\ \emph {et~al.}(2021)\citenamefont
  {Raymond}, \citenamefont {Knafo}, \citenamefont {Knebel}, \citenamefont
  {Kaneko}, \citenamefont {Brison}, \citenamefont {Flouquet}, \citenamefont
  {Aoki},\ and\ \citenamefont {Lapertot}}]{Raymond2021}%
  \BibitemOpen
  \bibfield  {author} {\bibinfo {author} {\bibfnamefont {S.}~\bibnamefont
  {Raymond}}, \bibinfo {author} {\bibfnamefont {W.}~\bibnamefont {Knafo}},
  \bibinfo {author} {\bibfnamefont {G.}~\bibnamefont {Knebel}}, \bibinfo
  {author} {\bibfnamefont {K.}~\bibnamefont {Kaneko}}, \bibinfo {author}
  {\bibfnamefont {J.-P.}\ \bibnamefont {Brison}}, \bibinfo {author}
  {\bibfnamefont {J.}~\bibnamefont {Flouquet}}, \bibinfo {author}
  {\bibfnamefont {D.}~\bibnamefont {Aoki}}, \ and\ \bibinfo {author}
  {\bibfnamefont {G.}~\bibnamefont {Lapertot}},\ }\href@noop {} {\bibfield
  {journal} {\bibinfo  {journal} {J. Phys. Soc. Jpn.}\ }\textbf {\bibinfo
  {volume} {90}},\ \bibinfo {pages} {113706} (\bibinfo {year}
  {2021})}\BibitemShut {NoStop}%
\bibitem [{\citenamefont {Eo}\ \emph {et~al.}()\citenamefont {Eo},
  \citenamefont {Saha}, \citenamefont {Kim}, \citenamefont {Ran}, \citenamefont
  {Horn}, \citenamefont {Hodovanets}, \citenamefont {Collini}, \citenamefont
  {Fuhrman}, \citenamefont {Nevidomskyy}, \citenamefont {Butch}, \citenamefont
  {Fuhrer},\ and\ \citenamefont {Paglione}}]{Eo2021}%
  \BibitemOpen
  \bibfield  {author} {\bibinfo {author} {\bibfnamefont {Y.~S.}\ \bibnamefont
  {Eo}}, \bibinfo {author} {\bibfnamefont {S.~R.}\ \bibnamefont {Saha}},
  \bibinfo {author} {\bibfnamefont {H.}~\bibnamefont {Kim}}, \bibinfo {author}
  {\bibfnamefont {S.}~\bibnamefont {Ran}}, \bibinfo {author} {\bibfnamefont
  {J.~A.}\ \bibnamefont {Horn}}, \bibinfo {author} {\bibfnamefont
  {H.}~\bibnamefont {Hodovanets}}, \bibinfo {author} {\bibfnamefont
  {J.}~\bibnamefont {Collini}}, \bibinfo {author} {\bibfnamefont {W.~T.}\
  \bibnamefont {Fuhrman}}, \bibinfo {author} {\bibfnamefont {A.~H.}\
  \bibnamefont {Nevidomskyy}}, \bibinfo {author} {\bibfnamefont {N.~P.}\
  \bibnamefont {Butch}}, \bibinfo {author} {\bibfnamefont {M.~S.}\ \bibnamefont
  {Fuhrer}}, \ and\ \bibinfo {author} {\bibfnamefont {J.}~\bibnamefont
  {Paglione}},\ }\href@noop {} {\bibinfo  {journal} {arXiv:2101.03102}\
  }\BibitemShut {NoStop}%
\bibitem [{\citenamefont {Tokunaga}\ \emph {et~al.}(2019)\citenamefont
  {Tokunaga}, \citenamefont {Sakai}, \citenamefont {Kambe}, \citenamefont
  {Hattori}, \citenamefont {Higa}, \citenamefont {Nakamine}, \citenamefont
  {Kitagawa}, \citenamefont {Ishida}, \citenamefont {Nakamura}, \citenamefont
  {Shimizu}, \citenamefont {Homma}, \citenamefont {Li}, \citenamefont {Honda},\
  and\ \citenamefont {Aoki}}]{Tokunaga2019}%
  \BibitemOpen
\bibfield  {journal} {  }\bibfield  {author} {\bibinfo {author} {\bibfnamefont
  {Y.}~\bibnamefont {Tokunaga}}, \bibinfo {author} {\bibfnamefont
  {H.}~\bibnamefont {Sakai}}, \bibinfo {author} {\bibfnamefont
  {S.}~\bibnamefont {Kambe}}, \bibinfo {author} {\bibfnamefont
  {T.}~\bibnamefont {Hattori}}, \bibinfo {author} {\bibfnamefont
  {N.}~\bibnamefont {Higa}}, \bibinfo {author} {\bibfnamefont {G.}~\bibnamefont
  {Nakamine}}, \bibinfo {author} {\bibfnamefont {S.}~\bibnamefont {Kitagawa}},
  \bibinfo {author} {\bibfnamefont {K.}~\bibnamefont {Ishida}}, \bibinfo
  {author} {\bibfnamefont {A.}~\bibnamefont {Nakamura}}, \bibinfo {author}
  {\bibfnamefont {Y.}~\bibnamefont {Shimizu}}, \bibinfo {author} {\bibfnamefont
  {Y.}~\bibnamefont {Homma}}, \bibinfo {author} {\bibfnamefont
  {D.}~\bibnamefont {Li}}, \bibinfo {author} {\bibfnamefont {F.}~\bibnamefont
  {Honda}}, \ and\ \bibinfo {author} {\bibfnamefont {D.}~\bibnamefont {Aoki}},\
  }\href@noop {} {\bibfield  {journal} {\bibinfo  {journal} {J. Phys. Soc.
  Jpn.}\ }\textbf {\bibinfo {volume} {88}},\ \bibinfo {pages} {073701}
  (\bibinfo {year} {2019})}\BibitemShut {NoStop}%
\bibitem [{\citenamefont {Xu}\ \emph {et~al.}(2019)\citenamefont {Xu},
  \citenamefont {Sheng},\ and\ \citenamefont {Yang}}]{Xu2019}%
  \BibitemOpen
  \bibfield  {author} {\bibinfo {author} {\bibfnamefont {Y.}~\bibnamefont
  {Xu}}, \bibinfo {author} {\bibfnamefont {Y.}~\bibnamefont {Sheng}}, \ and\
  \bibinfo {author} {\bibfnamefont {Y.-f.}\ \bibnamefont {Yang}},\ }\href@noop
  {} {\bibfield  {journal} {\bibinfo  {journal} {Phys. Rev. Lett.}\ }\textbf
  {\bibinfo {volume} {123}},\ \bibinfo {pages} {217002} (\bibinfo {year}
  {2019})}\BibitemShut {NoStop}%
\bibitem [{\citenamefont {Ishizuka}\ \emph {et~al.}(2019)\citenamefont
  {Ishizuka}, \citenamefont {Sumita}, \citenamefont {Daido},\ and\
  \citenamefont {Yanase}}]{Ishizuka2019}%
  \BibitemOpen
  \bibfield  {author} {\bibinfo {author} {\bibfnamefont {J.}~\bibnamefont
  {Ishizuka}}, \bibinfo {author} {\bibfnamefont {S.}~\bibnamefont {Sumita}},
  \bibinfo {author} {\bibfnamefont {A.}~\bibnamefont {Daido}}, \ and\ \bibinfo
  {author} {\bibfnamefont {Y.}~\bibnamefont {Yanase}},\ }\href@noop {}
  {\bibfield  {journal} {\bibinfo  {journal} {Phys. Rev. Lett.}\ }\textbf
  {\bibinfo {volume} {123}},\ \bibinfo {pages} {217001} (\bibinfo {year}
  {2019})}\BibitemShut {NoStop}%
\bibitem [{\citenamefont {Miao}\ \emph {et~al.}(2020)\citenamefont {Miao},
  \citenamefont {Liu}, \citenamefont {Xu}, \citenamefont {Kotta}, \citenamefont
  {Kang}, \citenamefont {Ran}, \citenamefont {Paglione}, \citenamefont
  {Kotliar}, \citenamefont {Butch}, \citenamefont {Denlinger},\ and\
  \citenamefont {Wray}}]{Miao2020}%
  \BibitemOpen
  \bibfield  {author} {\bibinfo {author} {\bibfnamefont {L.}~\bibnamefont
  {Miao}}, \bibinfo {author} {\bibfnamefont {S.}~\bibnamefont {Liu}}, \bibinfo
  {author} {\bibfnamefont {Y.}~\bibnamefont {Xu}}, \bibinfo {author}
  {\bibfnamefont {E.~C.}\ \bibnamefont {Kotta}}, \bibinfo {author}
  {\bibfnamefont {C.-J.}\ \bibnamefont {Kang}}, \bibinfo {author}
  {\bibfnamefont {S.}~\bibnamefont {Ran}}, \bibinfo {author} {\bibfnamefont
  {J.}~\bibnamefont {Paglione}}, \bibinfo {author} {\bibfnamefont
  {G.}~\bibnamefont {Kotliar}}, \bibinfo {author} {\bibfnamefont {N.~P.}\
  \bibnamefont {Butch}}, \bibinfo {author} {\bibfnamefont {J.~D.}\ \bibnamefont
  {Denlinger}}, \ and\ \bibinfo {author} {\bibfnamefont {L.~A.}\ \bibnamefont
  {Wray}},\ }\href@noop {} {\bibfield  {journal} {\bibinfo  {journal} {Phys.
  Rev. Lett.}\ }\textbf {\bibinfo {volume} {124}},\ \bibinfo {pages} {076401}
  (\bibinfo {year} {2020})}\BibitemShut {NoStop}%
\bibitem [{SM()}]{SM}%
  \BibitemOpen
  \href@noop {} {}\bibinfo {note} {See Supplemental Material for
  details.}\BibitemShut {Stop}%
\bibitem [{\citenamefont {Knebel}()}]{Knebel2022}%
  \BibitemOpen
  \bibfield  {author} {\bibinfo {author} {\bibfnamefont {G.}~\bibnamefont
  {Knebel}},\ }\href@noop {} {}\bibinfo {howpublished} {Private
  communication}\BibitemShut {NoStop}%
\bibitem [{\citenamefont {Niu}\ \emph {et~al.}(2020{\natexlab{b}})\citenamefont
  {Niu}, \citenamefont {Knebel}, \citenamefont {Braithwaite}, \citenamefont
  {Aoki}, \citenamefont {Lapertot}, \citenamefont {Seyfarth}, \citenamefont
  {Brison}, \citenamefont {Flouquet},\ and\ \citenamefont
  {Pourret}}]{Niu2020a}%
  \BibitemOpen
  \bibfield  {author} {\bibinfo {author} {\bibfnamefont {Q.}~\bibnamefont
  {Niu}}, \bibinfo {author} {\bibfnamefont {G.}~\bibnamefont {Knebel}},
  \bibinfo {author} {\bibfnamefont {D.}~\bibnamefont {Braithwaite}}, \bibinfo
  {author} {\bibfnamefont {D.}~\bibnamefont {Aoki}}, \bibinfo {author}
  {\bibfnamefont {G.}~\bibnamefont {Lapertot}}, \bibinfo {author}
  {\bibfnamefont {G.}~\bibnamefont {Seyfarth}}, \bibinfo {author}
  {\bibfnamefont {J.-P.}\ \bibnamefont {Brison}}, \bibinfo {author}
  {\bibfnamefont {J.}~\bibnamefont {Flouquet}}, \ and\ \bibinfo {author}
  {\bibfnamefont {A.}~\bibnamefont {Pourret}},\ }\href@noop {} {\bibfield
  {journal} {\bibinfo  {journal} {Phys. Rev. Lett.}\ }\textbf {\bibinfo
  {volume} {124}},\ \bibinfo {pages} {086601} (\bibinfo {year}
  {2020}{\natexlab{b}})}\BibitemShut {NoStop}%
\bibitem [{\citenamefont {Miyake}\ \emph {et~al.}(2021)\citenamefont {Miyake},
  \citenamefont {Shimizu}, \citenamefont {Sato}, \citenamefont {Li},
  \citenamefont {Nakamura}, \citenamefont {Homma}, \citenamefont {Honda},
  \citenamefont {Flouquet}, \citenamefont {Tokunaga},\ and\ \citenamefont
  {Aoki}}]{Miyake2021}%
  \BibitemOpen
  \bibfield  {author} {\bibinfo {author} {\bibfnamefont {A.}~\bibnamefont
  {Miyake}}, \bibinfo {author} {\bibfnamefont {Y.}~\bibnamefont {Shimizu}},
  \bibinfo {author} {\bibfnamefont {Y.~J.}\ \bibnamefont {Sato}}, \bibinfo
  {author} {\bibfnamefont {D.}~\bibnamefont {Li}}, \bibinfo {author}
  {\bibfnamefont {A.}~\bibnamefont {Nakamura}}, \bibinfo {author}
  {\bibfnamefont {Y.}~\bibnamefont {Homma}}, \bibinfo {author} {\bibfnamefont
  {F.}~\bibnamefont {Honda}}, \bibinfo {author} {\bibfnamefont
  {J.}~\bibnamefont {Flouquet}}, \bibinfo {author} {\bibfnamefont
  {M.}~\bibnamefont {Tokunaga}}, \ and\ \bibinfo {author} {\bibfnamefont
  {D.}~\bibnamefont {Aoki}},\ }\href@noop {} {\bibfield  {journal} {\bibinfo
  {journal} {J. Phys. Soc. Jpn.}\ }\textbf {\bibinfo {volume} {90}},\ \bibinfo
  {pages} {103702} (\bibinfo {year} {2021})}\BibitemShut {NoStop}%
\bibitem [{Note1()}]{Note1}%
  \BibitemOpen
  \bibinfo {note} {\label {note} The pertinence of the background subtractions
  done here is supported by the findings i) that the fields at the maxima of
  $\rho _{zz}$ versus $H$ extracted at constant temperatures coincide with the
  temperatures $T_{\Delta \rho _{zz}}^{max}$ at the maxima of $\Delta \rho
  _{zz}$ versus $T$ extracted at constant fields and ii) that the fields at the
  maxima of $\rho _{xx}$ versus $H$ extracted at constant temperatures coincide
  with the temperatures $T_{\Delta \rho _{xx}}^{max}$ at the maxima of $\Delta
  \rho _{xx}$ versus $T$ extracted at constant fields (see phase diagrams in
  Fig. \ref {Fig4}). Similar background-substraction procedure was done to
  analyze $\rho _{xx}$ data measured under pressure combined with magnetic
  fields in Ref. [\protect \rev@citealpnum {Valiska2021}].}\BibitemShut {Stop}%
\bibitem [{Note2()}]{Note2}%
  \BibitemOpen
  \bibinfo {note} {Details about the $T^2$ fits to the data are shown in the
  Supplemental Material for $\protect \mathbf {I}\parallel \protect \mathbf
  {c}$ and in the Supplementary Materials of Ref. [\protect \rev@citealpnum
  {Knafo2019}] for $\protect \mathbf {I}\parallel \protect \mathbf
  {a}$}\BibitemShut {NoStop}%
\bibitem [{\citenamefont {Thomas}\ \emph {et~al.}()\citenamefont {Thomas},
  \citenamefont {Stevens}, \citenamefont {Santos}, \citenamefont {Fender},
  \citenamefont {Bauer}, \citenamefont {Ronning}, \citenamefont {Thompson},
  \citenamefont {Huxley},\ and\ \citenamefont {Rosa}}]{Thomas2021}%
  \BibitemOpen
  \bibfield  {author} {\bibinfo {author} {\bibfnamefont {S.~M.}\ \bibnamefont
  {Thomas}}, \bibinfo {author} {\bibfnamefont {C.}~\bibnamefont {Stevens}},
  \bibinfo {author} {\bibfnamefont {F.~B.}\ \bibnamefont {Santos}}, \bibinfo
  {author} {\bibfnamefont {S.~S.}\ \bibnamefont {Fender}}, \bibinfo {author}
  {\bibfnamefont {E.~D.}\ \bibnamefont {Bauer}}, \bibinfo {author}
  {\bibfnamefont {F.}~\bibnamefont {Ronning}}, \bibinfo {author} {\bibfnamefont
  {J.~D.}\ \bibnamefont {Thompson}}, \bibinfo {author} {\bibfnamefont
  {A.}~\bibnamefont {Huxley}}, \ and\ \bibinfo {author} {\bibfnamefont
  {P.~F.~S.}\ \bibnamefont {Rosa}},\ }\href@noop {} {\bibinfo  {journal}
  {arXiv:2103.09194}\ }\BibitemShut {NoStop}%
\bibitem [{\citenamefont {Willa}\ \emph {et~al.}(2021)\citenamefont {Willa},
  \citenamefont {Hardy}, \citenamefont {Aoki}, \citenamefont {Li},
  \citenamefont {Wiecki}, \citenamefont {Lapertot},\ and\ \citenamefont
  {Meingast}}]{Willa2021}%
  \BibitemOpen
\bibfield  {journal} {  }\bibfield  {author} {\bibinfo {author} {\bibfnamefont
  {K.}~\bibnamefont {Willa}}, \bibinfo {author} {\bibfnamefont
  {F.}~\bibnamefont {Hardy}}, \bibinfo {author} {\bibfnamefont
  {D.}~\bibnamefont {Aoki}}, \bibinfo {author} {\bibfnamefont {D.}~\bibnamefont
  {Li}}, \bibinfo {author} {\bibfnamefont {P.}~\bibnamefont {Wiecki}}, \bibinfo
  {author} {\bibfnamefont {G.}~\bibnamefont {Lapertot}}, \ and\ \bibinfo
  {author} {\bibfnamefont {C.}~\bibnamefont {Meingast}},\ }\href@noop {}
  {\bibfield  {journal} {\bibinfo  {journal} {Phys. Rev. B}\ }\textbf {\bibinfo
  {volume} {104}},\ \bibinfo {pages} {205107} (\bibinfo {year}
  {2021})}\BibitemShut {NoStop}%
\bibitem [{Note3()}]{Note3}%
  \BibitemOpen
  \bibinfo {note} {In Ref. [\protect \rev@citealpnum {Willa2021}], it has been
  alternatively proposed that the magnetic susceptibility $\chi _a$, the
  electronic heat capacity, and the thermal expansion should not be compared to
  $\rho _{xx}$ or $\rho _{zz}$, but that they should be compared to $\partial
  \rho _{xx}/\partial T$, where a maximum at $\simeq 15$~K is also observed.
  Within this picture, a set of three energy scales may be needed to describe
  the transport and thermodynamic properties of UTe$_2$: the first one of
  $\simeq 15$~K already identified, the second one of $\simeq 7$~K defined at
  the maximum of $\partial \rho _{zz}/\partial T$, and the third one of $\simeq
  35$~K defined at the maximum of $\chi _b$. In a recent NMR investigation
  \cite {Tokunaga2022}, three temperatures scales were defined from the
  variation of the spin-spin-relaxation-rate $1/T_2$ measurements: $T_H=30$~K
  at the onset of low-temperature increase of $1/T_2$, $T_P=16$~K at a maximum
  of $1/T_2$, and $T_L=7$~K at the onset of a lower-temperature increase of
  $1/T_2$. These three temperatures were proposed to be respectively related
  with the temperatures $T_{\chi _b}^{max}$, $T^*$ and a third temperature
  $T_\mu =5$~K, below which the muon spin relaxation rate was found to increase
  \cite {Sundar2019}.}\BibitemShut {Stop}%
\bibitem [{\citenamefont {Tokunaga}\ \emph {et~al.}(2022)\citenamefont
  {Tokunaga}, \citenamefont {Sakai}, \citenamefont {Kambe}, \citenamefont
  {Haga}, \citenamefont {Tokiwa}, \citenamefont {Opletal}, \citenamefont
  {Fujibayashi}, \citenamefont {Kinjo}, \citenamefont {Kitagawa}, \citenamefont
  {Ishida}, \citenamefont {Nakamura}, \citenamefont {Shimizu}, \citenamefont
  {Homma}, \citenamefont {Li}, \citenamefont {Honda},\ and\ \citenamefont
  {Aoki}}]{Tokunaga2022}%
  \BibitemOpen
  \bibfield  {author} {\bibinfo {author} {\bibfnamefont {Y.}~\bibnamefont
  {Tokunaga}}, \bibinfo {author} {\bibfnamefont {H.}~\bibnamefont {Sakai}},
  \bibinfo {author} {\bibfnamefont {S.}~\bibnamefont {Kambe}}, \bibinfo
  {author} {\bibfnamefont {Y.}~\bibnamefont {Haga}}, \bibinfo {author}
  {\bibfnamefont {Y.}~\bibnamefont {Tokiwa}}, \bibinfo {author} {\bibfnamefont
  {P.}~\bibnamefont {Opletal}}, \bibinfo {author} {\bibfnamefont
  {H.}~\bibnamefont {Fujibayashi}}, \bibinfo {author} {\bibfnamefont
  {K.}~\bibnamefont {Kinjo}}, \bibinfo {author} {\bibfnamefont
  {S.}~\bibnamefont {Kitagawa}}, \bibinfo {author} {\bibfnamefont
  {K.}~\bibnamefont {Ishida}}, \bibinfo {author} {\bibfnamefont
  {A.}~\bibnamefont {Nakamura}}, \bibinfo {author} {\bibfnamefont
  {Y.}~\bibnamefont {Shimizu}}, \bibinfo {author} {\bibfnamefont
  {Y.}~\bibnamefont {Homma}}, \bibinfo {author} {\bibfnamefont
  {D.}~\bibnamefont {Li}}, \bibinfo {author} {\bibfnamefont {F.}~\bibnamefont
  {Honda}}, \ and\ \bibinfo {author} {\bibfnamefont {D.}~\bibnamefont {Aoki}},\
  }\href@noop {} {\bibfield  {journal} {\bibinfo  {journal} {J. Phys. Soc.
  Jpn.}\ }\textbf {\bibinfo {volume} {91}},\ \bibinfo {pages} {023707}
  (\bibinfo {year} {2022})}\BibitemShut {NoStop}%
\bibitem [{\citenamefont {Rossat-Mignod}\ \emph {et~al.}(1988)\citenamefont
  {Rossat-Mignod}, \citenamefont {Regnault}, \citenamefont {Jacoud},
  \citenamefont {Vettier}, \citenamefont {Lejay}, \citenamefont {Flouquet},
  \citenamefont {Walker}, \citenamefont {Jaccard},\ and\ \citenamefont
  {Amato}}]{RossatMignod1988}%
  \BibitemOpen
  \bibfield  {author} {\bibinfo {author} {\bibfnamefont {J.}~\bibnamefont
  {Rossat-Mignod}}, \bibinfo {author} {\bibfnamefont {L.}~\bibnamefont
  {Regnault}}, \bibinfo {author} {\bibfnamefont {J.}~\bibnamefont {Jacoud}},
  \bibinfo {author} {\bibfnamefont {C.}~\bibnamefont {Vettier}}, \bibinfo
  {author} {\bibfnamefont {P.}~\bibnamefont {Lejay}}, \bibinfo {author}
  {\bibfnamefont {J.}~\bibnamefont {Flouquet}}, \bibinfo {author}
  {\bibfnamefont {E.}~\bibnamefont {Walker}}, \bibinfo {author} {\bibfnamefont
  {D.}~\bibnamefont {Jaccard}}, \ and\ \bibinfo {author} {\bibfnamefont
  {A.}~\bibnamefont {Amato}},\ }\href@noop {} {\bibfield  {journal} {\bibinfo
  {journal} {J. Magn. Magn. Mater.}\ }\textbf {\bibinfo {volume} {76-77}},\
  \bibinfo {pages} {376} (\bibinfo {year} {1988})}\BibitemShut {NoStop}%
\bibitem [{Note4()}]{Note4}%
  \BibitemOpen
  \bibinfo {note} {Similar falls of two temperatures scales, $T_\chi ^{max}$ at
  the maximum of the magnetic susceptibility, and $T_0$ at the onset of an
  'hidden-order' phase transition, were observed in the vicinity of
  metamagnetism in the heavy-fermion paramagnet URu$_2$Si$_2$ [\protect
  \rev@citealpnum {Knafo2020}].}\BibitemShut {Stop}%
\bibitem [{\citenamefont {Raymond}\ \emph {et~al.}(1998)\citenamefont
  {Raymond}, \citenamefont {Regnault}, \citenamefont {Kambe}, \citenamefont
  {Flouquet},\ and\ \citenamefont {Lejay}}]{Raymond1998}%
  \BibitemOpen
  \bibfield  {author} {\bibinfo {author} {\bibfnamefont {S.}~\bibnamefont
  {Raymond}}, \bibinfo {author} {\bibfnamefont {L.~P.}\ \bibnamefont
  {Regnault}}, \bibinfo {author} {\bibfnamefont {S.}~\bibnamefont {Kambe}},
  \bibinfo {author} {\bibfnamefont {J.}~\bibnamefont {Flouquet}}, \ and\
  \bibinfo {author} {\bibfnamefont {P.}~\bibnamefont {Lejay}},\ }\href
  {\doibase 10.1088/0953-8984/10/11/002} {\bibfield  {journal} {\bibinfo
  {journal} {J. Phys.: Condens. Matter}\ }\textbf {\bibinfo {volume} {10}},\
  \bibinfo {pages} {2363} (\bibinfo {year} {1998})}\BibitemShut {NoStop}%
\bibitem [{\citenamefont {Flouquet}\ \emph {et~al.}(2004)\citenamefont
  {Flouquet}, \citenamefont {Haga}, \citenamefont {Haen}, \citenamefont
  {Braithwaite}, \citenamefont {Knebel}, \citenamefont {Raymond},\ and\
  \citenamefont {Kambe}}]{Flouquet2004}%
  \BibitemOpen
  \bibfield  {author} {\bibinfo {author} {\bibfnamefont {J.}~\bibnamefont
  {Flouquet}}, \bibinfo {author} {\bibfnamefont {Y.}~\bibnamefont {Haga}},
  \bibinfo {author} {\bibfnamefont {P.}~\bibnamefont {Haen}}, \bibinfo {author}
  {\bibfnamefont {D.}~\bibnamefont {Braithwaite}}, \bibinfo {author}
  {\bibfnamefont {G.}~\bibnamefont {Knebel}}, \bibinfo {author} {\bibfnamefont
  {S.}~\bibnamefont {Raymond}}, \ and\ \bibinfo {author} {\bibfnamefont
  {S.}~\bibnamefont {Kambe}},\ }\href@noop {} {\bibfield  {journal} {\bibinfo
  {journal} {J. Magn. Magn. Mater.}\ }\textbf {\bibinfo {volume} {272-276}},\
  \bibinfo {pages} {27 } (\bibinfo {year} {2004})}\BibitemShut {NoStop}%
\bibitem [{\citenamefont {Sato}\ \emph {et~al.}(2004)\citenamefont {Sato},
  \citenamefont {Koike}, \citenamefont {Katano}, \citenamefont {Metoki},
  \citenamefont {Kadowaki},\ and\ \citenamefont {Kawarazaki}}]{Sato2004}%
  \BibitemOpen
  \bibfield  {author} {\bibinfo {author} {\bibfnamefont {M.}~\bibnamefont
  {Sato}}, \bibinfo {author} {\bibfnamefont {Y.}~\bibnamefont {Koike}},
  \bibinfo {author} {\bibfnamefont {S.}~\bibnamefont {Katano}}, \bibinfo
  {author} {\bibfnamefont {N.}~\bibnamefont {Metoki}}, \bibinfo {author}
  {\bibfnamefont {H.}~\bibnamefont {Kadowaki}}, \ and\ \bibinfo {author}
  {\bibfnamefont {S.}~\bibnamefont {Kawarazaki}},\ }\href {\doibase
  10.1143/JPSJ.73.3418} {\bibfield  {journal} {\bibinfo  {journal} {J. Phys.
  Soc. Jpn.}\ }\textbf {\bibinfo {volume} {73}},\ \bibinfo {pages} {3418}
  (\bibinfo {year} {2004})}\BibitemShut {NoStop}%
\bibitem [{\citenamefont {Lester}\ \emph {et~al.}(2021)\citenamefont {Lester},
  \citenamefont {Ramos}, \citenamefont {Perry}, \citenamefont {Croft},
  \citenamefont {Laver}, \citenamefont {Bewley}, \citenamefont {Guidi},
  \citenamefont {Hiess}, \citenamefont {Wildes}, \citenamefont {Forgan} \emph
  {et~al.}}]{Lester2021}%
  \BibitemOpen
  \bibfield  {author} {\bibinfo {author} {\bibfnamefont {C.}~\bibnamefont
  {Lester}}, \bibinfo {author} {\bibfnamefont {S.}~\bibnamefont {Ramos}},
  \bibinfo {author} {\bibfnamefont {R.}~\bibnamefont {Perry}}, \bibinfo
  {author} {\bibfnamefont {T.}~\bibnamefont {Croft}}, \bibinfo {author}
  {\bibfnamefont {M.}~\bibnamefont {Laver}}, \bibinfo {author} {\bibfnamefont
  {R.}~\bibnamefont {Bewley}}, \bibinfo {author} {\bibfnamefont
  {T.}~\bibnamefont {Guidi}}, \bibinfo {author} {\bibfnamefont
  {A.}~\bibnamefont {Hiess}}, \bibinfo {author} {\bibfnamefont
  {A.}~\bibnamefont {Wildes}}, \bibinfo {author} {\bibfnamefont
  {E.}~\bibnamefont {Forgan}},  \emph {et~al.},\ }\href@noop {} {\bibfield
  {journal} {\bibinfo  {journal} {Nat. Comm.}\ }\textbf {\bibinfo {volume}
  {12}},\ \bibinfo {pages} {1} (\bibinfo {year} {2021})}\BibitemShut {NoStop}%
\bibitem [{\citenamefont {Mills}\ and\ \citenamefont
  {Lederer}(1966)}]{Mills1966}%
  \BibitemOpen
  \bibfield  {author} {\bibinfo {author} {\bibfnamefont {D.}~\bibnamefont
  {Mills}}\ and\ \bibinfo {author} {\bibfnamefont {P.}~\bibnamefont
  {Lederer}},\ }\href@noop {} {\bibfield  {journal} {\bibinfo  {journal} {J.
  Phys. Chem. Solids.}\ }\textbf {\bibinfo {volume} {27}},\ \bibinfo {pages}
  {1805} (\bibinfo {year} {1966})}\BibitemShut {NoStop}%
\bibitem [{\citenamefont {Jullien}\ \emph {et~al.}(1974)\citenamefont
  {Jullien}, \citenamefont {B\'eal-Monod},\ and\ \citenamefont
  {Coqblin}}]{Jullien1974}%
  \BibitemOpen
  \bibfield  {author} {\bibinfo {author} {\bibfnamefont {R.}~\bibnamefont
  {Jullien}}, \bibinfo {author} {\bibfnamefont {M.~T.}\ \bibnamefont
  {B\'eal-Monod}}, \ and\ \bibinfo {author} {\bibfnamefont {B.}~\bibnamefont
  {Coqblin}},\ }\href@noop {} {\bibfield  {journal} {\bibinfo  {journal} {Phys.
  Rev. B}\ }\textbf {\bibinfo {volume} {9}},\ \bibinfo {pages} {1441} (\bibinfo
  {year} {1974})}\BibitemShut {NoStop}%
\bibitem [{\citenamefont {Moriya}\ and\ \citenamefont
  {Takimoto}(1995)}]{Moriya1995}%
  \BibitemOpen
  \bibfield  {author} {\bibinfo {author} {\bibfnamefont {T.}~\bibnamefont
  {Moriya}}\ and\ \bibinfo {author} {\bibfnamefont {T.}~\bibnamefont
  {Takimoto}},\ }\href@noop {} {\bibfield  {journal} {\bibinfo  {journal} {J.
  Phys. Soc. Jpn.}\ }\textbf {\bibinfo {volume} {64}},\ \bibinfo {pages} {960}
  (\bibinfo {year} {1995})}\BibitemShut {NoStop}%
\bibitem [{\citenamefont {Rosch}(1999)}]{Rosch1999}%
  \BibitemOpen
  \bibfield  {author} {\bibinfo {author} {\bibfnamefont {A.}~\bibnamefont
  {Rosch}},\ }\href@noop {} {\bibfield  {journal} {\bibinfo  {journal} {Phys.
  Rev. Lett.}\ }\textbf {\bibinfo {volume} {82}},\ \bibinfo {pages} {4280}
  (\bibinfo {year} {1999})}\BibitemShut {NoStop}%
\bibitem [{\citenamefont {Cooper}\ \emph {et~al.}(1985)\citenamefont {Cooper},
  \citenamefont {Siemann}, \citenamefont {Yang}, \citenamefont {Thayamballi},\
  and\ \citenamefont {Benerjea}}]{Cooper1985}%
  \BibitemOpen
  \bibfield  {author} {\bibinfo {author} {\bibfnamefont {B.}~\bibnamefont
  {Cooper}}, \bibinfo {author} {\bibfnamefont {R.}~\bibnamefont {Siemann}},
  \bibinfo {author} {\bibfnamefont {D.}~\bibnamefont {Yang}}, \bibinfo {author}
  {\bibfnamefont {P.}~\bibnamefont {Thayamballi}}, \ and\ \bibinfo {author}
  {\bibfnamefont {A.}~\bibnamefont {Benerjea}},\ }\href@noop {} {\emph
  {\bibinfo {title} {{Handbook on the Physics and Chemistry of the Actinides,
  vol. 2}}}}\ (\bibinfo  {publisher} {edited by A.J. Freeman and G.H. Lander,
  North-Holland, Amsterdam},\ \bibinfo {year} {1985})\ Chap.~\bibinfo {chapter}
  {6}, p.\ \bibinfo {pages} {435}\BibitemShut {NoStop}%
\bibitem [{\citenamefont {Kinjo}\ \emph {et~al.}(2022)\citenamefont {Kinjo},
  \citenamefont {Fujibayashi}, \citenamefont {Nakamine}, \citenamefont
  {Kitagawa}, \citenamefont {Ishida}, \citenamefont {Tokunaga}, \citenamefont
  {Sakai}, \citenamefont {Kambe}, \citenamefont {Nakamura}, \citenamefont
  {Shimizu}, \citenamefont {Homma}, \citenamefont {Li}, \citenamefont {Honda},\
  and\ \citenamefont {Aoki}}]{Kinjo2022}%
  \BibitemOpen
  \bibfield  {author} {\bibinfo {author} {\bibfnamefont {K.}~\bibnamefont
  {Kinjo}}, \bibinfo {author} {\bibfnamefont {H.}~\bibnamefont {Fujibayashi}},
  \bibinfo {author} {\bibfnamefont {G.}~\bibnamefont {Nakamine}}, \bibinfo
  {author} {\bibfnamefont {S.}~\bibnamefont {Kitagawa}}, \bibinfo {author}
  {\bibfnamefont {K.}~\bibnamefont {Ishida}}, \bibinfo {author} {\bibfnamefont
  {Y.}~\bibnamefont {Tokunaga}}, \bibinfo {author} {\bibfnamefont
  {H.}~\bibnamefont {Sakai}}, \bibinfo {author} {\bibfnamefont
  {S.}~\bibnamefont {Kambe}}, \bibinfo {author} {\bibfnamefont
  {A.}~\bibnamefont {Nakamura}}, \bibinfo {author} {\bibfnamefont
  {Y.}~\bibnamefont {Shimizu}}, \bibinfo {author} {\bibfnamefont
  {Y.}~\bibnamefont {Homma}}, \bibinfo {author} {\bibfnamefont
  {D.}~\bibnamefont {Li}}, \bibinfo {author} {\bibfnamefont {F.}~\bibnamefont
  {Honda}}, \ and\ \bibinfo {author} {\bibfnamefont {D.}~\bibnamefont {Aoki}},\
  }\href@noop {} {\bibfield  {journal} {\bibinfo  {journal} {Phys. Rev. B}\
  }\textbf {\bibinfo {volume} {105}},\ \bibinfo {pages} {L140502} (\bibinfo
  {year} {2022})}\BibitemShut {NoStop}%
\bibitem [{\citenamefont {Sundar}\ \emph {et~al.}(2019)\citenamefont {Sundar},
  \citenamefont {Gheidi}, \citenamefont {Akintola}, \citenamefont {C\^ot\'e},
  \citenamefont {Dunsiger}, \citenamefont {Ran}, \citenamefont {Butch},
  \citenamefont {Saha}, \citenamefont {Paglione},\ and\ \citenamefont
  {Sonier}}]{Sundar2019}%
  \BibitemOpen
  \bibfield  {author} {\bibinfo {author} {\bibfnamefont {S.}~\bibnamefont
  {Sundar}}, \bibinfo {author} {\bibfnamefont {S.}~\bibnamefont {Gheidi}},
  \bibinfo {author} {\bibfnamefont {K.}~\bibnamefont {Akintola}}, \bibinfo
  {author} {\bibfnamefont {A.~M.}\ \bibnamefont {C\^ot\'e}}, \bibinfo {author}
  {\bibfnamefont {S.~R.}\ \bibnamefont {Dunsiger}}, \bibinfo {author}
  {\bibfnamefont {S.}~\bibnamefont {Ran}}, \bibinfo {author} {\bibfnamefont
  {N.~P.}\ \bibnamefont {Butch}}, \bibinfo {author} {\bibfnamefont {S.~R.}\
  \bibnamefont {Saha}}, \bibinfo {author} {\bibfnamefont {J.}~\bibnamefont
  {Paglione}}, \ and\ \bibinfo {author} {\bibfnamefont {J.~E.}\ \bibnamefont
  {Sonier}},\ }\href@noop {} {\bibfield  {journal} {\bibinfo  {journal} {Phys.
  Rev. B}\ }\textbf {\bibinfo {volume} {100}},\ \bibinfo {pages} {140502(R)}
  (\bibinfo {year} {2019})}\BibitemShut {NoStop}%
\bibitem [{\citenamefont {Knafo}\ \emph {et~al.}(2020)\citenamefont {Knafo},
  \citenamefont {Araki}, \citenamefont {Lapertot}, \citenamefont {Aoki},
  \citenamefont {Knebel},\ and\ \citenamefont {Braithwaite}}]{Knafo2020}%
  \BibitemOpen
  \bibfield  {author} {\bibinfo {author} {\bibfnamefont {W.}~\bibnamefont
  {Knafo}}, \bibinfo {author} {\bibfnamefont {S.}~\bibnamefont {Araki}},
  \bibinfo {author} {\bibfnamefont {G.}~\bibnamefont {Lapertot}}, \bibinfo
  {author} {\bibfnamefont {D.}~\bibnamefont {Aoki}}, \bibinfo {author}
  {\bibfnamefont {G.}~\bibnamefont {Knebel}}, \ and\ \bibinfo {author}
  {\bibfnamefont {D.}~\bibnamefont {Braithwaite}},\ }\href@noop {} {\bibfield
  {journal} {\bibinfo  {journal} {Nat. Phys.}\ }\textbf {\bibinfo {volume}
  {16}},\ \bibinfo {pages} {942} (\bibinfo {year} {2020})}\BibitemShut
  {NoStop}%
\end{thebibliography}

\begin{thebibliography}{11}%
\makeatletter
\providecommand \@ifxundefined [1]{%
 \@ifx{#1\undefined}
}%
\providecommand \@ifnum [1]{%
 \ifnum #1\expandafter \@firstoftwo
 \else \expandafter \@secondoftwo
 \fi
}%
\providecommand \@ifx [1]{%
 \ifx #1\expandafter \@firstoftwo
 \else \expandafter \@secondoftwo
 \fi
}%
\providecommand \natexlab [1]{#1}%
\providecommand \enquote  [1]{``#1''}%
\providecommand \bibnamefont  [1]{#1}%
\providecommand \bibfnamefont [1]{#1}%
\providecommand \citenamefont [1]{#1}%
\providecommand \href@noop [0]{\@secondoftwo}%
\providecommand \href [0]{\begingroup \@sanitize@url \@href}%
\providecommand \@href[1]{\@@startlink{#1}\@@href}%
\providecommand \@@href[1]{\endgroup#1\@@endlink}%
\providecommand \@sanitize@url [0]{\catcode `\\12\catcode `\$12\catcode
  `\&12\catcode `\#12\catcode `\^12\catcode `\_12\catcode `\%12\relax}%
\providecommand \@@startlink[1]{}%
\providecommand \@@endlink[0]{}%
\providecommand \url  [0]{\begingroup\@sanitize@url \@url }%
\providecommand \@url [1]{\endgroup\@href {#1}{\urlprefix }}%
\providecommand \urlprefix  [0]{URL }%
\providecommand \Eprint [0]{\href }%
\providecommand \doibase [0]{http://dx.doi.org/}%
\providecommand \selectlanguage [0]{\@gobble}%
\providecommand \bibinfo  [0]{\@secondoftwo}%
\providecommand \bibfield  [0]{\@secondoftwo}%
\providecommand \translation [1]{[#1]}%
\providecommand \BibitemOpen [0]{}%
\providecommand \bibitemStop [0]{}%
\providecommand \bibitemNoStop [0]{.\EOS\space}%
\providecommand \EOS [0]{\spacefactor3000\relax}%
\providecommand \BibitemShut  [1]{\csname bibitem#1\endcsname}%
\let\auto@bib@innerbib\@empty
%</preamble>
\bibitem [{\citenamefont {Xu}\ \emph {et~al.}(2019)\citenamefont {Xu},
  \citenamefont {Sheng},\ and\ \citenamefont {Yang}}]{Xu2019_sm}%
  \BibitemOpen
  \bibfield  {author} {\bibinfo {author} {\bibfnamefont {Y.}~\bibnamefont
  {Xu}}, \bibinfo {author} {\bibfnamefont {Y.}~\bibnamefont {Sheng}}, \ and\
  \bibinfo {author} {\bibfnamefont {Y.-f.}\ \bibnamefont {Yang}},\ }\href@noop
  {} {\bibfield  {journal} {\bibinfo  {journal} {Phys. Rev. Lett.}\ }\textbf
  {\bibinfo {volume} {123}},\ \bibinfo {pages} {217002} (\bibinfo {year}
  {2019})}\BibitemShut {NoStop}%
\bibitem [{\citenamefont {Ishizuka}\ \emph {et~al.}(2019)\citenamefont
  {Ishizuka}, \citenamefont {Sumita}, \citenamefont {Daido},\ and\
  \citenamefont {Yanase}}]{Ishizuka2019_sm}%
  \BibitemOpen
  \bibfield  {author} {\bibinfo {author} {\bibfnamefont {J.}~\bibnamefont
  {Ishizuka}}, \bibinfo {author} {\bibfnamefont {S.}~\bibnamefont {Sumita}},
  \bibinfo {author} {\bibfnamefont {A.}~\bibnamefont {Daido}}, \ and\ \bibinfo
  {author} {\bibfnamefont {Y.}~\bibnamefont {Yanase}},\ }\href@noop {}
  {\bibfield  {journal} {\bibinfo  {journal} {Phys. Rev. Lett.}\ }\textbf
  {\bibinfo {volume} {123}},\ \bibinfo {pages} {217001} (\bibinfo {year}
  {2019})}\BibitemShut {NoStop}%
\bibitem [{\citenamefont {Miao}\ \emph {et~al.}(2020)\citenamefont {Miao},
  \citenamefont {Liu}, \citenamefont {Xu}, \citenamefont {Kotta}, \citenamefont
  {Kang}, \citenamefont {Ran}, \citenamefont {Paglione}, \citenamefont
  {Kotliar}, \citenamefont {Butch}, \citenamefont {Denlinger},\ and\
  \citenamefont {Wray}}]{Miao2020_sm}%
  \BibitemOpen
  \bibfield  {author} {\bibinfo {author} {\bibfnamefont {L.}~\bibnamefont
  {Miao}}, \bibinfo {author} {\bibfnamefont {S.}~\bibnamefont {Liu}}, \bibinfo
  {author} {\bibfnamefont {Y.}~\bibnamefont {Xu}}, \bibinfo {author}
  {\bibfnamefont {E.~C.}\ \bibnamefont {Kotta}}, \bibinfo {author}
  {\bibfnamefont {C.-J.}\ \bibnamefont {Kang}}, \bibinfo {author}
  {\bibfnamefont {S.}~\bibnamefont {Ran}}, \bibinfo {author} {\bibfnamefont
  {J.}~\bibnamefont {Paglione}}, \bibinfo {author} {\bibfnamefont
  {G.}~\bibnamefont {Kotliar}}, \bibinfo {author} {\bibfnamefont {N.~P.}\
  \bibnamefont {Butch}}, \bibinfo {author} {\bibfnamefont {J.~D.}\ \bibnamefont
  {Denlinger}}, \ and\ \bibinfo {author} {\bibfnamefont {L.~A.}\ \bibnamefont
  {Wray}},\ }\href@noop {} {\bibfield  {journal} {\bibinfo  {journal} {Phys.
  Rev. Lett.}\ }\textbf {\bibinfo {volume} {124}},\ \bibinfo {pages} {076401}
  (\bibinfo {year} {2020})}\BibitemShut {NoStop}%
\bibitem [{\citenamefont {\={O}nuki}(2018)}]{Onuki2018_sm}%
  \BibitemOpen
  \bibfield  {author} {\bibinfo {author} {\bibfnamefont {Y.}~\bibnamefont
  {\={O}nuki}},\ }\href@noop {} {\emph {\bibinfo {title} {{Physics of Heavy
  Fermions}}}}\ (\bibinfo  {publisher} {World Scientific Publishing Company},\
  \bibinfo {year} {2018})\ Chap.~\bibinfo {chapter} {3}\BibitemShut {NoStop}%
\bibitem [{\citenamefont {Niu}\ \emph {et~al.}(2020)\citenamefont {Niu},
  \citenamefont {Knebel}, \citenamefont {Braithwaite}, \citenamefont {Aoki},
  \citenamefont {Lapertot}, \citenamefont {Seyfarth}, \citenamefont {Brison},
  \citenamefont {Flouquet},\ and\ \citenamefont {Pourret}}]{Niu2020a_sm}%
  \BibitemOpen
  \bibfield  {author} {\bibinfo {author} {\bibfnamefont {Q.}~\bibnamefont
  {Niu}}, \bibinfo {author} {\bibfnamefont {G.}~\bibnamefont {Knebel}},
  \bibinfo {author} {\bibfnamefont {D.}~\bibnamefont {Braithwaite}}, \bibinfo
  {author} {\bibfnamefont {D.}~\bibnamefont {Aoki}}, \bibinfo {author}
  {\bibfnamefont {G.}~\bibnamefont {Lapertot}}, \bibinfo {author}
  {\bibfnamefont {G.}~\bibnamefont {Seyfarth}}, \bibinfo {author}
  {\bibfnamefont {J.-P.}\ \bibnamefont {Brison}}, \bibinfo {author}
  {\bibfnamefont {J.}~\bibnamefont {Flouquet}}, \ and\ \bibinfo {author}
  {\bibfnamefont {A.}~\bibnamefont {Pourret}},\ }\href@noop {} {\bibfield
  {journal} {\bibinfo  {journal} {Phys. Rev. Lett.}\ }\textbf {\bibinfo
  {volume} {124}},\ \bibinfo {pages} {086601} (\bibinfo {year}
  {2020})}\BibitemShut {NoStop}%
\bibitem [{\citenamefont {Miyake}\ \emph {et~al.}(2019)\citenamefont {Miyake},
  \citenamefont {Shimizu}, \citenamefont {Sato}, \citenamefont {Li},
  \citenamefont {Nakamura}, \citenamefont {Homma}, \citenamefont {Honda},
  \citenamefont {Flouquet}, \citenamefont {Tokunaga},\ and\ \citenamefont
  {Aoki}}]{Miyake2019_sm}%
  \BibitemOpen
  \bibfield  {author} {\bibinfo {author} {\bibfnamefont {A.}~\bibnamefont
  {Miyake}}, \bibinfo {author} {\bibfnamefont {Y.}~\bibnamefont {Shimizu}},
  \bibinfo {author} {\bibfnamefont {Y.~J.}\ \bibnamefont {Sato}}, \bibinfo
  {author} {\bibfnamefont {D.}~\bibnamefont {Li}}, \bibinfo {author}
  {\bibfnamefont {A.}~\bibnamefont {Nakamura}}, \bibinfo {author}
  {\bibfnamefont {Y.}~\bibnamefont {Homma}}, \bibinfo {author} {\bibfnamefont
  {F.}~\bibnamefont {Honda}}, \bibinfo {author} {\bibfnamefont
  {J.}~\bibnamefont {Flouquet}}, \bibinfo {author} {\bibfnamefont
  {M.}~\bibnamefont {Tokunaga}}, \ and\ \bibinfo {author} {\bibfnamefont
  {D.}~\bibnamefont {Aoki}},\ }\href@noop {} {\bibfield  {journal} {\bibinfo
  {journal} {J. Phys. Soc. Jpn.}\ }\textbf {\bibinfo {volume} {88}},\ \bibinfo
  {pages} {063706} (\bibinfo {year} {2019})}\BibitemShut {NoStop}%
\bibitem [{\citenamefont {Miyake}\ \emph {et~al.}(2021)\citenamefont {Miyake},
  \citenamefont {Shimizu}, \citenamefont {Sato}, \citenamefont {Li},
  \citenamefont {Nakamura}, \citenamefont {Homma}, \citenamefont {Honda},
  \citenamefont {Flouquet}, \citenamefont {Tokunaga},\ and\ \citenamefont
  {Aoki}}]{Miyake2021_sm}%
  \BibitemOpen
  \bibfield  {author} {\bibinfo {author} {\bibfnamefont {A.}~\bibnamefont
  {Miyake}}, \bibinfo {author} {\bibfnamefont {Y.}~\bibnamefont {Shimizu}},
  \bibinfo {author} {\bibfnamefont {Y.~J.}\ \bibnamefont {Sato}}, \bibinfo
  {author} {\bibfnamefont {D.}~\bibnamefont {Li}}, \bibinfo {author}
  {\bibfnamefont {A.}~\bibnamefont {Nakamura}}, \bibinfo {author}
  {\bibfnamefont {Y.}~\bibnamefont {Homma}}, \bibinfo {author} {\bibfnamefont
  {F.}~\bibnamefont {Honda}}, \bibinfo {author} {\bibfnamefont
  {J.}~\bibnamefont {Flouquet}}, \bibinfo {author} {\bibfnamefont
  {M.}~\bibnamefont {Tokunaga}}, \ and\ \bibinfo {author} {\bibfnamefont
  {D.}~\bibnamefont {Aoki}},\ }\href@noop {} {\bibfield  {journal} {\bibinfo
  {journal} {J. Phys. Soc. Jpn.}\ }\textbf {\bibinfo {volume} {90}},\ \bibinfo
  {pages} {103702} (\bibinfo {year} {2021})}\BibitemShut {NoStop}%
\bibitem [{\citenamefont {Braithwaite}\ \emph {et~al.}(2019)\citenamefont
  {Braithwaite}, \citenamefont {Vali\v{s}ka}, \citenamefont {Knebel},
  \citenamefont {Lapertot}, \citenamefont {Brison}, \citenamefont {Pourret},
  \citenamefont {Zhitomirsky}, \citenamefont {Flouquet}, \citenamefont
  {Honda},\ and\ \citenamefont {Aoki}}]{Braithwaite2019_sm}%
  \BibitemOpen
  \bibfield  {author} {\bibinfo {author} {\bibfnamefont {D.}~\bibnamefont
  {Braithwaite}}, \bibinfo {author} {\bibfnamefont {M.}~\bibnamefont
  {Vali\v{s}ka}}, \bibinfo {author} {\bibfnamefont {G.}~\bibnamefont {Knebel}},
  \bibinfo {author} {\bibfnamefont {G.}~\bibnamefont {Lapertot}}, \bibinfo
  {author} {\bibfnamefont {J.-P.}\ \bibnamefont {Brison}}, \bibinfo {author}
  {\bibfnamefont {A.}~\bibnamefont {Pourret}}, \bibinfo {author} {\bibfnamefont
  {M.~E.}\ \bibnamefont {Zhitomirsky}}, \bibinfo {author} {\bibfnamefont
  {J.}~\bibnamefont {Flouquet}}, \bibinfo {author} {\bibfnamefont
  {F.}~\bibnamefont {Honda}}, \ and\ \bibinfo {author} {\bibfnamefont
  {D.}~\bibnamefont {Aoki}},\ }\href@noop {} {\bibfield  {journal} {\bibinfo
  {journal} {Commun. Phys.}\ }\textbf {\bibinfo {volume} {2}},\ \bibinfo
  {pages} {147} (\bibinfo {year} {2019})}\BibitemShut {NoStop}%
\bibitem [{\citenamefont {Knebel}\ \emph {et~al.}(2020)\citenamefont {Knebel},
  \citenamefont {Kimata}, \citenamefont {Vali\v{s}ka}, \citenamefont {Honda},
  \citenamefont {Li}, \citenamefont {Braithwaite}, \citenamefont {Lapertot},
  \citenamefont {Knafo}, \citenamefont {Pourret}, \citenamefont {Sato},
  \citenamefont {Shimizu}, \citenamefont {Kihara}, \citenamefont {Brison},
  \citenamefont {Flouquet},\ and\ \citenamefont {Aoki}}]{Knebel2020_sm}%
  \BibitemOpen
  \bibfield  {author} {\bibinfo {author} {\bibfnamefont {G.}~\bibnamefont
  {Knebel}}, \bibinfo {author} {\bibfnamefont {M.}~\bibnamefont {Kimata}},
  \bibinfo {author} {\bibfnamefont {M.}~\bibnamefont {Vali\v{s}ka}}, \bibinfo
  {author} {\bibfnamefont {F.}~\bibnamefont {Honda}}, \bibinfo {author}
  {\bibfnamefont {D.}~\bibnamefont {Li}}, \bibinfo {author} {\bibfnamefont
  {D.}~\bibnamefont {Braithwaite}}, \bibinfo {author} {\bibfnamefont
  {G.}~\bibnamefont {Lapertot}}, \bibinfo {author} {\bibfnamefont
  {W.}~\bibnamefont {Knafo}}, \bibinfo {author} {\bibfnamefont
  {A.}~\bibnamefont {Pourret}}, \bibinfo {author} {\bibfnamefont {Y.~J.}\
  \bibnamefont {Sato}}, \bibinfo {author} {\bibfnamefont {Y.}~\bibnamefont
  {Shimizu}}, \bibinfo {author} {\bibfnamefont {T.}~\bibnamefont {Kihara}},
  \bibinfo {author} {\bibfnamefont {J.-P.}\ \bibnamefont {Brison}}, \bibinfo
  {author} {\bibfnamefont {J.}~\bibnamefont {Flouquet}}, \ and\ \bibinfo
  {author} {\bibfnamefont {D.}~\bibnamefont {Aoki}},\ }\href@noop {} {\bibfield
   {journal} {\bibinfo  {journal} {J. Phys. Soc. Jpn.}\ }\textbf {\bibinfo
  {volume} {89}},\ \bibinfo {pages} {053707} (\bibinfo {year}
  {2020})}\BibitemShut {NoStop}%
\bibitem [{\citenamefont {Vali{\v{s}}ka}\ \emph {et~al.}(2021)\citenamefont
  {Vali{\v{s}}ka}, \citenamefont {Knafo}, \citenamefont {Knebel}, \citenamefont
  {Lapertot}, \citenamefont {Aoki},\ and\ \citenamefont
  {Braithwaite}}]{Valiska2021_sm}%
  \BibitemOpen
  \bibfield  {author} {\bibinfo {author} {\bibfnamefont {M.}~\bibnamefont
  {Vali{\v{s}}ka}}, \bibinfo {author} {\bibfnamefont {W.}~\bibnamefont
  {Knafo}}, \bibinfo {author} {\bibfnamefont {G.}~\bibnamefont {Knebel}},
  \bibinfo {author} {\bibfnamefont {G.}~\bibnamefont {Lapertot}}, \bibinfo
  {author} {\bibfnamefont {D.}~\bibnamefont {Aoki}}, \ and\ \bibinfo {author}
  {\bibfnamefont {D.}~\bibnamefont {Braithwaite}},\ }\href@noop {} {\bibfield
  {journal} {\bibinfo  {journal} {Phys. Rev. B}\ }\textbf {\bibinfo {volume}
  {104}},\ \bibinfo {pages} {214507} (\bibinfo {year} {2021})}\BibitemShut
  {NoStop}%
\bibitem [{\citenamefont {Li}\ \emph {et~al.}(2021)\citenamefont {Li},
  \citenamefont {Nakamura}, \citenamefont {Honda}, \citenamefont {Sato},
  \citenamefont {Homma}, \citenamefont {Shimizu}, \citenamefont {Ishizuka},
  \citenamefont {Yanase}, \citenamefont {Knebel}, \citenamefont {Flouquet},\
  and\ \citenamefont {Aoki}}]{Li2021_sm}%
  \BibitemOpen
  \bibfield  {author} {\bibinfo {author} {\bibfnamefont {D.}~\bibnamefont
  {Li}}, \bibinfo {author} {\bibfnamefont {A.}~\bibnamefont {Nakamura}},
  \bibinfo {author} {\bibfnamefont {F.}~\bibnamefont {Honda}}, \bibinfo
  {author} {\bibfnamefont {Y.~J.}\ \bibnamefont {Sato}}, \bibinfo {author}
  {\bibfnamefont {Y.}~\bibnamefont {Homma}}, \bibinfo {author} {\bibfnamefont
  {Y.}~\bibnamefont {Shimizu}}, \bibinfo {author} {\bibfnamefont
  {J.}~\bibnamefont {Ishizuka}}, \bibinfo {author} {\bibfnamefont
  {Y.}~\bibnamefont {Yanase}}, \bibinfo {author} {\bibfnamefont
  {G.}~\bibnamefont {Knebel}}, \bibinfo {author} {\bibfnamefont
  {J.}~\bibnamefont {Flouquet}}, \ and\ \bibinfo {author} {\bibfnamefont
  {D.}~\bibnamefont {Aoki}},\ }\href {\doibase 10.7566/JPSJ.90.073703}
  {\bibfield  {journal} {\bibinfo  {journal} {J. Phys. Soc. Jpn.}\ }\textbf
  {\bibinfo {volume} {90}},\ \bibinfo {pages} {073703} (\bibinfo {year}
  {2021})}\BibitemShut {NoStop}%
\end{thebibliography}

%merlin.mbs apsrev4-1.bst 2010-07-25 4.21a (PWD, AO, DPC) hacked
%Control: key (0)
%Control: author (8) initials jnrlst
%Control: editor formatted (1) identically to author
%Control: production of article title (-1) disabled
%Control: page (0) single
%Control: year (1) truncated
%Control: production of eprint (0) enabled
%

\end{document}